\documentclass[aps,onecolumn]{revtex4-1}
\usepackage{eurosym}
\usepackage{amsfonts}
\usepackage{amsmath}
\usepackage{amsopn}
\usepackage{amssymb}
\usepackage{graphicx}
\usepackage{graphics}
\usepackage{color}
\usepackage{verbatim}

\begin{document}

\title{Vortex solitons: Old results and new perspectives}
\author{Boris A. Malomed}
\affiliation{Department of Physical Electronics, School of Electrical Engineering,
Faculty of Engineering, and Center for Light-Matter Interaction, Tel Aviv
University, Tel Aviv 69978, Israel}

\begin{abstract}
A comparative review is given of some well-known and some recent results
obtained in studies of two- and three-dimensional (2D and 3D) solitons, with
emphasis on states carrying embedded vorticity. Physical realizations of
multidimensional solitons in atomic Bose-Einstein condensates (BECs) and
nonlinear optics are briefly discussed too. Unlike 1D solitons, which are
typically stable, 2D and 3D ones are vulnerable to instabilities induced by
the occurrence of the critical and supercritical collapse, respectively, in
the 2D and 3D models with the cubic self-focusing nonlinearity. Vortex
solitons are subject to a still stronger splitting instability. For this
reason, a central problem is looking for physical settings in which 2D and
3D solitons may be stabilized. The review addresses in detail two
well-established topics, \textit{viz}., the stabilization of vortex solitons
by means of competing nonlinearities, or by trapping potentials
(harmonic-oscillator and spatially-periodic ones). The former topic includes
a new addition, closely related to the recent breakthrough, \textit{viz}.,
the prediction and creation of robust \textit{quantum droplets}. Two other
topics included in the review outline new schemes which were recently
elaborated for the creation of stable vortical solitons in BEC. One scheme
relies on the use of the spin-orbit coupling (SOC) in binary condensates
with cubic intrinsic attraction, making it possible to predict stable 2D and
3D solitons, which couple or mix components with vorticities $S=0$ and $\pm
1 $ (\textit{semi-vortices} (SVs) or \textit{mixed modes} (MMs),
respectively). In this system, the situation is drastically different in the
2D and 3D geometries. In 2D, the SOC helps to create a ground state (GS,
which does not exist otherwise), represented by stable SV or MM solitons,
whose norm falls below the threshold value at which the critical collapse
sets in. In the 3D geometry, the supercritical collapse does not allow one
to create a GS, but metastable solitons of the SV and MM types can be
constructed. Another new scheme makes it possible to create stable 2D\
vortex-ring solitons with arbitrarily high $S$ in a binary BEC with
components coupled by microwave radiation. Some other topics are addressed
briefly, such as vortex solitons in dissipative media, and attempts to
create vortex solitons in experiments.
\end{abstract}

\maketitle

\textbf{List of acronyms}: 1D -- one-dimensional; 2D -- two dimensional; 3D
-- three-dimensional; BEC -- Bose-Einstein condensate; CGLE -- complex
Ginzburg-Landau equation; CQ -- cubic-quintic (nonlinearity), FF --
fundamental frequency; GPE -- Gross-Pitaevskii equation; GS -- ground state;
GVD -- group-velocity dispersion; HO -- harmonic oscillator (potential); HV
-- hidden vorticity; LHY -- Lee-Huang-Yang (correction to the mean-field
dynamics of BEC); MF -- mean field; MM -- mixed mode; NLSE -- nonlinear Schr%
\"{o}dinger equation; OL -- optical lattice; PT -- parity-time (symmetry);
QD -- quantum droplet; SH -- second harmonic; SOC -- spin-orbit coupling; SV
-- semi-vortex; TF -- Thomas-Fermi (approximation); TS -- Townes' soliton;
VA -- variational approximation; VAV -- vortex-antivortex (composite state in
a two-component system); VK -- Vakhitov-Kolokolov (stability
criterion)

\section{Introduction: The topic and objectives of the article}

\subsection{The concept of vortex solitons}

Since term \textquotedblleft soliton" was coined by Zabuski and Kruskal \cite%
{ZK}, it is commonly applied to localized self-trapped modes which
spontaneously emerge in diverse media as a result of the balance between the
diffraction and/or material dispersion affecting the propagation of linear
waves, and self-focusing (self-attractive) material nonlinearity.
Theoretical and experimental studies of solitons have grown into a huge
research area penetrating many disciplines, in physics, mathematics, and
beyond. The largest share of these works have been dealing with
one-dimensional (1D) solitons \cite{KA,Peyrard}, the dominant areas in
current studies of solitons being nonlinear optics \cite{KA} and
Bose-Einstein condensates (BECs) in ultracold atomic gases \cite%
{Brazh,BEC-sol1,Morsch,PhysicaD,BEC-sol2,BEC-sol3}. The extension of the
concept of solitons to two- and three-dimensional (2D and 3D) geometry is a
highly nontrivial generalization (multidimensional optical solitons, which
are self-trapped (localized) both in the spatial and temporal directions,
are often called spatiotemporal solitons \cite{old-review}, alias
\textquotedblleft light bullets" \cite{Silb}). A well-known problem impeding
the studies of multidimensional solitons is that, while a majority of 1D
solitons, such as ones described by the Korteweg - de Vries \cite{Kruskal},
nonlinear Schr\"{o}dinger equation \cite{ZS}, sine-Gordon \cite{AKNS}, and
Landau-Lifshitz \cite{Borovik,Mikhailov,Rodin} equations, may be produced as
solutions of integrable \cite{Zakharov,Newell,Ablowitz} or nearly integrable
\cite{1989} models, only few 2D equations are integrable (most essential
among them are Kadomtsev-Petviashvili equations of two types \cite{Dryuma},
with opposite signs of the dispersion, only one of them, the so-called KP-I
equation, giving rise to 2D solitons), and there are no integrable 3D
equations which would find any realization in physics. The lack of (nearly)
integrable models obviously makes the theoretical study of multidimensional
solitons more difficult. Another fundamental issue is that the most common
cubic self-focusing nonlinearity tends to create 2D and 3D solitons which
are subject to instability driven by the \textit{wave collapse} (alias
\textit{blowup}) that occurs in the same equations, i.e., spontaneous
formation of singularities, driven by the self-focusing effect \cite%
{Askaryan,Townes}, after a finite evolution time \cite{Berge,SulemSulem,Gadi}%
. The collapse induced by the cubic nonlinearity is \textit{critical} in 2D,
which means that it sets in when the norm of the underlying wave field
exceeds a certain finite critical value, and \textit{supercritical} in 3D,
where an initial state with an arbitrarily small norm may blow up due to the
collapse. A clear experimental demonstration of the collapse of an optical
\textit{vortex beam} (in water), which is accompanied by spontaneous
breaking of the azimuthal symmetry of the collapsing beam, and is obviously
relevant to the topic under the consideration, was reported in Ref. \cite%
{Gaeta}).

On the other hand, solitons created in 2D and 3D geometries offer a great
variety of new possibilities. Most significant, vorticity may be embedded in
both 2D and 3D localized states, thus creating vortex solitons, which is the
main subject of the present review. The intrinsic vorticity is characterized
by an integer winding number, alias topological charge, $S$
(\textquotedblleft spin"), which is defined through a total change of the
phase, $\Delta \varphi =2\pi S$, accumulated in the course of a trip along a
closed trajectory surrounding the vortex' pivot (phase singularity), placed
at $r=0$, where $r$ is the radial coordinate. The presence of the vorticity
implies that, similar to the commonly known asymptotic structure of the
Bessel functions, $J_{|S|}(r)$, the local amplitude of the wave field(s) in
vortex solitons must vanish $\sim r^{|S|}$ at $r\rightarrow 0$ (here $|S|$
is written as integer $S$ may be negative). Due to this peculiarity, the
vortex soliton features a \textquotedblleft hole" in the center, i.e., the
localized mode is shaped as a ring (or, quite frequently, as a broad
annulus) in 2D, and a torus (\textquotedblleft donut") in 3D. Nevertheless,
an exception is known: in the 2D model based on the Gross-Pitaevskii
equation (GPE) for a BEC with the quintic self-defocusing interaction and an
external potential $U=-U_{0}r^{-2}$, $U_{0}>0$, which pulls atoms to the
center, stable vortex states are possible in which the amplitude does not
vanish at $r\rightarrow 0$, but instead \emph{diverges} $\sim r^{-1/2}$ \cite%
{pull}. This states is a result of the balance between the pull to the
center and quintic self-repulsion. Indeed, the absolute values of the wave
function in a vortex state with winding number $S\neq 0$ cannot be finite at
$r=0$, as its phase, $\varphi =S\theta $ (where $\theta $ is the angular
coordinate) is not defined at $r=0$. The usual solution of this problem is
resorting to solutions with the amplitude vanishing at $r=0$. However, an
alternative solution is possible too, with the amplitude \emph{diverging} at
$r\rightarrow 0$, as in Ref. \cite{pull}. Indeed, it is commonly known that
the Bessel equation, which produces the asymptotic form $\sim r^{|S|}$ of
the solution for the vortex' amplitude, also has the singular solution
(given by the Neumann function), $\sim r^{-|S|}$. In the 2D setting, the
latter solutions with any $|S|\geq 1$ is unphysical, as it gives rise to a
divergent norm. Nonetheless, the above-mentioned specific singularity, $\sim
r^{-1/2}$, which is produced by the interplay of the vortex structure with
the pulling potential and quintic self-defocusing, is \emph{integrable} (the
corresponding\ integral norm converges at $r\rightarrow 0$), hence this
solution is a physically relevant one.

In 3D, solitons with more sophisticated topological structures are known
too, in the form of \textit{skyrmions }\cite%
{skyrmion1,skyrmion2,skyrmion3,semi1,semi2,VPG-skyrmion}, \textit{hopfions},
\cite{hopfion,Yasha}, \textit{knots }\cite{super,Sutcliffe,Radu}, etc.,
which may carry two independent topological charges. In particular, hopfions
may be realized in terms of a single complex wave field, which forms a
twisted vortical torus, with the phase of the wave function winding both
along the torus-forming ring, which determines the overall vorticity of the
hopfion, and along the ring in the torus' cross section, which determines
the intrinsic twist of the hopfion \cite{Yasha}. In addition to physics of
nuclear matter and the classical field theory, which are the origin of the
Skyrme model and diverse 3D states generated by it \cite%
{skyrmion1,skyrmion2,skyrmion3,Sutcliffe,Radu,field-theory}, these complex
3D modes find important realizations in ferromagnets \cite%
{ferro,ferro-Sutcliffe}, semiconductors \cite{semi1,semi2}, superconductors
\cite{super}, and in various configurations of BEC.

In comparison with the fundamental (zero-vorticity) solitons in the same
geometries, the creation of 2D and 3D vortex rings and tori is a still more
challenging objective, because, in addition to the above-mentioned
collapse-driven instability, they are subject to an even stronger azimuthal
instability, which tends to break the axially symmetric ring or torus into
fragments, each one being, roughly speaking, a fundamental soliton, see
illustration below in Fig. \ref{splitting} \cite{Minsk1,Minsk2}. Therefore,
theoretical prediction and experimental realization of physically relevant
settings that may support stable fundamental and, especially, vortex
solitons in 2D and 3D geometries is a challenging objective, which has drawn
much interest in the course of the two last decades. It was the subject of
several reviews, which were published both relatively long ago and more
recently, being chiefly focused on selected aspects of the broad topic \cite%
{old-review,Dum,me,NatureRev}.

The present article aims to provide a review specifically focused on vortex
solitons. Because this theme is a vast one too, the review addresses some
selected aspects in detail, and others in a brief form, as a comprehensive
review might easily grow to the size of a book. Two first topics selected
for the presentation are well-established ones. They rely on the
stabilization of vortex states by competing nonlinearities \cite%
{Manolo,Pego,nine,CQ2comp2D,FWM,S=1and2chi2chi32D,S=3and4chi2chi32D,chi2chi33D}%
, or in trapping potentials \cite{2D,Sadhan,Ueda,Ueda2,Dum2D,Dum3D} (3D and
2D harmonic-oscillator (HO) potentials are considered in detail, and
spatially periodic lattices, which were a subject of several earlier
published reviews, are addressed in a brief form). The former topic includes
a new addition, \textit{viz}., the consideration of the recently created
\cite{Tarr1}-\cite{Ing2} \textit{quantum droplets} (QDs). Models of QDs make
it possible to predict robust vortex solitons with the help of specific
competing nonlinearities \cite{3DLHY,GZ-ln,necklace,HS}, as well as of
necklace-shaped circular chains of QDs, which may carry the angular momentum
\cite{necklace} (unlike fundamental QDs, creation of such modes has not yet
been reported).

Two other topics in the review present entirely recent developments. These
are schemes for the creation of stable 2D and 3D solitons in models of
binary BEC, with its two components interacting by means of the spin-orbit
coupling (SOC) \cite{we,Sherman2,HP}, or through a resonant microwave field
\cite{Jieli2}. The latter section also briefly outlines results for vortex
solitons obtained in other models with effectively nonlocal nonlinearities.
The section of the article preceding the concluding one touches upon some
other topics which are relevant to the broad theme of vortex solitons, such
as settings with spatially modulated local nonlinearities, discrete media,
and, in a somewhat more detailed form, nonlinear dissipative and $\mathcal{PT%
}$-symmetric systems. The same section also briefly mentions the state of
the art as concerns experimental creation of vortex solitons (thus far, they
were only observed as transient states, in very specific settings).

\subsection{A related field: \textquotedblleft dark" vortices supported by a
finite background}

A broad topic related to bright vortex solitons, i.e., localized states with
embedded vorticity, is the prediction and creation of solitons in media with
self-repulsive nonlinearities. These are 2D (and 3D) vortex states, embedded
in flat modulationally stable background fields. These states may be
considered as 2D counterparts of dark solitons, which are well-known
solutions of the integrable 1D nonlinear Schr\"{o}dinger equation \cite{ZS}.
In fact, these delocalized vortex modes are most frequently called
\textquotedblleft vortices", in the context of optics and BEC alike. They
are well-known objects, studied in detail theoretically and created in
experiments, prior to the work on bright vortex solitons. Results obtained
for the \textquotedblleft dark vortices", starting from the earliest ones
\cite{Coullet,Neu,Swartz1}, have been collected in many publications dealing
with various settings in photonics \cite%
{Swartz2,Soskin,vort-review-early,Vyslo,Bessel-beam-DiTrapani,vort-in-liq-cryst,vort-in-communications}
(including exciton-polariton condensates \cite%
{polaritons1,polaritons,polaritons2}, where the vortices exist in
dissipative media) and BEC \cite{Aftalion,
Cornell-vortex,Cornell2,Cornell3,Hall,TsuKasa,BEC-exper,Fetter}, as well as
in quantum Fermi gases \cite{Fermi,Fermi-review}; see also book \cite{black}
for an overview of the theme. The studies of vortices in this context are
also known as \textquotedblleft singular optics", with pivots of the
vortices considered as singularities of optical fields; many publications on
the latter theme were collected in a special issue of Journal of Optics \cite%
{Editorial} (see also book \cite{Gbur}).

It is relevant to mention too that vortices are closely related to the
concept of the optical angular momentum \cite%
{Allen,opt-angular-mom,Bliokh-Nori}, and the possibility of storage of the
angular momentum in ultracold gases \cite{angular-momentum-storage}.

As concerns the stability of the \textquotedblleft dark vortices", in the 2D
geometry the stability of the vortex with $S=1$ is secured by the
conservation of the associated angular momentum, while multiple vortices,
with $S\geq 2$, are usually unstable against splitting into $S$ unitary
vortices \cite{Neu}. The instability is qualitatively explained by the fact
that two separated vortices with equal topological charges repel each other.
In the 3D geometry, even the vortex with $S=1$ may be unstable against
spontaneous bending of its pivotal line, as predicted theoretically \cite%
{Svidzinsky,bending,bending2} and observed experimentally \cite%
{Dalibard-bending,Bagnato-bending}.

Lastly, it is relevant to mention a 2D two-component system which gives rise
to stable bound states with a zero-vorticity component filling the central
\textquotedblleft hole" of the dark-vortex structure induced in the other
component \cite{Berloff}.

\section{Earlier results and recent additions to them: competing
nonlinearities and trapping potentials}

\subsection{Unstable vortex solitons in media with quadratic and saturable
nonlinearities}

As said above, the first objective in the studies of multidimensional
solitons is securing their stability. One possibility is to consider media
with collapse-free nonlinearities. In optics, they may be represented by
quadratic (alias $\chi ^{(2)}$, or second-harmonic-generating) terms \cite%
{chi1,chi2,chi3,chi4}, which do not lead to collapse in 2D and 3D geometries
\cite{Rubi,HaoHe}. This circumstance stimulated the experimental creation of
fundamental spatiotemporal solitons in a quasi-2D $\chi ^{(2)}$ setting (the
localization in the third direction was provided by a waveguiding structure,
which was necessary, as this direction was \textquotedblleft sacrificed" for
inducing sufficiently strong group-velocity dispersion (GVD), that was
necessary for self-trapping in the temporal direction) \cite{Frank1,Frank2}.
There are two basic forms of the quadratic nonlinearity, which are usually
referred to as \textquotedblleft Types I and II". They correspond,
respectively, to the degenerate two-wave system and full three-wave one. The
latter system is modeled by three $\chi ^{(2)}$-coupled equations for the
paraxial propagation of two components $u_{1,2}^{\mathrm{(FF)}}$ of the
fundamental-frequency (FF) wave, which represent two mutually orthogonal
polarizations of light, and amplitude $u^{\mathrm{(SH)}}$ of the
second-harmonic (SH) wave \cite{chi1,chi2,chi3,chi4}:%
\begin{gather}
i\frac{\partial }{\partial z}u_{j}^{\mathrm{(FF)}}+\left( -1\right) ^{j}%
\frac{b}{2}u_{j}^{\mathrm{(FF)}}  \notag \\
+\frac{1}{2}\left( \frac{\partial ^{2}}{\partial x^{2}}+\frac{\partial }{%
\partial y^{2}}+D_{\mathrm{FF}}\frac{\partial ^{2}}{\partial \tau ^{2}}%
\right) u_{j}^{\mathrm{(FF)}}+u^{\mathrm{(SH)}}\left( u_{3-j}^{\mathrm{(FH)}%
}\right) ^{\ast }=0,  \label{FF} \\
2i\left( \frac{\partial }{\partial z}+ic\frac{\partial }{\partial \tau }%
\right) u^{\mathrm{(SH)}}-qu^{\mathrm{(SH)}}  \notag \\
+\frac{1}{2}\left( \frac{\partial ^{2}}{\partial x^{2}}+\frac{\partial }{%
\partial y^{2}}+D_{\mathrm{SH}}\frac{\partial ^{2}}{\partial \tau ^{2}}%
\right) u^{\mathrm{(SH)}}+u_{1}^{\mathrm{(FF)}}u_{2}^{\mathrm{(FF)}}=0,
\label{SH}
\end{gather}%
with $j=1,2$, where $\ast $ stands for the complex conjugate, $z$ is the
propagation distance, the paraxial-diffraction operator $(1/2)\left(
\partial ^{2}/\partial x^{2}+\partial ^{2}/\partial y^{2}\right) $ acts on
functions of transverse coordinates $\left( x,y\right) $,
\begin{equation}
\tau \equiv t-z/V_{\mathrm{gr}}  \label{tau}
\end{equation}%
is the reduced time, which combines the usual temporal variable, $t$, and
the propagation distance, $V_{\mathrm{gr}}$ being the group velocity of the
FF's carrier wave \cite{KA}, and walkoff coefficient $c$ is proportional to
the mismatch between group velocities of the SH and FF waves. Further, real $%
b$ accounts for the phase birefringence of the SH components, $D_{\mathrm{%
FF,SH}}$ are group-velocity-dispersion (GVD) coefficients of the FF and SH
waves, and real $q$ is the phase mismatch between them, while the
diffraction and $\chi ^{(2)}$ coefficients are scaled to be $1$. The system
creates bright spatiotemporal solitons in the case of the anomalous GVD,
i.e., $D_{\mathrm{FF,SH}}>0$. The degenerate Type-I system is obtained from
Eqs. (\ref{FF}) and (\ref{SH}) by setting
\begin{equation}
b=0,u_{1}^{\mathrm{(FF)}}=u_{2}^{\mathrm{(FF)}}\equiv u^{\mathrm{(FF)}}/%
\sqrt{2}.  \label{b=0}
\end{equation}

Further analysis has demonstrated that the $\chi ^{(2)}$ nonlinearity of
either type (I or II), while readily maintaining stable fundamental 2D and
3D solitons, cannot stabilize solitons with embedded vorticities $S_{1,2}^{%
\mathrm{(FF)}}$ and $S^{\mathrm{(SH)}}=S_{1}^{\mathrm{(FF)}}+S_{2}^{\mathrm{%
(FF)}}$ of the FF and SH components against splitting, even in the 2D
settings, for any nonzero values of $S_{1,2}^{\mathrm{(FF)}}$ (in the
two-wave system, one has the single value of $S^{\mathrm{(FF)}}$ and,
accordingly, $S^{\mathrm{(SH)}}=2S^{\mathrm{(FF)}}$). This negative result
was predicted theoretically for the two-wave \cite%
{Dima1,Petrov1,Petrov2,Dima2} and three-wave \cite{3wave,Herve-3waves}
systems in 2D (which implies the consideration of $\tau $-independent
solutions of Eqs. (\ref{FF}) and (\ref{SH})), and demonstrated
experimentally for the former one \cite{Petrov}. The instability of bright
vortex states makes it possible to realize experimentally their controllable
splitting in a set of fundamental solitons, by means of interaction with an
additional seed beam, which was taken too as a vortical one \cite{stimulated}%
. Nevertheless, in some cases the instability may be very weak \cite%
{Herve-3waves}, which may allow one to produce quasi-stable vortex solitons.

Another option is to resort to nonlinearities which impose saturation on the
growth of the cubic term at large intensities of the wave field(s). The
simplest realization is represented by the following generalized nonlinear
Schr\"{o}dinger equation (NLSE), written in terms of the propagation of an
optical wave, with complex amplitude $u\left( z,y;z\right) $ in a bulk
waveguide \cite{Rypdal}:%
\begin{equation}
i\frac{\partial u}{\partial z}+\frac{1}{2}\left( \frac{\partial ^{2}}{%
\partial x^{2}}+\frac{\partial }{\partial y^{2}}\right) u+\frac{|u|^{2}u}{%
1+|u|^{2}/U_{0}^{2}}=0,  \label{satur}
\end{equation}%
where positive constant $U_{0}^{2}$ characterizes the saturation intensity.
In fact, one may set $U_{0}=1$, by means of rescaling $u\equiv U_{0}\tilde{u}
$, $\left( x,y\right) \equiv \left( \tilde{x},\tilde{y}\right) /U_{0}$, $%
z\equiv \tilde{z}/U_{0}^{2}$, hence Eq. (\ref{satur}) may be written in the
parameter-free form. The nonlinear term in this equation adequately models,
in particular, optical properties of warm atomic gases \cite{Tikho}. The
temporal variable does not appear in Eq. (\ref{satur}), as it governs the
evolution of the monochromatic wave in the \textit{spatial domain} \cite{KA}.

A general form of localized vortex-soliton solutions to Eq. (\ref{satur}) is%
\begin{equation}
u\left( x,y,z\right) =\exp \left( ikz+iS\theta \right) U(r),  \label{vortex}
\end{equation}%
where $\left( r,\theta \right) $ are polar coordinates in the $\left(
x,y\right) $ plane, $k>0$ is a real propagation constant, which is a free
parameter of the vortex-soliton family, integer $S$ is the vorticity
(winding number), and real amplitude $U(r)$ obeys the ordinary differential
equation,%
\begin{equation}
\frac{d^{2}U}{dr^{2}}+\frac{1}{r}\frac{dU}{dr}-\frac{S^{2}}{r^{2}}U+\frac{%
2U^{3}}{1+U^{2}}=2kU,  \label{ODE}
\end{equation}%
supplemented by the above-mentioned boundary condition at $r\rightarrow 0$,%
\begin{equation}
U\sim r^{|S|},  \label{r-->0}
\end{equation}%
and subject to the condition of the exponential localization at $%
r\rightarrow \infty $:%
\begin{equation}
U\sim r^{-1/2}\exp \left( -\sqrt{2k}r\right)  \label{r-->infty}
\end{equation}%
(pre-exponential factor $r^{-1/2}$ in Eq. (\ref{r-->infty}) is essentially
the same as in the standard asymptotic approximation for cylindrical
functions at $r\rightarrow \infty $).

However, both the experiment \cite{Tikho} and numerical analysis \cite%
{Tikho,Dima1} have demonstrated that all the vortex solitons generated by
Eqs. (\ref{vortex})-(\ref{r-->infty}) are unstable. In particular, the
simplest one, with $S=1$, spontaneously splits in two fragments, which are
close to fundamental (zero-vorticity) solitons.

\subsection{The stabilization of two- and three-dimensional vortex solitons
by cubic-quintic and other competing nonlinearities}

\subsubsection{Formulation of the model}

The NLSE of the cubic-quintic (CQ) type appears if the saturable
nonlinearity in Eq. (\ref{satur}) is replaced by its truncated expansion:%
\begin{equation}
i\frac{\partial u}{\partial z}+\frac{1}{2}\frac{\partial ^{2}u}{\partial
\tau ^{2}}+\frac{1}{2}\left( \frac{\partial ^{2}}{\partial x^{2}}+\frac{%
\partial }{\partial y^{2}}\right) u+|u|^{2}u-|u|^{4}u=0.  \label{CQ}
\end{equation}%
Equation (\ref{CQ}) is written in the spatiotemporal \ form, including the
GVD term (its sign corresponds to the anomalous dispersion, because normal
dispersion cannot support bright spatiotemporal solitons \cite{KA}), $\tau $
being the same reduced time as in Eqs. (\ref{FF}) and (\ref{SH}). All free
coefficients are scaled out in the normalized form of Eq. (\ref{CQ}).

The CQ terms, as written in Eq. (\ref{CQ}), adequately model the nonlinear
response in specific optical media. In particular, the use of this
nonlinearity in a bulk waveguide filled by liquid carbon disulfide has made
it possible to create robust fundamental (zero-vorticity) 2D spatial
solitons \cite{Cid-robust}. Further, a recently developed technique makes it
possible to design desirable coefficients of cubic, quintic, and
higher-order nonlinearities in colloidal suspensions of metallic
nanoparticles \cite{Cid-review}. In particular, stable 2D fundamental
solitons were created in a medium with an effective quintic-septimal
nonlinearity, where the usual cubic term was negligible \cite{Cid2}.

The 1D reduction of Eq. (\ref{CQ}), for solutions which do not depend on $x$
and $y$, admits well-known exact soliton solutions \cite{Pushkarovs,Enns},%
\begin{equation}
u_{\mathrm{sol}}(z,\tau )=2e^{ikz}\sqrt{\frac{k}{1+\sqrt{1-16k/3}\cosh
\left( 2\sqrt{2k}\tau \right) }},  \label{Pushk}
\end{equation}%
which exist and are completely stable in interval%
\begin{equation}
0<k<3/16.  \label{3/16}
\end{equation}%
3D vortex-soliton solutions are produced by the substitution of a
generalization of ansatz (\ref{vortex}),%
\begin{equation}
u\left( x,y,z,\tau \right) =\exp \left( ikz+iS\theta \right) U(r,\tau ),
\label{3Dvortex}
\end{equation}%
in Eq. (\ref{CQ}), where $U(r,\tau )$ is a real even function of $\tau $
obeying an accordingly modified version of Eq. (\ref{ODE}):%
\begin{equation}
\frac{\partial ^{2}U}{\partial r^{2}}+\frac{1}{r}\frac{\partial U}{\partial r%
}-\frac{S^{2}}{r^{2}}U+\frac{\partial ^{2}U}{\partial \tau ^{2}}%
+2U^{3}-2U^{5}=2kU.  \label{U}
\end{equation}%
In the 3D case, boundary conditions (\ref{r-->0}) and (\ref{r-->infty}) are
supplemented by the condition of the exponential localization of the
solution at $|\tau |\rightarrow \infty $: $U\sim \exp \left( -\sqrt{2k}|\tau
|\right) $, while being subject at $r\rightarrow 0$ to the same condition (%
\ref{r-->0}) as in 2D.

\subsubsection{The shape and stability of 2D vortex solitons}

Vortex solitons in the 2D version of the NLSE with the CQ nonlinearity were
introduced in work \cite{Drits}. The existence of stable vortex solitons
with $S=1$ in this model was revealed by work \cite{Manolo}.

Numerical solution of Eq. (\ref{U}) without the temporal derivative produces
vortex-soliton\ radial profiles with the propagation constant belonging to
the same interval (\ref{3/16}) as in the 1D case, because Eq. (\ref{U})
takes an asymptotically 1D form at $r\rightarrow \infty $. In the limit of $%
k\rightarrow 3/16$, the 2D solitons feature a \textquotedblleft flat-top"
shape in a circular area of large radius,
\begin{equation}
R\approx \left( 1/\sqrt{6}\right) \ln \left( \left( 3/16-k\right)
^{-1}\right) .  \label{R}
\end{equation}%
It is filled by the solution with a nearly constant amplitude, close to the
largest possible value, $U_{\max }=\sqrt{3}/2$ (it is a constant solution of
Eq. (\ref{U}) for $k=3/16$), with a \textquotedblleft hole" of radius $r_{0}$
produced by the vorticity around the pivot (placed at $r=0$), see Figs. 1(a)
in \cite{Manolo} and 4-6 in \cite{Pego}. The hole's radius may be estimated
by means of the Thomas-Fermi (TF) approximation, which neglects derivatives
in Eq. (\ref{U}), thus yielding two solutions: $U=0$, and one produced by
equation
\begin{equation}
U^{4}-U^{2}+\left( k+\frac{S^{2}}{2r^{2}}\right) =0.  \label{TF}
\end{equation}%
Obviously, relevant roots of Eq. (\ref{TF}), with real $U^{2}>0$, do not
exist at%
\begin{equation}
r<\left( r_{0}\right) _{\mathrm{TF}}=\frac{S}{\sqrt{1/2-2k}},  \label{r0}
\end{equation}%
which determines the radius of the vorticity-drilled \textquotedblleft hole"
in the TF approximation (at $r<r_{0}$, this approximation admits solely the $%
U=0$ solution). For $k\rightarrow 3/16$, Eq. (\ref{r0}) yields $r_{0}\approx
2\sqrt{2}S$, which is indeed much smaller than the outer radius (\ref{TF})
of the flat-top-shaped soliton. The TF approximation can be applied to
vortex solutions in other contexts too \cite{Fetter}.

The stability of the 2D vortex solitons was for the first time addressed, in
the framework of Eq. (\ref{CQ}) (without the time-derivative term) in work
\cite{Manolo} by means of direct simulations. It was found that a part of
the vortex-soliton family with $S=1$ is stable, and, furthermore, such
solitons, if set in motion, may collide quasi-elastically. In a rigorous
form, the stability of the vortex-soliton solutions was investigated in work
\cite{Pego}, through numerical computation of stability eigenvalues for
small perturbations added to the stationary states. Thus, stability regions
in the existence interval (\ref{3/16}) were identified for all values of
winding number $S$, in the form of
\begin{equation}
k_{\min }(S)<k<3/16=0.1875  \label{kmin}
\end{equation}%
(all fundamental solitons with $S=0$ are stable). The numerically found
stability boundaries are \cite{Pego}:%
\begin{equation}
k_{\min }(1)=0.1487,k_{\min }(2)=0.1619,k_{\min }(3)=0.1700,k_{\min
}(4)=0.1769,k_{\min }(5)=0.1806.  \label{kkmin}
\end{equation}%
Thus, with the increase of the winding number from $1$ to $5$ the relative
width of the stability regions, with respect to the existence interval (\ref%
{3/16}), drops from $\approx 21\%$ to $\approx 4\%$.

The 2D version of the same NLSE (\ref{CQ}) can be realized in optics as the
equation for the spatiotemporal propagation in a planar waveguide, which
implies dropping coordinate $y$ in Eq. (\ref{CQ}), while keeping temporal
variable $\tau $. Thus one may define \textit{spatiotemporal vortices} \cite%
{spatiotemp-vort1,spatiotemp-vort2} as modes given by ansatz (\ref{vortex})
in which $\left( r,\theta \right) $ are realized as the polar coordinates in
the $\left( x,\tau \right) $ plane. The spatiotemporal vortex soliton may be
excited in the planar waveguide by an input whose amplitude and phase are
patterned according to Eq. (\ref{vortex}), taken at $z=0$ \cite%
{spatiotemp-vort1}.

Lastly, it is possible to introduce a two-dimensional NLSE/GPE with
specially selected coefficients depending on the radial coordinate, which
makes it possible to produce vortex solitons with various more sophisticated
shapes \cite{NonlinDyn,NonlinDyn2}.

\subsubsection{3D vortex solitons}

\begin{figure}[tbp]
\begin{center}
\includegraphics[height=8cm]{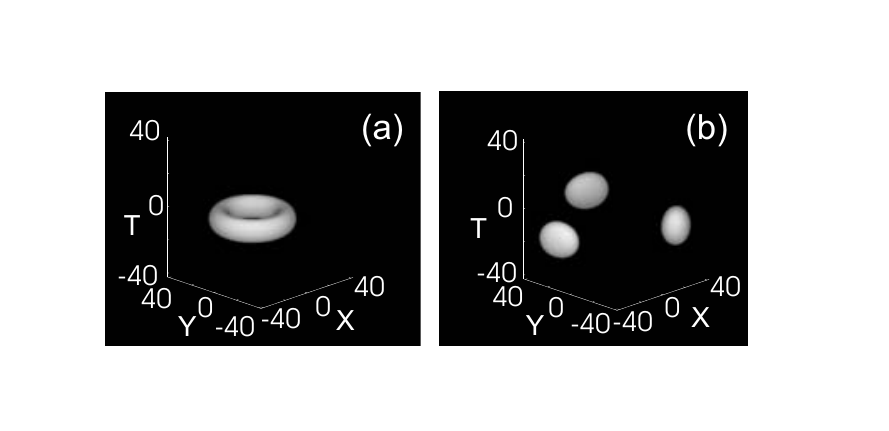}
\end{center}
\caption{Spontaneous splitting of an unstable 3D vortex torus, with winding
number $S=2$ and propagation constant $k=0.09$, in the framework of the NLSE
(\protect\ref{CQ}) with the CQ nonlinearity, as per Ref. \protect\cite{nine}%
. Panels (a) and (b) display, respectively, the intensity isosurface of the
stationary solution (the input at $z=0$), and the result of its evolution at
$z=250$. In this figure, the coordinates are related to those in Eq. (%
\protect\ref{CQ}) by $\left( X,Y,T\right) \equiv \protect\sqrt{2}\left(
x,y,t\right) $.}
\label{splitting}
\end{figure}

In the framework of 3D NLSE\ (\ref{CQ}), the stability of vortex solitons,
shaped as vortex tori, was investigated in work \cite{nine}. In this case
too, 3D solitons exist in interval (\ref{3/16}). The main numerical finding
is that they are stable in a part of this interval, defined as in Eq. (\ref%
{kmin}), with $k_{\min }^{\mathrm{(3D)}}(S=1)\approx 0.13$ \cite{nine}. No
stability interval was found for 3D vortex solitons with $S\geq 2$. Unstable
vortex tori spontaneously split into fragments, which seem as fundamental ($%
S=0$) solitons, see Fig. \ref{splitting}.

\subsubsection{Stable vortex solitons supported by competing quadratic and
cubic nonlinearities.}

As mentioned above, systems of equations (\ref{FF}) and (\ref{SH}), which
model the co-propagation of fundamental-frequency and second-harmonic waves,
coupled by $\chi ^{(2)}$ terms, produce stable solitons solely with zero
vorticity, but no stable vortex solitons. Two-component vortex solitons in
such a system (reduced to two equations by substitution (\ref{b=0})) may be
stabilized if it includes the repulsive cubic self-interaction. In 2D, this
was demonstrated for vortex solitons with $S=1$ and $2$, with the relative
size of the stability regions $\approx 8\%$ and $\approx 5\%$, respectively
\cite{S=1and2chi2chi32D}. Small stability regions, with the respective size $%
\approx 3\%$ and $\approx 1.5\%$, were found for vortex solitons with $S=3$
and $4$ as well \cite{S=3and4chi2chi32D}. Finally, 3D vortex solitons with $%
S=1$ may also be stable in the system with the competing quadratic and
self-defocusing cubic interactions, the corresponding relative size of the
stability domain being $\approx 10\%$ , while all the 3D solitons with $%
S\geq 2$ are unstable, as well as in the CQ model \cite{chi2chi33D}.

\subsubsection{Interactions between multidimensional solitons}

Interactions between stable 2D and 3D solitons have also been theoretically
investigated. In particular, an effective potential for the interaction
between far separated 2D identical solitons with chemical potential $\mu <0$%
, which carry vorticity $S$, was derived in Ref. \cite{interaction}:%
\begin{equation}
U\left( R,\delta \right) =C(-1)^{S+1}R^{-1//2}\exp \left( -\sqrt{-2\mu }%
R\right) \cos \delta ,  \label{Pot}
\end{equation}%
where $R$ and $\delta $ are the distance and phase shift between the
solitons, and $C$ is a positive constant (the interaction potential was
derived in Ref. \cite{interaction} in a more general form, which applies as
well to models based on complex Ginzburg-Landau equations (CGLEs), that
include dissipation and gain, see Eq. (\ref{GL}) below). Interactions
between vortex solitons were investigated by means of direct simulations too
\cite{SKA-int1,SKA-int2}.

\subsection{A new realization of competing nonlinearities: Stable 2D and 3D
vortex quantum droplets with embedded vorticity}

\subsubsection{Gross-Pitaevskii equations with the Lee-Huang-Yang (LHY)
corrections}

Recent theoretical and experimental works with binary BEC have revealed a
fascinating possibility of the creation of stable 3D and 2D soliton-shaped
objects, which are stabilized by the LHY corrections to the mean-field (MF)
theory, which account for the effect of quantum fluctuations around MF
states \cite{LeeHY}. An appropriate system is provided by a binary BEC with
self-repulsion in each component (in this case the LHY correction is
relevant) and the MF attraction between the components. In this setting, the
normalized GPE for identical wave functions of the two components in 3D is
\cite{Petrov-QD}%
\begin{equation}
i\frac{\partial \psi }{\partial t}=-\frac{1}{2}\nabla _{\mathrm{3D}}^{2}\psi
-|\psi |^{2}\psi +|\psi |^{3}\psi ,  \label{LHY}
\end{equation}%
where the self-focusing cubic term implies that the cross-attraction is
slightly stronger than self-repulsion, while the defocusing quartic term
represents the LHY corrections. Similar to other models with competing
attractive and repulsive nonlinearities, Eq. (\ref{LHY}) readily generates a
family of stable fundamental (zero-vorticity) soliton-like modes, which are
often called \textquotedblleft quantum droplets" (QDs), as they represent
self-trapped modes filled by an ultradilute superfluid \cite{Petrov-QD,Astra}%
. The creation of stable 3D fundamental solitons, and collisions between
moving ones, by GPE which includes both the LHY term and an additional
self-repulsive quintic one was recently addressed in Ref. \cite{SKA}.

Soon after being predicted, QDs were experimentally created in a binary BEC
of $^{39}$K atoms. The experimentally observed shape may be nearly
two-dimensional, if the condensate is strongly confined in one direction
\cite{Tarr1,Tarr2}, or isotropic, if the trapping potential is weak \cite%
{Ing,Ing2}.

The reduction of the effective dimension from 3 to 2 by means of a tightly
binding potential applied in one direction entails the replacement of the
cubic-quartic nonlinearity in Eq. (\ref{LHY}) by the single cubic term
multiplied by a logarithmic factor \cite{Astra}:%
\begin{equation}
i\frac{\partial \psi }{\partial t}=-\frac{1}{2}\nabla _{\mathrm{2D}}^{2}\psi
+\ln \left( |\psi |^{2}\right) |\psi |^{2}\psi .  \label{ln}
\end{equation}%
The Hamiltonian corresponding to this equation is%
\begin{equation}
H=\frac{{1}}{2}\int \int \left[ \left\vert \nabla _{\mathrm{2D}}\psi
\right\vert ^{2}+|\psi |^{4}\ln \left( \frac{|\psi |^{2}}{\sqrt{e}}\right) %
\right] dxdy,  \label{Hlog}
\end{equation}%
where $e=\allowbreak 2.\,\allowbreak 718...$ is the base of natural
logarithms.

\subsubsection{Vortex solitons in the 2D setting}

Obviously, the nonlinearity in Eq. (\ref{ln}) is attractive at small values
of the normalized density, $|\psi |^{2}<1$, for which the logarithm is
negative, and it reverses its sign to repulsive at $|\psi |^{2}>1$, i.e.,
this nonlinearity seems as a competing one. In particular, similar to the
above-mentioned model with the CQ nonlinearity, Eq. (\ref{ln}) gives rise to
broad 2D solitons with the flat-top shape, featuring a nearly constant value
of the density in its intrinsic area determined by minimization of the
energy density, as defined by Eq. (\ref{Hlog}): $n_{\mathrm{flat-top}}=1/%
\sqrt{e}\approx 0.6065$. In this case, a crude approximation for the radius
of the inner \textquotedblleft hole" created by the embedded vorticity is
provided by the TF approximation, which neglects term $\nabla _{\mathrm{2D}%
}^{2}\psi $ in Eq. (\ref{ln}): $\left( r_{0}\right) _{\mathrm{TF}}=\sqrt{%
e/(2-\sqrt{e})}S\approx 2.8S$, cf. Eq. (\ref{r0}). Comparison with numerical
findings demonstrates that this approximation produces reasonable results
for $1\leq S\leq 4$ \cite{GZ-ln}.

\begin{figure}[t]
{\includegraphics[width=0.8\columnwidth]{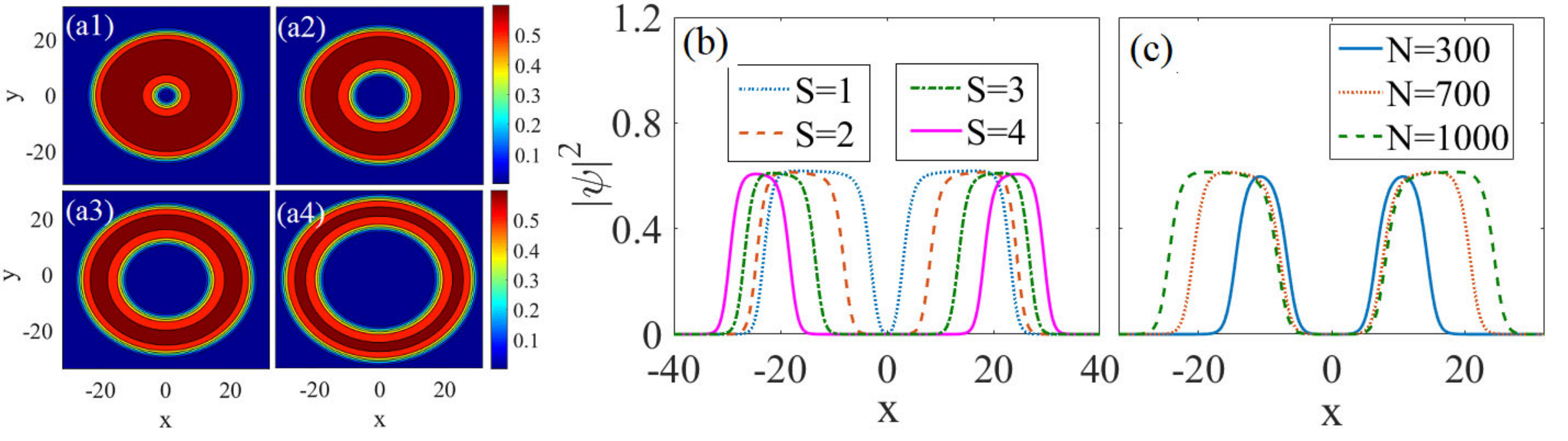}}
\caption{Panels (a1)-(a4) display density patterns of QDs (quantum droplets)
with embedded vorticities $S=1$,$2$,$3$,$4$ and fixed norm $N=1000$, as per
Ref. \protect\cite{GZ-ln}. The first three QDs are stable, while the one
corresponding to $S=4$ is subject to the splitting instability. (b)
Cross-sections of the density patterns from panels (a1)-(a4). (c) The same
as in (b), for for $S=1$ and different values of $N$. All vortex QDs shown
in (c) are stable.}
\label{ln_vortices}
\end{figure}

Numerical and approximate analytical consideration of Eq. (\ref{ln}) has
produced completely stable families of fundamental ($S=0$) solitons, and
partly stable families of solitary vortices, up to $S=5$ \cite{GZ-ln}. They
chiefly feature the flat-top shape, although become closer to narrow annuli
near the stability boundary. Typical examples of the vortex solitons are
displayed in Fig. \ref{ln_vortices}. They %exists for values of the total
%norm, $N$, exceeding a minimum value, $N_{\min }$, and
are stable provided that their norm exceeds a %still larger
threshold value, $N>N_{\mathrm{thr}}$%5>N_{\min }$
(actually, it is the double norm, defined for the binary modes with equal
components). The vortex solitons are unstable against spontaneous splitting
at $N<N_{\mathrm{thr}}$. For $1\leq S\leq 5$, values $N_{\mathrm{thr}}$ are
collected in Table I. For $S$ large enough (actually, for $S\geq 3$), the
dependence of $N_{\mathrm{thr}}$ on $S$ can be predicted in an approximate
analytical form, which is based on comparison of values of the energy
(Hamiltonian, see Eq. (\ref{Hlog})) of the unsplit vortex and a set of far
separated fragments produced by the splitting. The instability does not
occur if the splitting would imply the increase of the energy, which yields
the following prediction for the stability boundary \cite{GZ-ln}:%
\begin{equation}
N_{\mathrm{thr}}^{\mathrm{(analyt)}}\approx 6N^{4}  \label{^4}
\end{equation}%
(in fact, coefficient $6$ is a fitting factor, while the prediction produces
scaling $N_{\mathrm{thr}}\sim S^{4}$).
\begin{table}[tbp]
\centering%
\begin{tabular}{|l|l|l|l|l|l|}
\hline
$S$ & $1$ & $2$ & $3$ & $4$ & $5$ \\ \hline
$N_{\mathrm{thr}}$ & $60$ & $200$ & $510$ & $1380$ & $3550$ \\
$N_{\mathrm{thr}}^{\mathrm{(analyt)}}$ & n/a & n/a & $486$ & $1536$ & $3750$
\\ \hline
\end{tabular}%
\caption{Table I: The middle line shows numerically found minimum values of
the norm, $N_{\mathrm{thr}}$, necessary for the stability of 2D vortex QDs
(quantum droplets) with winding number $S$, as per Ref. \protect\cite{GZ-ln}%
. The bottom line presents results produced by the analytical approximation (%
\protect\ref{^4}), which is relevant for $S\geq 3$. }
\end{table}

It is relevant to mention that the 2D model, written in an explicitly
two-component form, admits vortex states with the \textit{hidden vorticity}
(HV), i.e., identical amplitude profiles and chemical potential of the
components but opposite vorticities, $S_{1}=-S_{2}$. It was found in Ref.
\cite{GZ-ln} that the HV modes with $S_{1}=-S_{2}=1$ have their limited
stability region too, while they are completely unstable for $%
S_{1}=-S_{2}\geq 2$.

\subsubsection{``Invisible" splitting instability of multiple vortices}

As shown above, QD families with multiple vorticity keep their partial
stability up to $S=5$. Nevertheless, it was demonstrated in Ref. \cite{GZ-ln}
that all multiple-vortex states, with $S\geq 2$, demonstrate virtually
invisible but, strictly speaking, existing \textit{structural instability}
against splitting. This means that a specially selected small perturbation,
without initiating any dynamical instability, may split the pivot (phase
singularity) of the multiple vortex with topological charge $S$ into sets of
$S$ or $S+2$ phase singularities corresponding to unitary vortices, although
the splitting remains almost invisible, as it occurs in the broad central
\textquotedblleft hole" induced by the multiple vorticity, where absolute
values of the wave function remain extremely small (in the case of the
splitting into $S+2$ phase singularities, one of them belongs to a unitary
antivortex, so that the total vorticity is conserved). It is worthy to note
that the structural instability is not specific to the model of the LHY
type, but is a generic effect, relevant to all models which support
dynamically stable higher-order vortex solitons. While this effect seems
intuitively obvious, it makes sense to produce it in an explicit analytical
form, following Ref. \cite{GZ-ln}.

To demonstrate the structural instability against splitting, one can place
the pivot of the state with $S\geq 2$ at the origin ($x=y=0$), and add a
specific perturbation mode carrying its own intrinsic vorticity $s=1$, with
small complex amplitude $-\varepsilon $, thus producing a perturbed
configuration,%
\begin{gather}
\psi _{\mathrm{pert}}\left( x,y\right) \approx \left( x+iy\right)
^{S}-\varepsilon \left( x+iy\right)  \notag \\
\equiv \left( x+iy\right) \left[ \left( x+iy\right) ^{S-1}-\varepsilon %
\right] .  \label{pert1}
\end{gather}%
Phase singularities of unitary vortices, into which the original vortex is
split by the perturbation, are identified as zeros of $\left\vert \psi _{%
\mathrm{pert}}\left( x,y\right) \right\vert $. \ As seen from Eq. (\ref%
{pert1}), they are located at the origin, $x_{\mathrm{piv}}^{(1)}=y_{\mathrm{%
piv}}^{(1)}=0$, and at $\left( S-1\right) $ additional points, produced by $%
\left( S-1\right) $ branches of the root of degree $1/\left( S-1\right) $:
\begin{equation}
x_{\mathrm{piv}}^{(1+j)}+iy_{\mathrm{piv}}^{(1+j)}=\varepsilon
^{1/(S-1)},~j=1,...,S-1.  \label{j}
\end{equation}

Another option is to consider a small disturbance with intrinsic vorticity $%
s=-1$ (instead of $+1$, which was taken above):%
\begin{equation}
\psi _{\mathrm{pert}}\left( x,y\right) =\left( x+iy\right) ^{S}-\varepsilon
\left( x-iy\right) .  \label{pert2}
\end{equation}%
In this case, the pivots are located at points defined by equation $\left(
x+iy\right) ^{S}=\varepsilon \left( x-iy\right) $, i.e., $\left( x+iy\right)
^{S+1}=\varepsilon r^{2}$, which yields a set of $S+2$ roots, \textit{viz}.,%
\begin{equation}
x_{\mathrm{piv}}^{(k)}+iy_{\mathrm{piv}}^{(k)}=\varepsilon
^{1/(S+1)}\left\vert \varepsilon \right\vert ^{2/\left( S^{2}-1\right)
},~k=1,...,S+1,  \label{k}
\end{equation}%
plus the central pivot at the origin.

\begin{figure}[t]
{\includegraphics[width=0.75\columnwidth]{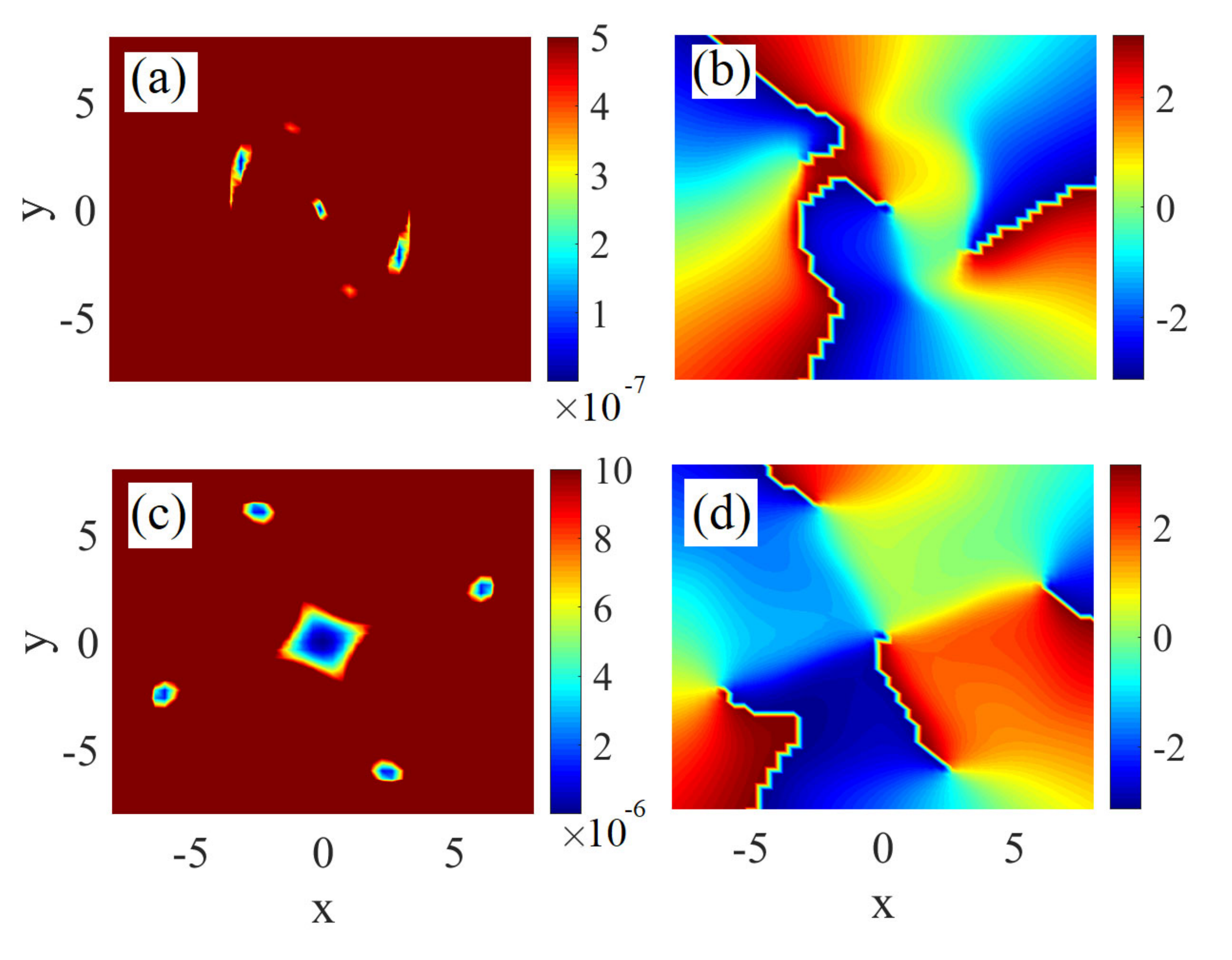}}
\caption{(a) A zoom (in domain $\left\vert x,y\right\vert \leq 8$) of the
density pattern, $\left\vert \protect\psi \left( x,y\right) \right\vert ^{2}$%
, for a QD with $(S,N)=(3,1000)$, which was initially perturbed as per Eq. (%
\protect\ref{pert1}), with $s=+1$ and $\protect\varepsilon =0.0013+0.0023i$.
The pattern is produced by the simulation of Eq. (\protect\ref{GPE}) up to $%
t=5000$. (b) The corresponding phase pattern clearly identifies three
unitary vortices, whose pivots (phase singularities) are located at zeros of
the local amplitude. (c) and (d): The same as in (a) and (b), but with the
initial perturbation carrying vorticity $s=-1$, as per Eq. (\protect\ref%
{pert2}). In this case, the amplitude and phase patterns demonstrate
splitting in a set of five pivots: one with winding number $-1$ located in
the middle, surrounded by four satellites with winding numbers $+1$. Note
the extremely small scale of the local amplitude, $\left\vert \protect\psi %
\left( x,y\right) \right\vert \sim 10^{-7}$ in (a) and (c). On the normal
scale, these splitting patterns remain invisible. The results are borrowed
from Ref. \protect\cite{GZ-ln}.}
\label{pivots}
\end{figure}

Verification of this consideration by direct simulations of Eq. (\ref{ln})
is displayed in Fig. \ref{pivots}. Strong magnification of the picture
observed in the nearly empty ``hole" confirms the splitting of the pivot
with multiplicity $S$ into $S$ or $S+2$ sets of unitary-vortex pivots, under
the action of the small initial perturbation with its intrinsic vorticity $%
s=+1$ or $s=-1$, respectively (the vortex soliton with $S=1$ does not
undergo the splitting). On the other hand, the splitting remains virtually
invisible on the normal scale of $\left\vert \psi \left( x,y\right)
\right\vert $, hence it does not imply actual instability of the solitary
states with multiple vorticity.

\subsubsection{3D vortex solitons}

\begin{figure}[t]
{\includegraphics[width=0.5\columnwidth]{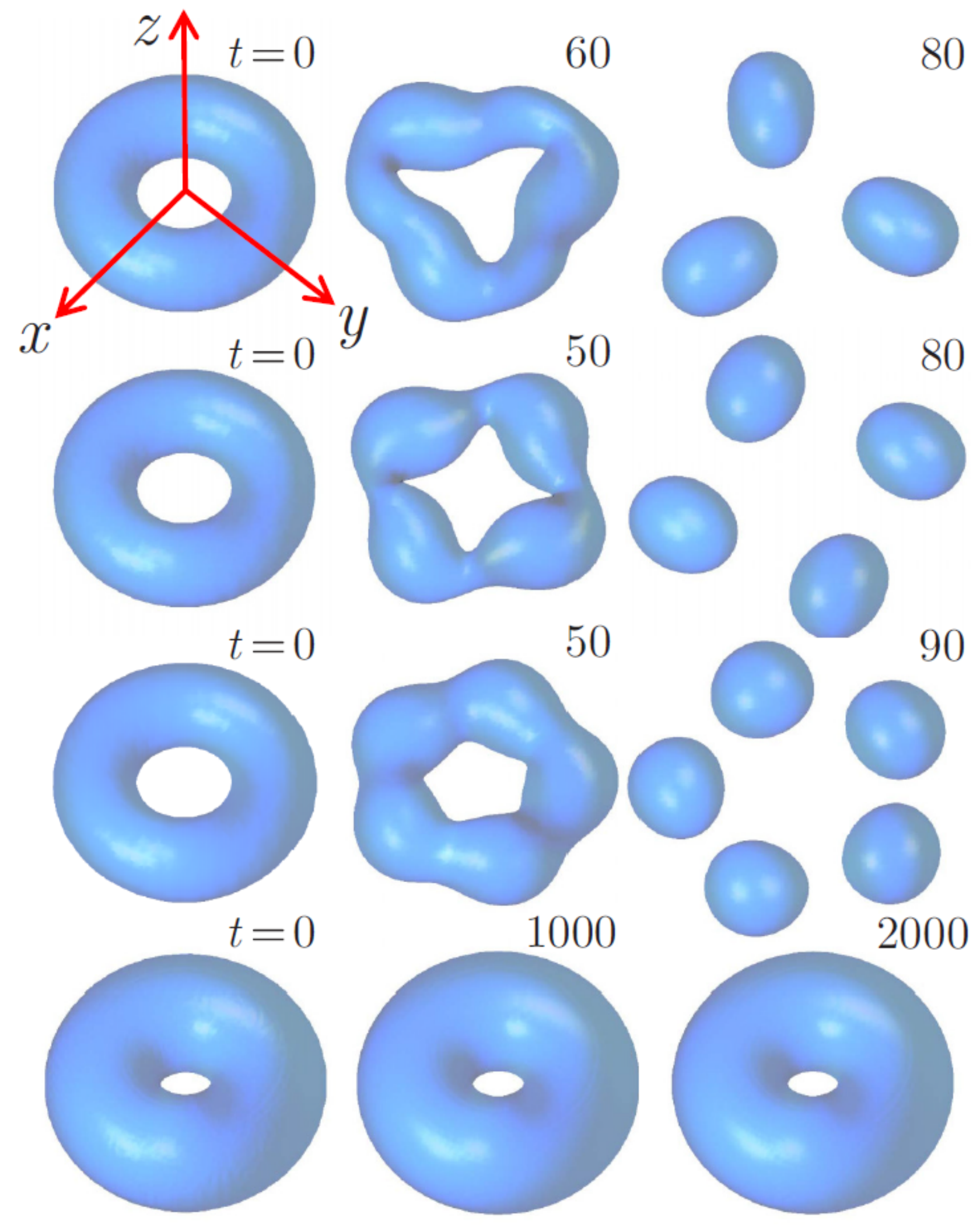}}
\caption{The evolution in time $t$ of three-dimensional QDs (quantum
droplets) with vorticity $S=2$, as produced by simulations of Eq. (\protect
\ref{LHY}) reported in Ref. \protect\cite{3DLHY}, and shown by means of
surfaces $\left\vert \protect\psi \left( x,y,z\right) \right\vert -\mathrm{%
const}$. The first three rows display examples of splitting of the same
unstable 3D QD (quantum droplet) into sets of three, four, or five
fragments, initiated by different eigenmodes of small perturbations
initially added to the stationary QD. A random small perturbation causes the
splitting into four fragments, as the respective instability eigenmode has
the largest growth rate. The bottom row shows the evolution of a stable QD
with $S=2$ and with a larger norm.}
\label{3DQD}
\end{figure}

Numerical solutions of three-dimensional Eq. (\ref{LHY}) for solitons with
embedded vorticity were presented in Ref. \cite{3DLHY}. Being supported by
the competing set of the cubic and quartic terms, the solitons again tend to
feature a flat-top shape. An analytical estimate for the stability of the 3D
vortex soliton against the spontaneous splitting, which is based, as in the
2D case, on the comparison of energies of the unsplit state and the set of
far separated fragments, yields scaling $S_{\min }\sim S^{6}$, cf. the 2D
result given by Eq. (\ref{^4}). In accordance with this prediction, a
conspicuous stability region was found, in a numerical form, for $S=1$,
while for $S=2$ only vortex solitons with a very large norm may be stable.
Examples of the evolution of stable and unstable 3D QDs with $S=2$ are
displayed in Fig. \ref{3DQD}.

\begin{figure}[t]
{\includegraphics[width=0.6\columnwidth]{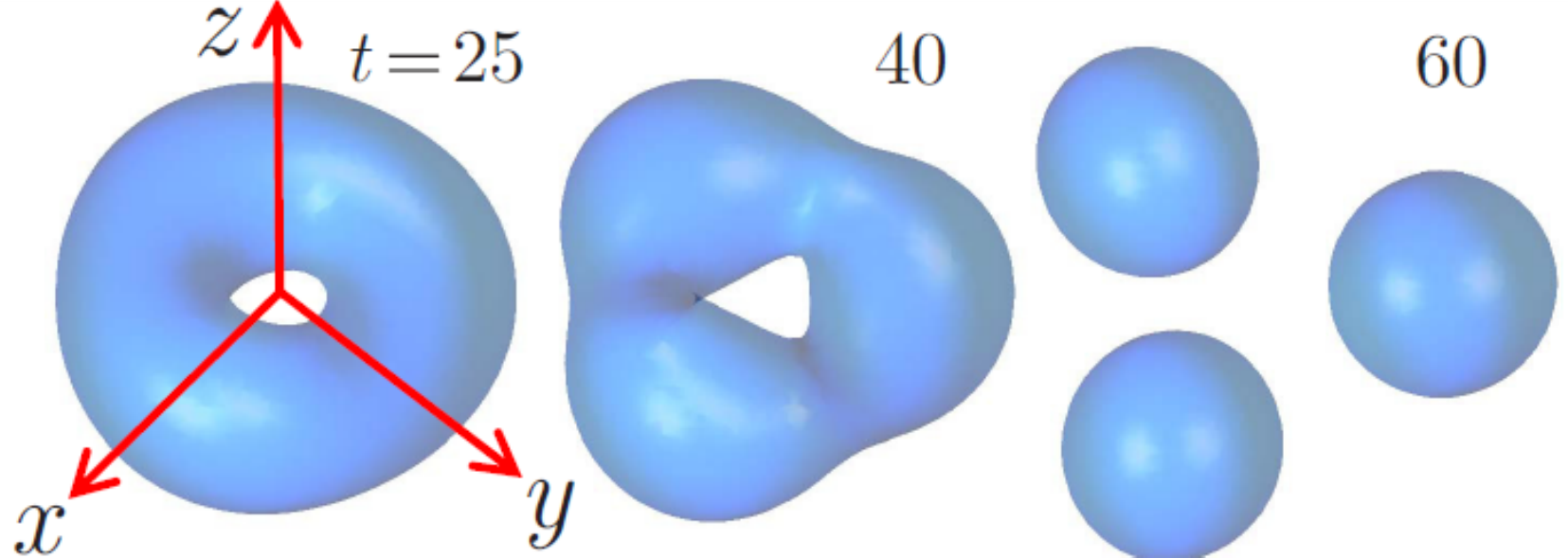}}
\caption{Spontaneous splitting of an HV (hidden-vorticity) QD with
vorticities of its components $S_{1}=-S_{2}=1$ $S=2$, as produced by
simulations of the full two-component version of Eq. (\protect\ref{LHY}) in
Ref. \protect\cite{3DLHY}, and shown by means of constant-density surfaces
of one component (for the second one, the picture seems identical).}
\label{HV_splitting}
\end{figure}

All the HV states turn out to be unstable in the 3D model against
spontaneous splitting, see an example in Fig. \ref{HV_splitting}.

Lastly, it is relevant to mention that the creation of QDs was predicted and
experimentally realized in dipolar BECs, making use of the attractive
long-range interactions and LHY effect \cite{dip1,dip2}. However, the
theoretical analysis has demonstrated than all QDs with embedded vorticity
are unstable in that setting \cite{Macri}.

\subsection{Trapping potentials for vortex solitons}

\subsubsection{The harmonic-oscillator (HO) trap}

The use of trapping potentials, which are natural ingredients of
experimentally relevant settings in BEC and optics alike, makes it possible
to stabilize both fundamental and vortex solitons. The model is written as
the normalized GPE \cite{Pit} for the mean-field wave function $\psi
(x,y,z,t)$:%
\begin{equation}
i\frac{\partial \psi }{\partial t}=-\frac{1}{2}\nabla ^{2}\psi +U(x,y,z)\psi
-|\psi |^{2}\psi .  \label{GPE}
\end{equation}%
Here $U$ is the trapping potential, and the negative sign in front of the
cubic term implies self-attraction, as above.\ Actually, a
quasi-two-dimensional trapping potential, i.e., $U=U\left( x,y\right) $, may
be sufficient to support stable 3D solitons \cite%
{low-dim,low-dim2,Herve-lowdim}.

In optics, time $t$ in Eq. (\ref{GPE}) is replaced by the propagation
distance, $z$, while the original coordinate $z$ is replaced by the temporal
variable $\tau $, cf. Eq. (\ref{CQ}). The effective potential in the optical
waveguide may be solely two-dimensional, being proportional to the local
change of the refractive index, $U\left( x,y\right) \sim -\delta n(x,y)$.

For axially and spherically symmetric potentials, $U=U\left( r,z\right) $,
where $\left( r,z,\theta \right) $ is the set of cylindrical coordinates,
solutions to Eq.~(\ref{GPE}) are looked for in the form similar to that
defined above in Eq. (\ref{3Dvortex}):
\begin{equation}
\psi =\exp \left( -i\mu t+iS\theta \right) R(r,z),  \label{vort}
\end{equation}%
with real chemical potential $\mu $, and real amplitude function $R\left(
r,z\right) $ satisfying the stationary equation which is similar to Eq. (\ref%
{U}):%
\begin{equation}
\mu u=-\frac{1}{2}\left( \frac{\partial ^{2}}{\partial r^{2}}+\frac{1}{r}%
\frac{\partial }{\partial r}+\frac{\partial ^{2}}{\partial z^{2}}-\frac{S^{2}%
}{r^{2}}\right) R+U\left( r,z\right) R-R^{3}.  \label{u}
\end{equation}

Most relevant for the realization in BEC is the harmonic-oscillator (HO)
trapping potential \cite{Pit},%
\begin{equation}
U\left( x,y,z\right) =\frac{1}{2}\left( x^{2}+y^{2}+\Omega ^{2}z^{2}\right) ,
\label{HO}
\end{equation}%
where $\Omega ^{2}$ accounts for anisotropy of the trap, the limits of $%
\Omega ^{2}\gg 1$ and $\Omega ^{2}\ll 1$ corresponding, respectively, to
nearly 2D (\textquotedblleft pancake-shaped" \cite{pancake}) and nearly
one-dimensional (\textquotedblleft cigar-shaped" \cite{Randy-NJP})
configurations. Equation (\ref{GPE}) can be derived from the Hamiltonian,
\begin{equation}
H=\frac{1}{2}\int \int \int \left[ \left( \left\vert \frac{\partial \psi }{%
\partial x}\right\vert ^{2}+\left\vert \frac{\partial \psi }{\partial y}%
\right\vert ^{2}+\left\vert \frac{\partial \psi }{\partial z}\right\vert
^{2}\right) +(x^{2}+y^{2}+\Omega ^{2}z^{2})|\psi |^{2}-|\psi |^{4}\right]
dxdydz,  \label{E}
\end{equation}%
which is a dynamical invariant (conserved quantity) of Eq. (\ref{GPE}). Two
other dynamical invariants of Eq. (\ref{GPE}) are the total norm,
\begin{equation}
N=\int \int \int \left\vert \psi (x,y,z)\right\vert ^{2}dxdydz,  \label{N}
\end{equation}%
and $z$-component of the angular momentum,
\begin{equation}
M_{z}=i\int \int \int \left( y\frac{\partial \psi }{\partial x}-x\frac{%
\partial \psi }{\partial y}\right) \psi ^{\ast }dxdydz,  \label{M}
\end{equation}%
whose value in stationary state (\ref{vort}) is
\begin{equation}
M_{z}=SN.  \label{MN}
\end{equation}%
Note that, unlike the model-specific Hamiltonian, the definitions of the
norm and angular momentum, given by Eqs. (\ref{N}) - (\ref{MN}) or their 2D
counterparts are universal.

\subsubsection{Lattice (spatially-periodic) trapping potentials}

Many experiments in BEC and photonics made use of spatially periodic
potentials, which are represented by optical lattices (OLs) for atomic
condensates \cite{Brazh,Morsch}, by photonic crystals steering the
transmission of light in optics \cite{PhCr,JY}, or by similar periodic
structures acting on exciton-polariton fields in semiconductor microcavities
\cite{Kriz}. In addition to assembling the traditional photonic-crystal
structures, similar multi-channel waveguides can be produced by burning
multi-core patterns in bulk silica \cite{Jena}, and, on the other hand,
virtual (rewritable) structures can be created in photorefractive materials,
illuminating them by counterpropagating laser beams in the ordinary
polarization, which interfere linearly and induce an effective lattice,
while solitons are built by extraordinarily-polarized beams, which are
subject to the action of saturable nonlinearity \cite{MSegev}.

It is relevant to stress that, while the HO potential, considered above,
maintains bound states in the absence of any nonlinearity, just as
eigenmodes of this potential in quantum mechanics, no self-trapping is
possible in periodic potentials. As the basic model, one can take the 3D\
GPE with the full 3D OL potential, or its quasi-2D version (as mentioned
above, the 2D potential may be sufficient to create stable vortex solitons
in the fully 3D form \cite{low-dim,low-dim2,Herve-lowdim}):%
\begin{equation}
i\frac{\partial \psi }{\partial t}=-\frac{1}{2}\nabla ^{2}\psi -\varepsilon %
\left[ \cos \left( kx\right) +\cos \left( ky\right) +\sigma \cos \left(
kz\right) \right] \psi -g|\psi |^{2}\psi .  \label{OL}
\end{equation}%
Here, $\sigma =1$ or $0$, for the full 3D or quasi-2D lattice, respectively,
and $g=+1$ and $-1$ corresponds, severally, to the self-focusing and
defocusing signs of the cubic nonlinearity. Note that the defocusing
nonlinearity may create gap solitons in the presence of the lattice
potential \cite{Brazh,Morsch,Yang-book}. In optics, as mentioned above, $t$
is replaced by the propagation coordinate, $z$, while original coordinate $z$
is replaced by the reduced-time variable (\ref{tau}), and solely the 2D
potential, with $\sigma =0$ in Eq. \ref{OL}, is relevant. The sign of the
cubic nonlinearity is usually $g=+1$ (self-focusing) in optical media. GPE (%
\ref{OL}) conserves two dynamical invariants, \textit{viz}., the total
norm/energy, defined as in Eq. (\ref{N}), and the Hamiltonian,%
\begin{equation}
H=\int \int \int \left[ \frac{1}{2}\left( \left\vert \frac{\partial \psi }{%
\partial x}\right\vert ^{2}+\left\vert \frac{\partial \psi }{\partial y}%
\right\vert ^{2}+\left\vert \frac{\partial \psi }{\partial z}\right\vert
^{2}\right) -\varepsilon \left[ \cos \left( kx\right) +\cos \left( ky\right)
+\sigma \cos \left( kz\right) \right] |\psi |^{2}-\frac{g}{2}|\psi |^{4}%
\right] dxdydz,  \label{H-OL}
\end{equation}%
cf. Eq. (\ref{E}).

\subsection{Vortex solitons in the HO trap}

\subsubsection{3D vortex modes and their stability}

Before displaying numerical findings, it is relevant to outline approximate
analytical results which are available in the present setting. In the linear
limit, Eq. (\ref{GPE}) is tantamount to the quantum-mechanical Schr\"{o}%
dinger equation for the 3D anisotropic HO. In the Cartesian coordinates, the
corresponding 3D eigenfunctions are built as
\begin{equation}
\psi _{jkl}(x,y,z,t)=e^{-i\mu _{0}t}\Phi _{j}(x)\Phi _{k}(y)\Phi _{l}\left(
\sqrt{\Omega }z\right) ,  \label{xyz}
\end{equation}%
where $\Phi _{j}$, $\Phi _{k}$ and $\Phi _{l}$ are stationary wave functions
of 1D harmonic oscillators with quantum numbers $j,k,l$, which correspond to
energy eigenvalues $j+1/2$, $k+1/2$ and $\left( l+1/2\right) \Omega $,
respectively, the chemical potential being
\begin{equation}
\mu _{0}=j+k+1+\left( l+1/2\right) \Omega ~.  \label{linear}
\end{equation}%
The states which carry over into ones (\ref{vort}) with vorticity $S$ in the
nonlinear model are constructed as combinations of factorized wave functions
(\ref{xyz}) with $l=0$ and $j+k=S$. Since the correction to $\mu $ from the
self-attractive nonlinearity is negative, this restriction and Eq. (\ref%
{linear}) with $l=0$ impose a bound on $\mu $,
\begin{equation}
\mu \leq \mu _{0}=S+1+(1/2)\Omega .  \label{bound}
\end{equation}%
In particular, the eigenfunctions of the linear model, with vorticities $S=1$
and $S=2$, are written, in terms of the cylindrical coordinates, $\left(
r,\theta ,z\right) $, as
\begin{equation}
\psi _{\mathrm{linear}}^{(S=1)}=\psi _{100}+i\psi _{010}\equiv r\exp \left[
-\left( 2+\Omega /2\right) it+i\theta -\left( r^{2}+z^{2}\right) /2\right] ,
\end{equation}%
\begin{equation}
\psi _{\mathrm{linear}}^{(S=2)}=\psi _{200}-\psi _{020}+2i\psi _{110}\equiv
r^{2}\exp \left[ -\left( 3+\Omega /2\right) it+2i\theta -\left(
r^{2}+z^{2}\right) /2\right] .
\end{equation}

\begin{figure}[t]
\begin{center}
\includegraphics[width=10cm]{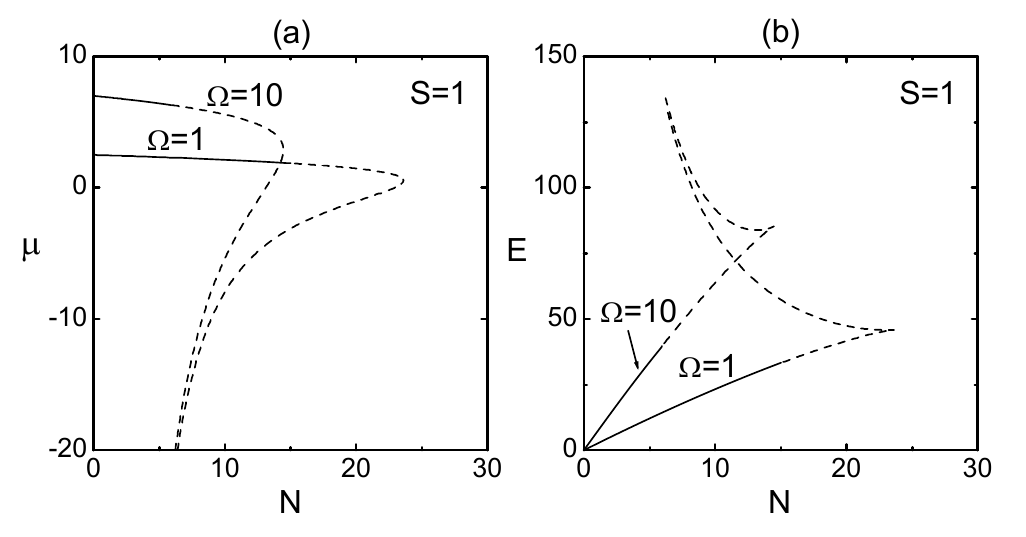}
\end{center}
\caption{(a) The chemical potential and (b) energy, defined as per Eq. (%
\protect\ref{E}), versus norm $N$ for the 3D vortex mode with $S=1$, trapped
in (anisotropic) HO potential (\protect\ref{HO}) with $\Omega =10$\ and $1$,
as found from numerical solution of Eq. (\protect\ref{u}). Solid and dotted
portions of the curves denote, respectively, stable and unstable parts of
the solution families, as identified by a numerical solution of the
eigenvalue problem based on Bogoliubov - de Gennes equations (\protect\ref%
{growth}) in Ref. \protect\cite{Dum2D}.}
\label{in_HO}
\end{figure}

Numerical solution of Eq. (\ref{u}) generates $\mu (N)$ and $E(N)$
dependences for families of vortical trapped modes with $S=1$. They are
displayed in Fig. \ref{in_HO} for $\Omega =10$ and $1$, which correspond to
the pancake-like and isotropic trapping configurations, respectively (for
the cigar-shaped trapped modes, that correspond, e.g., to $\Omega =0.1$, the
dependences, which are not shown here, are very close to those for $\Omega
=1 $ \cite{Dum3D}). An obvious feature of the figure is the presence of a
largest norm, corresponding to the turning point of the $\mu (N)$ curve,
which bounds the existence of the trapped modes. This limitation is caused
by the possibility of the supercritical collapse in the 3D model \cite%
{Berge,SulemSulem,Gadi}: if the norm is too large, the self-attraction
cannot be balanced by the gradient part of energy (\ref{E}). The largest
norm in this model was numerically found, as a function $\Omega $, in Ref.
\cite{Sadhan}. Another implication of the possibility of the supercritical
collapse is that all the stable modes found in the 3D model are actually
metastable ones. They cannot play the role of the GS, which does not exist
in the model admitting the supercritical collapse \cite{HP}.

The stability of stationary solutions produced by Eq. (\ref{u}) was
identified through the computation of eigenvalues $\lambda $ of small
perturbations \cite{Dum2D,Dum3D}. To this end, a perturbed solution to Eq. (%
\ref{GPE}) is looked for as
\begin{equation}
\psi (x,y,z,t)=[R(r,z)+u(r,z)\exp (\lambda t+iL\theta )+v^{\ast }(r,z)\exp
(\lambda ^{\ast }t-iL\theta )]\exp \left( iS\theta -i\mu t\right) ,
\label{pert}
\end{equation}%
where $\left( u,v\right) $ are eigenmodes of infinitesimal perturbations
corresponding to integer values of azimuthal index $L$, and $\ast $ stands
for the complex conjugate. The substitution of ansatz (\ref{pert}) in Eq. (%
\ref{GPE}) and linearization lead to\ the Bogoliubov - de Gennes equations
\cite{Pit},%
\begin{eqnarray}
\left( i\lambda +\mu \right) u+\frac{1}{2}\left[ \frac{\partial ^{2}}{%
\partial r^{2}}+\frac{1}{r}\frac{\partial }{\partial r}+\frac{\partial ^{2}}{%
\partial z^{2}}-\frac{(S+L)^{2}}{r^{2}}u-\rho ^{2}\right] u+R^{2}(v+2u) &=&0,
\notag \\
\left( -i\lambda +\mu \right) v+\frac{1}{2}\left[ \frac{\partial ^{2}}{%
\partial r^{2}}+\frac{1}{r}\frac{\partial }{\partial r}+\frac{\partial ^{2}}{%
\partial z^{2}}-\frac{(S-L)^{2}}{r^{2}}v-\rho ^{2}\right] v+R^{2}(u+2v) &=&0,
\label{growth}
\end{eqnarray}%
supplemented by the boundary conditions demanding that $u(r,z)$ and $v(r,z)$
decay exponentially at $r\rightarrow \infty $ and $\left\vert z\right\vert
\rightarrow \infty $, and decay as $r^{\left\vert S\pm L\right\vert }$ at $%
\rho \rightarrow 0$. The underlying stationary mode is stable if all
eigenvalues $\lambda $ produced by numerical solution of Eq. (\ref{growth})
\cite{Ueda} are pure imaginary. Results of the numerical stability analysis
are included in Fig. \ref{in_HO}, where stable and unstable portions of the
vortex-soliton families are designated.

Note that the stability region of the vortex mode with $S=1$ is essentially
smaller than formally predicted by the celebrated Vakhitov-Kolokolov (VK)
criterion, $d\mu /dN<0$, which provides a necessary, but generally, not
sufficient condition for the stability of self-trapped modes \cite%
{VK,Berge,Gadi} (it cannot detect instabilities accounted for by complex
eigenvalues $\lambda $ -- in particular, the splitting instability of the
modes with $S\geq 1$). In the region where the VK criterion holds, but the
vortices are unstable, they are vulnerable to the splitting instability
induced by perturbations with $L=2$ and $3$ in Eq. (\ref{pert}). The vortex
modes with $S\geq 2$ were found to be completely unstable \cite{Dum2D,Dum3D}.

\begin{figure}[t]
\begin{center}
\includegraphics[width=10cm]{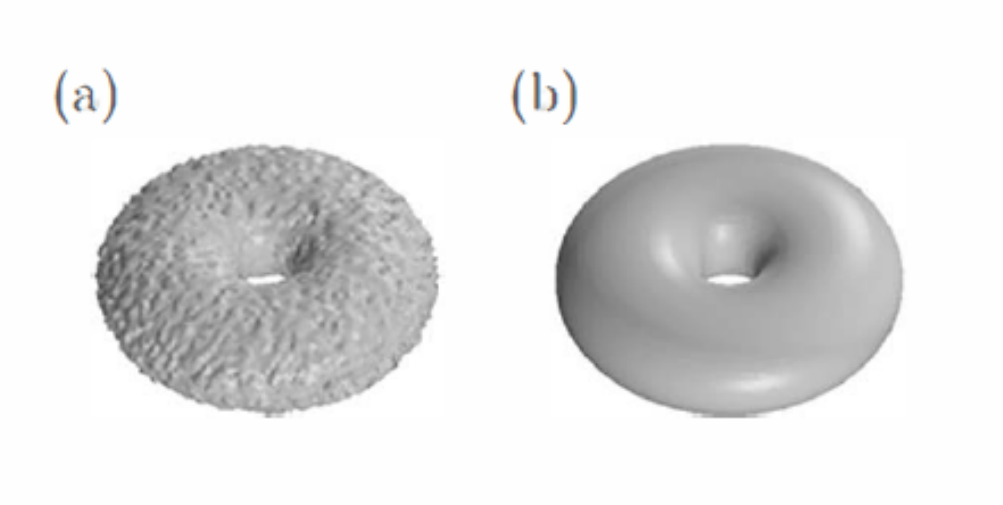}
\end{center}
\caption{Self-cleaning of a stable 3D vortex torus with $S=1$, produced in
\protect\cite{Dum3D} by simulations of the model based on Eq. (\protect\ref%
{GPE}) with isotropic ($\Omega =1$) potential (\protect\ref{HO}), after the
application of a random perturbation at the amplitude level of $10\%$.
Panels (a) and (b) display, severally, the initially perturbed shape of the
vortex state at $t=0$, and its shape at $t=120$. The unperturbed vortex has
chemical potential $\protect\mu =2$ and norm $N=12.55$.}
\label{cleaning}
\end{figure}

The predictions for the stability based on the computation of the stability
eigenvalues were verified by direct simulations of Eq. (\ref{GPE}), starting
with stationary modes to which small arbitrary perturbations were added. The
robustness of stable vortices is illustrated by Fig. \ref{cleaning}, which
demonstrates that they absorb the perturbations and clean themselves up.

Lastly, it is relevant to mention the consideration of a binary model with
the repulsive cubic nonlinear terms (both intra- and inter-component ones)
reported in Ref. \cite{Radik}. By means of systematic simulations and a
semi-analytical approximation based on a finite-mode Galerkin decomposition
of the wave functions, nontrivial two-component stationary states and
dynamical regimes have been found in that system.

\subsubsection{2D trapped vortex modes}

\paragraph{The one-component setting}

As said above, taking the limit of $\Omega \rightarrow \infty $ in HO
potential (\ref{HO}) leads to the reduction of the 3D model to its 2D
version, with Eqs. (GPE), (\ref{u}), (\ref{E}), (\ref{N}), (\ref{M}), and (%
\ref{growth}) carrying over into their 2D counterparts, and the effective 2D
HO potential taking the form of $U\left( x,y\right) =(1/2)\left(
x^{2}+y^{2}\right) $. In particular, stationary solutions of the 2D version
of Eq. (\ref{GPE}) are looked for as $\psi =\exp \left( -i\mu t+iS\theta
\right) R(r)$, cf. Eq. (\ref{vort}). In the limit form of the linear
equation, the chemical potential is given by the usual eigenvalue for the 2D
linear Schr\"{o}dinger equations with this potential:%
\begin{equation}
\mu _{0}^{\mathrm{(2D)}}=S+1,  \label{S+1}
\end{equation}%
cf. Eq. (\ref{bound}).

The shape and stability of nonlinear 2D modes trapped in the HO potential
was studied in a number of works \cite{2D,2D2,2D3,2D4,Dum2D}. The results,
produced by the computation of stability eigenvalues in the framework of the
2D Bogoliubov - de Gennes equations (\ref{growth}), demonstrate that the
family of the fundamental ($S=0$) trapped modes is stable in its entire
existence region,%
\begin{equation}
0\leq N<N_{\mathrm{TS}}\approx 5.85,  \label{Nmax}
\end{equation}%
where $N_{\mathrm{TS}}$ is the numerically found value of the norm of the
\textit{Townes soliton }(TS), which determines the threshold for the onset
of the critical collapse in the 2D NLSE \cite{Berge,SulemSulem,Gadi}. In the
absence of trapping potentials, the family of TSs is degenerate, as, at all
values of $\mu $, they assume the single value of the norm, which is exactly
equal to $N_{\mathrm{TS}}$. The degeneracy is a consequence of the specific
scaling invariance of the NLSE in 2D \cite{Berge,SulemSulem,Gadi}. The
trapping potential introduces a characteristic spatial scale (the usual HO
length, or the spatial period, in the case of lattice potentials), which
breaks the scaling invariance and thus lifts the degeneracy, making $N$ a
function of $\mu $. In fact, Eq. (\ref{Nmax}) demonstrates that the
degeneracy is lifted so that the soliton's norm falls \emph{below} the
collapse-onset threshold. This circumstance lends the trapped modes with $%
S=0 $ protection against the collapse, i.e., \emph{stability}, and actually
makes them the system's GS, which did not exist in the absence of the
trapping potential. This is a major difference from the stabilization
mechanism provided by the trapping potential in 3D, where, as said above,
the GS cannot exist (as the supercritical collapse occurs at any value of
the norm), only metastability of the trapped modes being possible.

The family of TSs with embedded vorticity $S=1$ is also degenerate in the
free space, admitting a single value of the norm,
\begin{equation}
N_{\mathrm{TS}}^{(S=1)}\approx 24.1  \label{S=1}
\end{equation}%
\cite{Minsk2} (it is relevant to mention that the first generalization of
the TS, in the form of higher-order radial states with $S=0$, was introduced
in Ref. \cite{Yankauskas}, soon after the concept of the TS was established;
while such states are unstable, they can be stabilized by means of the
``management" technique, which makes the coefficient of the cubic
nonlinearity a periodically varying function of $t$ \cite{VPG-management}).
Recently, an analytical approximation was developed, which produces values $%
N_{\mathrm{TS}}^{(S)}$ for $S\geq 1$ with a good accuracy \cite{Jieli2}, see
Eq. (\ref{beta_max}) and Table II below.

The HO trapping potential stabilizes the modes with $S=1$ in interval%
\begin{equation}
0\leq N<7.79\approx 0.32N_{\mathrm{TS}}^{(S=1)},  \label{main}
\end{equation}%
the corresponding stability region in terms of the chemical potential being $%
1.276\equiv \mu _{\mathrm{cr}}<\mu \equiv \mu _{0}^{\mathrm{(2D)}%
}(S=1)\equiv 2$ (the right edge of the region is determined by Eq. (\ref{S+1}%
)) \cite{Dum2D}. Trapped 2D\ vortices with $S\geq 2$ they remain completely
unstable.

The partial stability of the family of trapped vortex modes with $S=1$,
predicted through the computation of the corresponding eigenvalues, was
corroborated by direct simulations of the perturbed evolution. The
simulations have also revealed a noteworthy dynamical regime for the trapped
vortices with $S=1$ in interval
\begin{equation}
0.32N_{\mathrm{TS}}^{(S=1)}\approx 7.79<N<10.30\approx 0.43N_{\mathrm{TS}%
}^{(S=1)},  \label{intermediate}
\end{equation}%
adjacent to one given by Eq. (\ref{main}). In interval (\ref{intermediate}),
the evolution of the unstable vortex is time-periodic, as shown in Fig. \ref%
{periodic}: it spits into two fragments which then recombine back into the
vortex, keeping the vorticity of the configuration in the course of the
cycles, while the splitting orientation slowly rotates in the $\left(
x,y\right) $ plane. The vortices with still larger values of the norm, $%
N>10.30$, also split in two fragments, which, however, fail to recombine.
Instead, each one quickly blows up, i.e., collapses.
\begin{figure}[t]
\includegraphics[width=13cm]{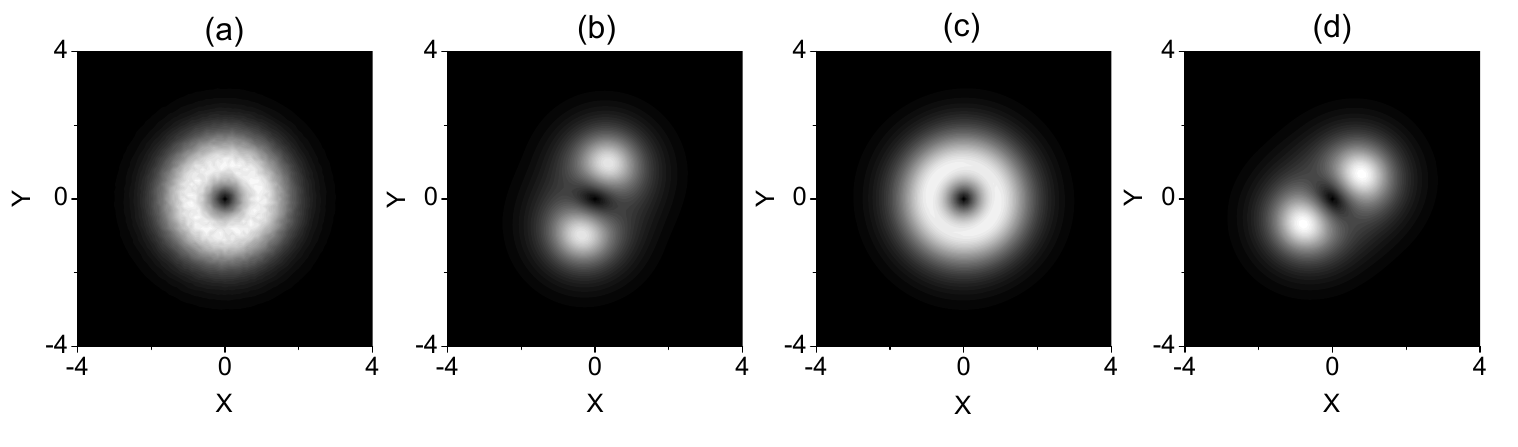}
\caption{Periodic evolution of an initially perturbed 2D vortex with $S=1$, $%
\protect\mu =1.2$ and $N=8.48$, which belongs to interval (\protect\ref%
{intermediate}), as per Ref. \protect\cite{Dum2D}. In this dynamical regime,
the trapped mode periodically splits into two fragments and recombines,
while its vorticity is conserved: (a) $t=0$, (b) $t=100$, (c) $t=140$, and
(d) $t=180$. It is relevant to compare this dynamical state with its
counterpart in the dissipative model based on Eq. (\protect\ref{GL2}), in
which the unstable vortex with $S=1$ splits into a stably rotating dipole,
see Fig. \protect\ref{dipole} below.}
\label{periodic}
\end{figure}

A similar scenario of the instability development of 3D vortex solitons with
embedded vorticity $S=1$, trapped in the three-dimensional HO potential, was
revealed by simulations reported in Refs. \cite{Ueda,Ueda2}.

\paragraph{Two-component systems}

A noteworthy generalization of the above analysis\ was performed for a
system of two nonlinearly coupled fields, which may be realized as a binary
BEC, or as co-propagation of two optical beams in a bulk waveguide \cite%
{Brtka}. The respective coupled two-dimensional GPE/NLSE system is
\begin{subequations}
\begin{align}
i\frac{\partial \psi _{1}}{\partial t}& =\left[ -\frac{1}{2}\nabla ^{2}+%
\frac{1}{2}\left( x^{2}+y^{2}\right) -\left( |\psi _{1}|^{2}+\eta \left\vert
\psi _{2}\right\vert ^{2}\right) \right] \psi _{1},  \notag \\
&  \label{system} \\
i\frac{\partial \psi _{2}}{\partial t}& =\left[ -\frac{1}{2}\nabla ^{2}+%
\frac{1}{2}\left( x^{2}+y^{2}\right) -\left( |\psi _{2}|^{2}+\eta \left\vert
\psi _{1}\right\vert ^{2}\right) \right] \psi _{2},  \notag
\end{align}%
where $\eta $ is the relative strength of the attraction ($\eta >0$) or
repulsion ($\eta <0$) between the components. Note that this system
conserves the norm separately in each component,
\end{subequations}
\begin{equation}
N_{1,2}=\int \int \left\vert \psi _{1,2}\left( x,y\right) \right\vert
^{2}dxdy.  \label{N12}
\end{equation}

In addition to the symmetric states with equal vorticities of both
components, this system gives rise to composite modes with the HV (hidden
vorticity), formed by the components with equal norms $N/2$ and opposite
vorticities, $S_{1,2}=\pm 1$. The HV modes are stable, roughly, in the
region of $N<7$, $-1<\eta <0.2$. A related problem is the study of compound
states in which one component is fundamental ($S=0$) and the other carries
vorticity \cite{VPG-mixed}.

Another relevant two-component system models the dynamics of BEC\ in
parallel 2D layers, coupled by hopping of atoms between them, with the HO
trapping potential acting in each of them \cite{Viskol}:
\begin{subequations}
\begin{align}
i\frac{\partial \psi _{1}}{\partial t}& =\left[ -\frac{1}{2}\nabla ^{2}+%
\frac{1}{2}\left( x^{2}+y^{2}\right) -|\psi _{1}|^{2}\right] \psi
_{1}-\kappa \psi _{2},  \notag \\
&  \label{kappa} \\
i\frac{\partial \psi _{2}}{\partial t}& =\left[ -\frac{1}{2}\nabla ^{2}+%
\frac{1}{2}\left( x^{2}+y^{2}\right) -|\psi _{2}|^{2}\right] \psi
_{2}-\kappa \psi _{1},  \notag
\end{align}%
where $\kappa $ is the linear-coupling constant. Unlike system (\ref{system}%
), Eqs. (\ref{kappa}) conserve only the total norm, $N\equiv N_{1}+N_{2}$
(see Eq. (\ref{N12})), and the system does not admit HV states. A specific
effect produced by this system is the \textit{spontaneous symmetry breaking}
of two-component vortex modes with $S=1$, which gives rise to stable modes
with different amplitudes of the two components, see a typical example in
Fig. \ref{asymmetric_vortex}.
\begin{figure}[t]
\includegraphics[width=13cm]{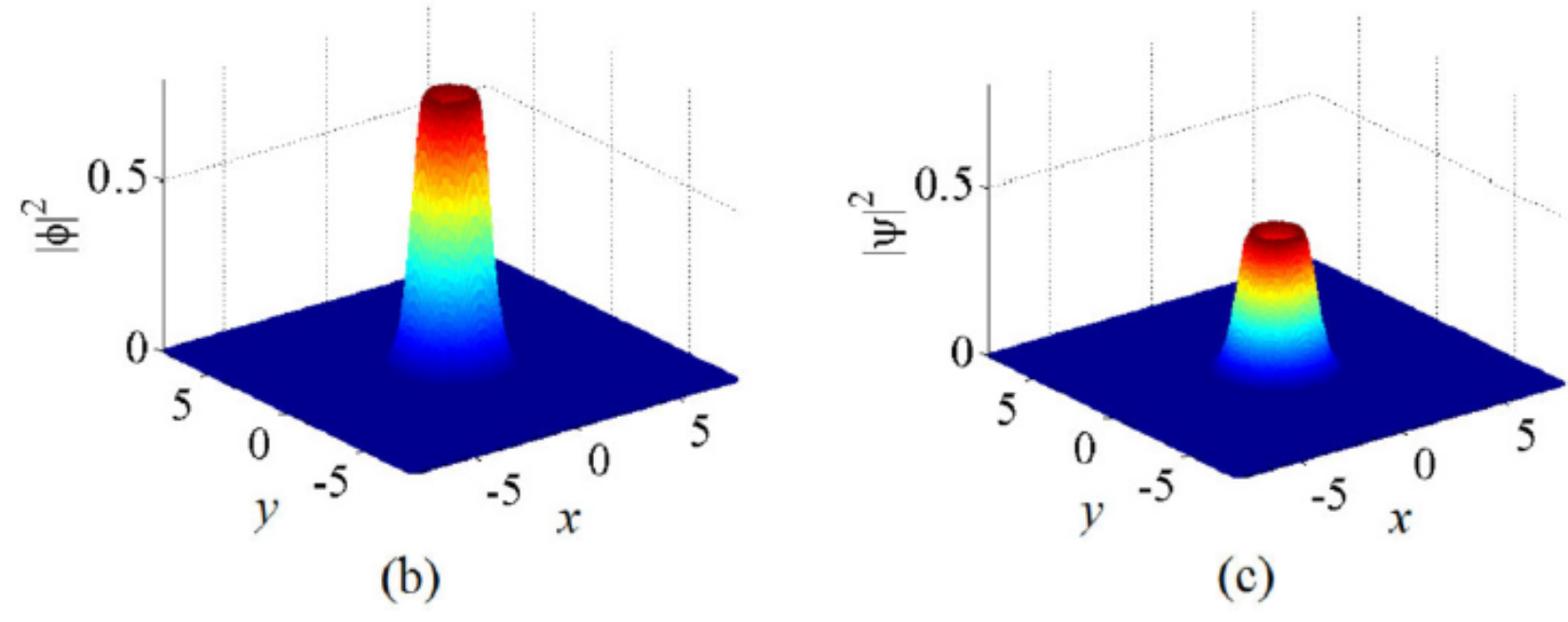}
\caption{Density profiles of two components of a stable asymmetric vortex
state with $S=1$ and total norm $N\equiv N_{1}+N_{2}=8.8$, produced by Eqs. (%
\protect\ref{kappa}) with $\protect\kappa =0.4$, as per Ref. \protect\cite%
{Viskol}.}
\label{asymmetric_vortex}
\end{figure}

The asymmetry of the binary vortex state is characterized by parameter
\end{subequations}
\begin{equation}
\theta \equiv \left( N_{1}-N_{2}\right) /\left( N_{1}+N_{2}\right) .
\label{theta}
\end{equation}%
A standard diagram for the symmetry-breaking \textit{bifurcation}
(transition to an asymmetric state) is usually displayed in the form of the $%
\theta (N)$ dependence. For the model based on Eqs. (\ref{kappa}), it is
plotted in Fig. \ref{symm-breaking}, for $\kappa =0.15$. It is seen that,
with the increase of the total norm, the symmetric vortex becomes unstable
at a critical (bifurcation) point, $N_{\mathrm{cr}}^{(S=1)}$. The
bifurcation gives rise to two mutually symmetric branches of stable
asymmetric vortex states, characterized by dependence $\pm \theta (N)$, a
typical example of which is shown in Fig. \ref{symm-breaking} (in the
figure, only the branch with $\theta >0$ is presented). Numerical findings
presented in Ref. \cite{Viskol} suggest a dependence of $N_{\mathrm{cr}%
}^{(S=1)}$ on the linear-coupling constant $\kappa $, which may be well
fitted by a simple linear relation,
\begin{equation}
N_{\mathrm{cr}}^{(S=1)}=0.57+19.06\kappa .  \label{fit}
\end{equation}%
\begin{figure}[t]
\includegraphics[width=6cm]{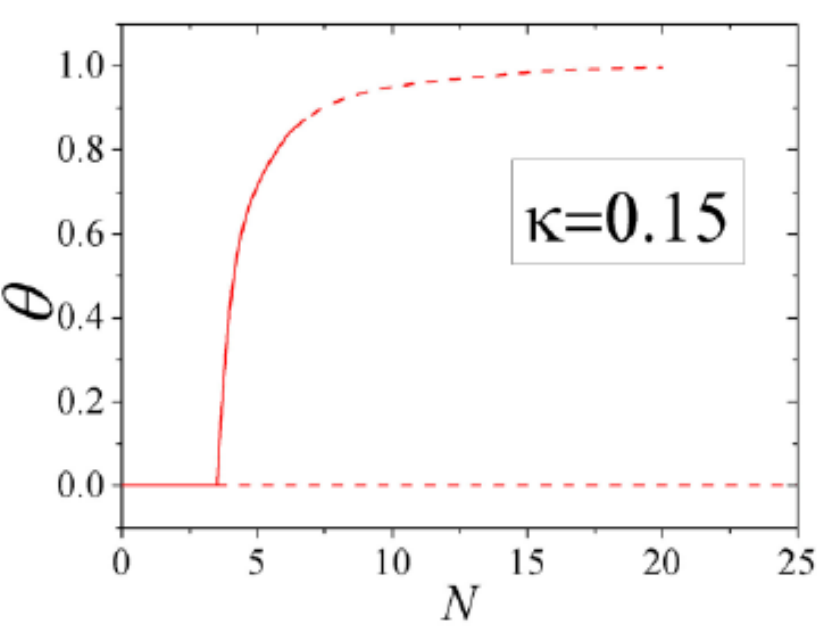}
\caption{The symmetry-breaking bifurcation diagram for two-component vortex
solitons produced by Eqs. (\protect\ref{kappa}), as per Ref. \protect\cite%
{Viskol}. The asymmetry parameter (\protect\ref{theta}) is shown as a
function of the total norm, $N$. Dashed lines represent unstable vortex
solitons.}
\label{symm-breaking}
\end{figure}

At still larger values of $N$, the asymmetric states, with one dominating
component, are destroyed by instability which is, generally, similar to that
outlined above for the single-component model. In Fig. \ref{symm-breaking},
solitons belonging to the dashed portion of the asymmetric branch are
subject to the latter instability.

\subsection{Vortex solitons supported by a spatially periodic (lattice)
potential}

\begin{figure}[t]
\includegraphics[width=13cm]{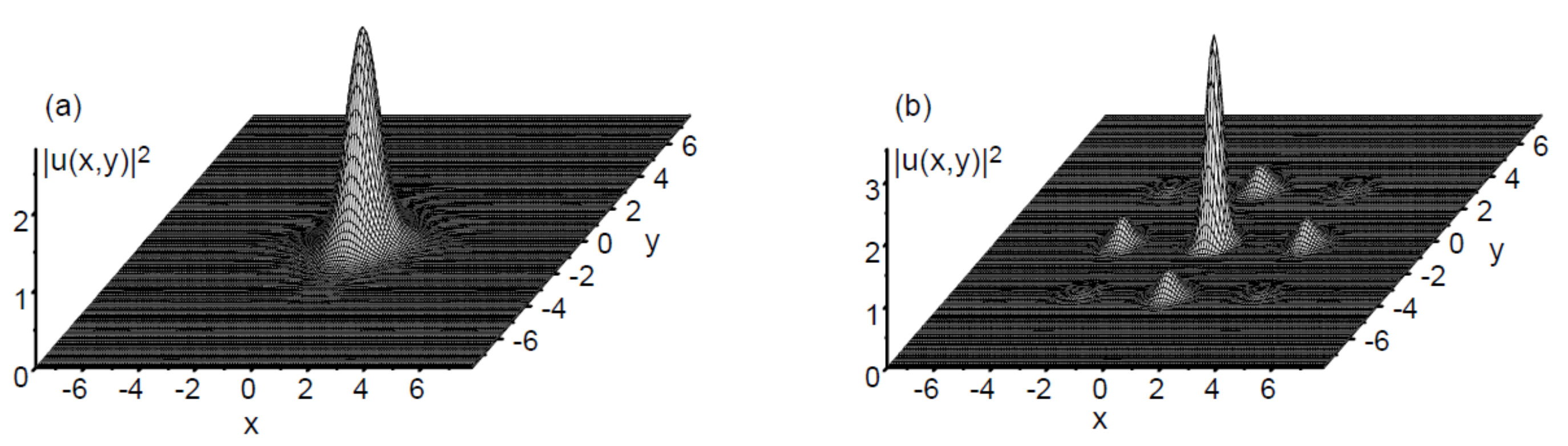}
\caption{Panel (b): an example of a stable on-site-centered vortex soliton,
with winding number $S=1$, supported by the lattice potential in the 2D
variant of Eq. (\protect\ref{OL}), with $k=2$, $\protect\varepsilon =5$, and
$g=1/2$, as per Ref. \protect\cite{lattice1}. The plot displays the density
distribution, $\left\vert \protect\psi \left( x,y\right) \right\vert ^{2}~$%
(the inset additionally shows its 1D cross section along the $x$ axis). The
2D norm of the soliton is $N=2\protect\pi $. The vorticity is represented by
the phase pattern (not shown here), with phase shifts $\protect\pi /2$
between four main peaks, which corresponds to the global phase circulation $2%
\protect\pi $. For comparison, panel (a) displays an example of a stable
fundamental soliton ($S=0$, with norm $N=10$), supported by the same model.}
\label{lattice-1}
\end{figure}

\begin{figure}[t]
\includegraphics[width=5cm]{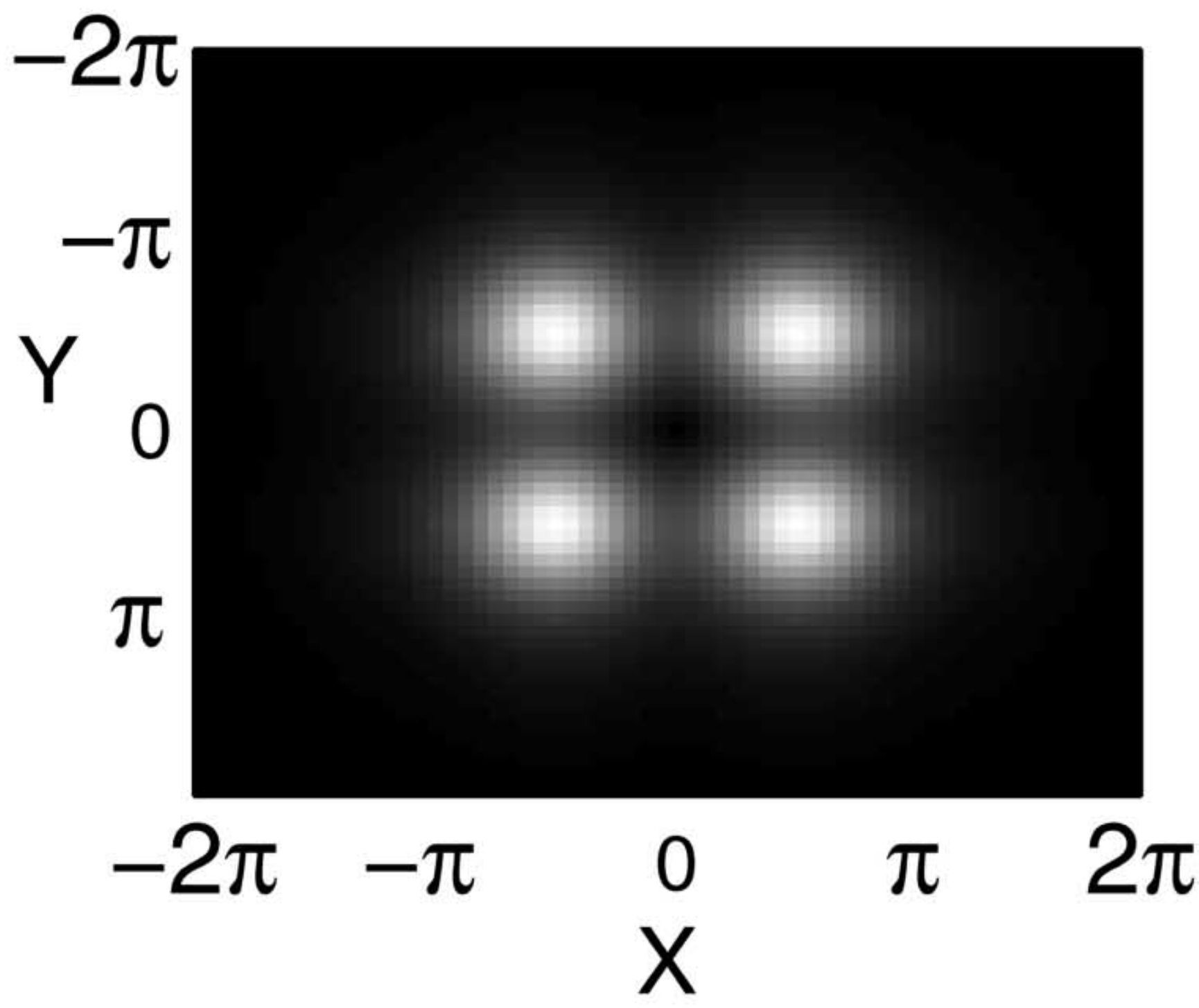}
\caption{An example of the density pattern in an off-site-centered vortex
soliton with $S=1$, supported by the 2D variant of Eq. (\protect\ref{OL}),
as per Ref. \protect\cite{lattice2}.}
\label{lattice-2}
\end{figure}

Creation and stabilization of solitons by means of lattice potentials was
addressed in many works, see reviews \cite%
{Brazh,Morsch,big,Vyslo1,Vyslo2,Yukalov,RMP,Entropy}, therefore these
results are presented here in a brief form. The possibility to stabilize 2D
solitons with embedded vorticity by means of the lattice potential was first
demonstrated in Refs. \cite{lattice1} and \cite{lattice2,lattice3}. As shown
in Figs. \ref{lattice-1} and \ref{lattice-2}, these vortex states do not
seem as familiar annulus-shaped (alias \textit{crater-shaped}) objects (cf.
Figs. \ref{splitting}(a), \ref{ln_vortices}(a1-a4), and \ref{periodic}%
(a,c)), but rather as circular chains of local peaks, in terms of the local
density. Vorticity $S$ is represented by the phase structure of these modes,
which feature the phase circulation of $2\pi S$ corresponding to a round
trip along a path encircling the mode's pivot. It is seen that the basic
structure of the lattice vortex with $S=1$ is represented by the set of four
density peaks, of two different types: rhombic, with a nearly empty central
site, which is displayed in Fig. \ref{lattice-1}, and a densely packed
square-shaped one, which does not include an empty central site, in Fig. \ref%
{lattice-2}. These two types of lattice vortices are usually called
on-site-centered and off-site centered ones, the latter type usually being
essentially more stable.

In addition to the basic on- and off-cite-centered vortex solitons with the
natural symmetric shape, the analysis has also revealed a possibility of the
existence of asymmetric ones, built as triangular or stretched rhombic and
rectangular multi-peak patterns \cite{Asymm}. A related result is a
possibility of replication of a vortex mode, originally created in one
potential well of a 2D double-well configuration, by building a twin vortex
in the adjacent well \cite{replication}.

\begin{figure}[t]
\includegraphics[width=10cm]{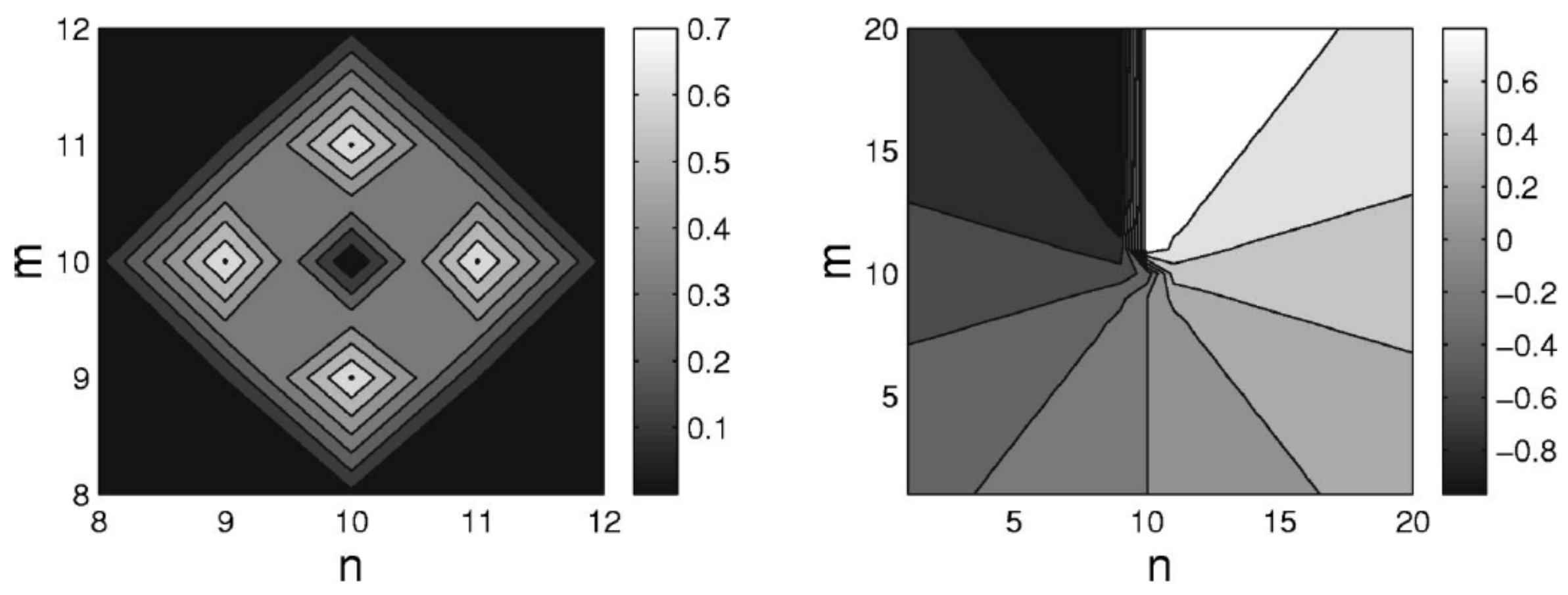}
\caption{Density and phase patterns of a stable discrete vortex soliton with
$S=1$, $\protect\mu =-0.32$ and $C=0.05$, generated by Eqs. (\protect\ref%
{DNLSE}) and (\protect\ref{stat-DNLSE}), as per Ref. \protect\cite{discrvort}%
.}
\label{discr-latt}
\end{figure}

In the limit of a very deep lattice potential, i.e., $\varepsilon
\rightarrow \infty $ in Eq. (\ref{OL}), the 2D version of Eq. (\ref{OL})
carries over into the discrete NLSE \cite{PGK-book}:%
\begin{equation}
i\frac{\partial \psi _{m,n}}{\partial t}+C\left( \psi _{m+1,n}+\psi
_{m-1,n}+\psi _{m,n+1}+\psi _{m,n-1}-4\psi _{m,n}\right) +\left\vert \psi
_{m,n}\right\vert ^{2}\psi _{m,n}=0,  \label{DNLSE}
\end{equation}%
where $\left( m,n\right) $ is the set of discrete coordinates replacing $%
\left( x,y\right) $, and real $C>0$ is a coefficient of the intersite
coupling in the discrete lattice. Stationary states are represented by
solutions to Eq. (\ref{DNLSE}) in the form of of%
\begin{equation}
\psi _{m,n}\left( t\right) =\exp \left( -i\mu t\right) u_{m,n},
\label{stat-DNLSE}
\end{equation}%
where the stationary lattice field $u_{m,n}$ is real for fundamental
discrete solitons, and complex for vortex solitons \cite{discrvort}. An
example of a numerically generated stable vortex discrete soliton with $S=1$
is displayed in Fig. \ref{discr-latt}. In discrete models, on-site-centered
vortex solitons also tend to be essentially more stable than their
off-site-centered counterparts \cite{discrvort}.

\begin{figure}[t]
\includegraphics[width=12cm]{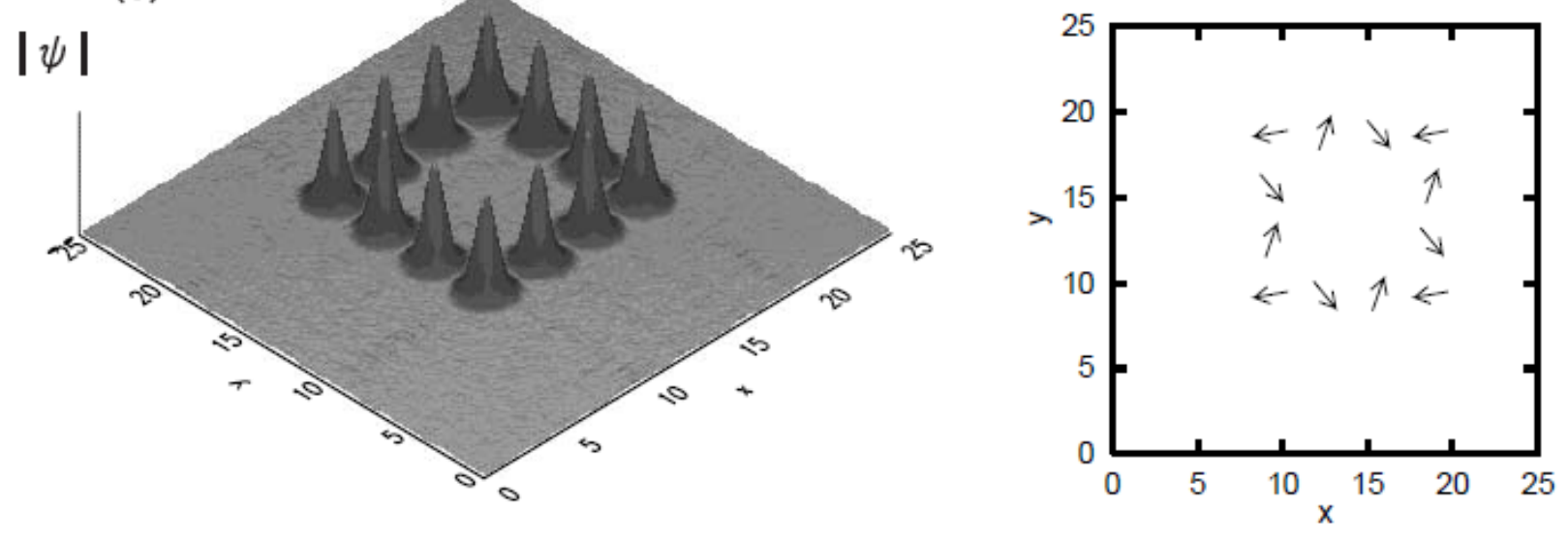}
\caption{A \emph{stable} higher-order vortex soliton, with winding number $%
S=4$ and total norm $N=33.4$, produced by the 2D variant of Eq. (\protect\ref%
{OL}) with $k=2$, $\protect\varepsilon =3$, and $g=1$, as per Ref.
\protect\cite{HS-EPL}. The left and right panels display, respectively, the
profile of $\left\vert \protect\psi \left( x,y\right) \right\vert $ and the
pattern of the phase vector with components $\left\{ \mathrm{Re}\left(
\protect\psi \left( x,y\right) \right) ,\mathrm{Im}\left( \protect\psi %
\left( x,y\right) \right) \right\} /\left\vert \protect\psi \left(
x,y\right) \right\vert $, at some moment of time. The total phase
circulation in this pattern is $8\protect\pi $, which corresponds to $S=4$.}
\label{S=4}
\end{figure}

The discrete NLSE supports vortex solitons with higher winding numbers, $%
S\geq 2$, but they are completely unstable \cite{discrvort}. Getting back to
the NLSE (\ref{OL}), stable higher-order vortex solitons were produced in
Ref. \cite{HS-EPL}. An example of a \emph{stable} circular-chain soliton
with $S=4$, built of $12$ local peaks, is shown in Fig. \ref{S=4}.

Furthermore, the 2D version of Eq. (\ref{OL}) supports more complex \textit{%
supervortex} complexes with two independent vorticities, local one $s$ and
global $S$ \cite{HS-EPL} (see also Ref. \cite{Radik2}). As shown in Fig. \ref%
{supervortex}, the supervortex is built as a circular chains of compact
local vortices with winding number $s=1$, each squeezed into one cell of the
lattice potential, and global vorticity $S$ imprinted onto the chain. Thus,
the supervortices with global vorticities $\pm S$ and fixed local one $s$ are%
\emph{\ different states}. In particular, for $\varepsilon =10$ in Eq. (\ref%
{OL}) and $s=1$, the supervortices are stable for $S=\pm 1$ and $\pm 2$,
unstable for $|S|>3$, and marginally stable for $S=\pm 3$ \cite{discrvort}.
\begin{figure}[t]
\includegraphics[width=12cm]{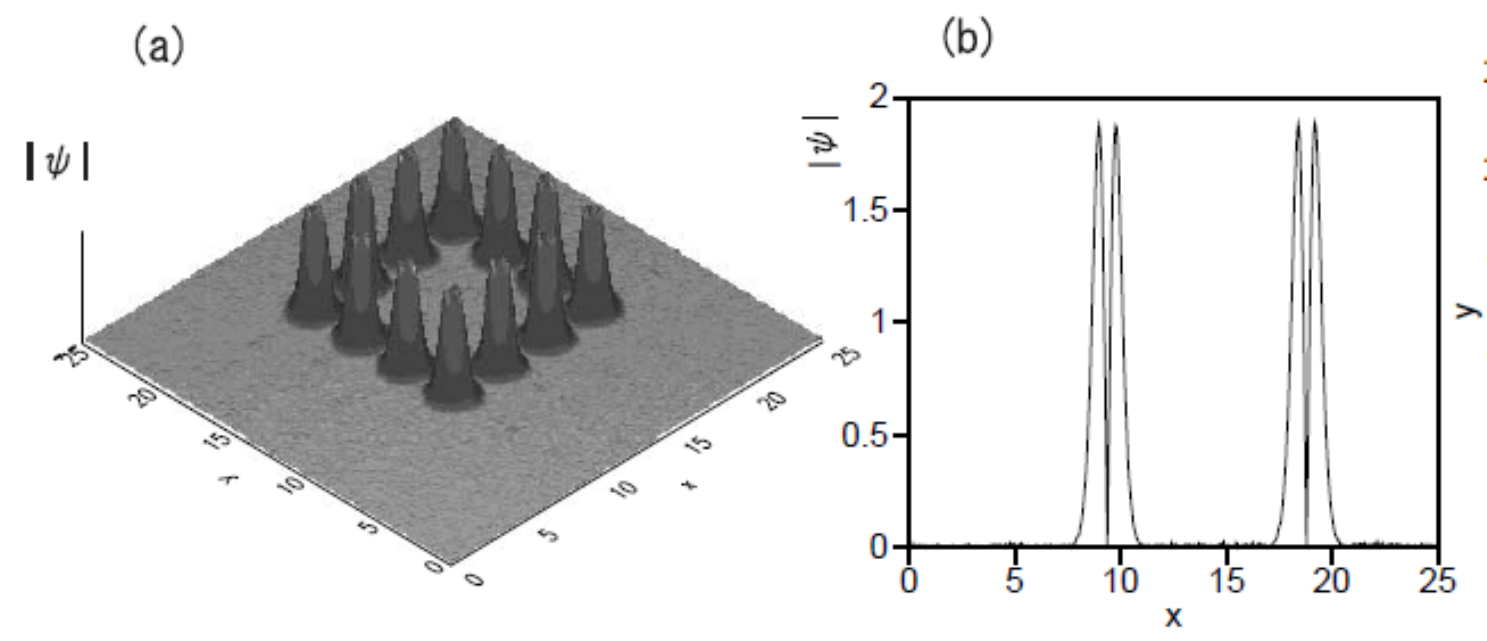}
\caption{A stable \textit{supervortex} complex, with local and global
vorticities $s=S=1$, and total norm $N=55.5$, produced by the 2D variant of
Eq. (\protect\ref{OL}) with $k=2$, $\protect\varepsilon =10$, and $g=1$, as
per Ref. \protect\cite{HS-EPL}. Panels (a) and (b) display, respectively,
the 2D profile of $\left\vert \protect\psi \left( x,y\right) \right\vert $
and the same in the 1D cross section, $\left\vert \protect\psi \left(
x,0\right) \right\vert $ (the latter one directly shows that each local
constituent of the complex is a compact vortex with the inner hole).}
\label{supervortex}
\end{figure}

Lastly, the full 3D form of the model based on Eq. (\ref{OL}) with the
quasi-2D lattice potential, $\sigma =0$, gives rise to stable 3D solitons
with embedded vorticities $S=1$ and $2$, which are also built as chains of
local soliton-like objects in the $\left( x,y\right) $ plane, each being
self-trapped in the transverse direction, see an example in Fig. \ref%
{Q2D-vortex}.
\begin{figure}[t]
\includegraphics[width=8cm]{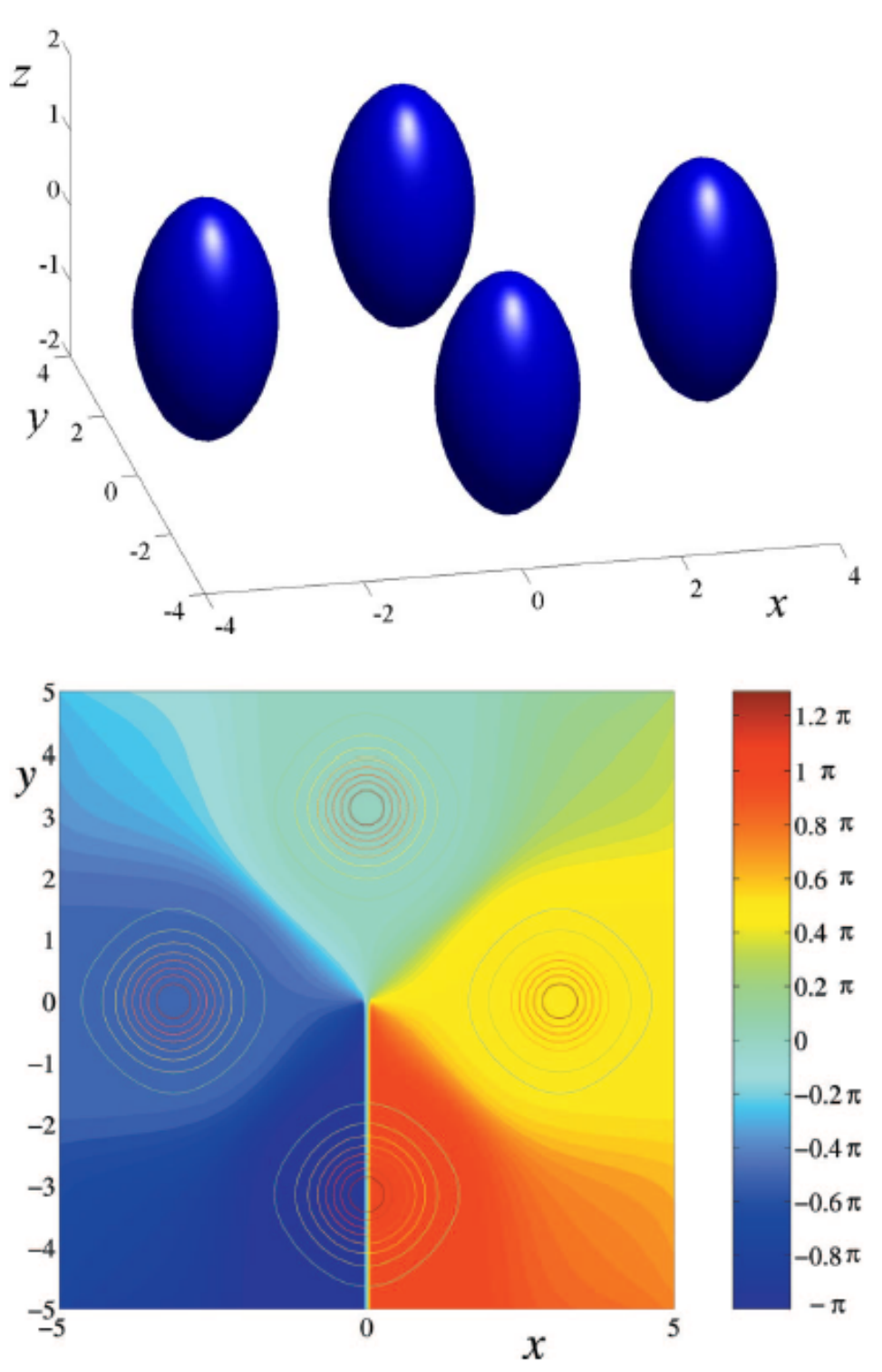}
\caption{A stable 3D vortex soliton with $S=1$ and total 3D norm $N=22.3$,
produced by Eq. (\protect\ref{OL}) in its 3D form, but with the quasi-2D
lattice potential, i.e., $\protect\sigma =0$, and $k=2$, $\protect%
\varepsilon =1.25$, $g=1$, as per Ref. \protect\cite{Herve-lowdim}. The top
and bototm panels display, respectively, the 3D amplitude profile, defined
by $\left\vert \protect\psi \left( x,y,z\right) \right\vert =1.45$, and the
phase pattern in the midplane, $z=0$, superimposed on the contour plot of $%
\left\vert \protect\psi \left( x,y,0\right) \right\vert $. }
\label{Q2D-vortex}
\end{figure}

\section{Stabilization of 2D and 3D semi-vortex and mixed-mode solitons by
the spin-orbit coupling (SOC)}

This section aims to summarize recent findings which put forward a
completely new approach to the stabilization of vorticity-carrying 2D and 3D
solitons in free space (without the use of any external potential), realized
in models of the binary (pseudo-spinor) atomic BEC with the SOC. The
presentation is chiefly based on Refs. \cite{we}, \cite{HP}, and \cite%
{Sherman2}.

\subsection{The models}

A great deal of attention has been lately drawn to the use of ultracold
quantum gases, both bosonic and fermionic, as \textit{simulators} of
various\ fundamental effects that were previously predicted and/or
discovered experimentally in much more complex settings of condensed-matter
physics \cite{simulator}. In particular, much interest has been recently
attracted to the implementation of the (pseudo-) SOC in atomic BEC, as an
efficient emulation of the fundamental SOC in semiconductors, where the
direct SOC is induced by the\ coupling of the electron's magnetic moment to
the magnetic field generated by the intrinsic electrostatic field of the
underlying ionic lattice, in the reference frame moving along with the
electron \cite{Dresselhaus,Rashba}.The experimental implementation of the
pseudo-SOC was proposed \cite{Campbell} and realized in the condensate of $%
^{87}$Rb atoms, using appropriately designed laser illumination and magnetic
fields \cite{socbec,socbec2,socbec3,socbec4}. Parallel to the experiments,
many theoretical studies on this topic have been carried out \cite%
{theory-SOC}-\cite{Fukuoka2}, \cite{we,HP}, see also reviews \cite{rf:1}-%
\cite{rf:15}. The SOC emulation in the atomic condensate is provided by
mapping the spinor wave function of semiconductor electrons into a
two-component pseudo-spinor wave function of the binary BEC composed of
atoms in two different hyperfine states. Namely, a pair of states of the $%
^{87}$Rb atom, $\left\vert \psi _{+}\right\rangle =\left\vert
F=1,m_{F}=0\right\rangle $ and $\left\vert \psi _{-}\right\rangle
=\left\vert F=1,m_{F}=-1\right\rangle $, were used to map the spin-up and
spin-down electron's wave function into them \cite%
{socbec,socbec2,socbec3,socbec4}. Thus, the dynamics of the \emph{fermionic}
wave functions of electrons in the semiconductor may be emulated by the
mean-field dynamics of the \emph{bosonic} gas.

The consideration of the interplay of the SOC, which is, essentially, linear
mixing between the two components of the spatially inhomogeneous binary BEC,
and the intrinsic nonlinearity in the bosonic condensate has made it
possible to predict diverse nonlinear patterns strongly affected or created
by the SOC, including 1D solitons \cite{Konotop}-\cite{Konotop5}, 2D gap
solitons supported by OL potentials \cite{gap-sol}, and 2D vortices and
vortex lattices, in forms specific to the spin-orbit-coupled BEC \cite%
{Fukuoka}-\cite{Fukuoka10}.

The 2D model of the binary SOC BEC is based on the following system of
coupled GPEs for two components of the pseudo-spinor wave function,$\Psi
\equiv \left\{ \psi _{+},\psi _{-}\right\} $ \cite{we}:
\begin{gather}
\left[ i\frac{\partial }{\partial t}+\frac{1}{2}\nabla ^{2}+i\lambda \left(
-\sigma _{y}\frac{\partial }{\partial x}+\sigma _{x}\frac{\partial }{%
\partial y}\right) \right.  \notag \\
\left. +\left(
\begin{array}{cc}
|\psi _{+}|^{2}+\eta |\psi _{-}|^{2} & 0 \\
0 & |\psi _{-}|^{2}+\eta |\psi _{+}|^{2}%
\end{array}%
\right) \right] \left(
\begin{array}{c}
\psi _{+} \\
\psi _{-}%
\end{array}%
\right) =0,  \label{R2D}
\end{gather}%
where $\sigma _{x,y,z}$ are the Pauli matrices, $\lambda $ is a real
coefficient of the SOC of the Rashba type (a combination with the SOC of the
Dresselhaus type, which is modeled by combination $\sigma _{x}\partial
/\partial x-\sigma _{y}\partial /\partial y$, instead of $-\sigma
_{y}\partial /\partial x+\sigma _{x}\partial /\partial y$ in Eq. (\ref{R2D}%
), is not considered here, as it tends to destroy 2D solitons \cite{Sherman2}%
), $\eta $ is the relative strength of the cross-attraction between the
components (cf. Eq. (\ref{system})), while the strength of the
self-attraction is normalized to be $1$. Coefficient $1/\lambda $ has the
dimension of length, defining a fixed scale which breaks the scale
invariance of the NLSE/GPE in the free 2D space and thus makes it possible,
as mentioned above, to create stable solitons with norms falling below the
collapse threshold.

Stationary solutions of Eq. (\ref{R2D}) for 2D solitons with real chemical
potential $\mu $ are looked for as $\psi _{\pm }=\exp \left( -i\mu t\right)
u_{\pm }\left( x,y\right) $, where complex stationary wave functions are
determined by equations%
\begin{gather}
\mu u_{+}=-\frac{1}{2}\nabla ^{2}u_{+}-(|u_{+}|^{2}+\eta |u_{-}|^{2})u_{+}+
\notag \\
\left( \frac{\partial u_{-}}{\partial x}-i\frac{\partial u_{-}}{\partial y}%
\right) ,  \label{+} \\
\mu u_{-}=-\frac{1}{2}\nabla ^{2}u_{-}-(|u_{-}|^{2}+\eta |u_{+}|^{2})u_{-}-
\notag \\
\left( \frac{\partial u_{+}}{\partial x}+i\frac{\partial u_{+}}{\partial y}%
\right) .  \label{-}
\end{gather}%
Dynamical invariants of Eqs. (\ref{R2D}) are the same total norm $N$ as
defined above, Hamiltonian, and linear momentum:%
\begin{gather}
H=\int \int \left\{ \frac{1}{2}\left( |\nabla \psi _{+}|^{2}+|\nabla \psi
_{-}|^{2}\right) -\frac{1}{2}\left( |\psi _{+}|^{4}+|\psi _{-}|^{4}\right)
-\eta |\psi _{+}|^{2}|\psi _{-}|^{2}\right.  \notag \\
\left. +\lambda \left[ \psi _{+}^{\ast }\left( \frac{\partial \psi _{-}}{%
\partial x}-i\frac{\partial \psi _{-}}{\partial y}\right) +\psi _{-}^{\ast
}\left( -\frac{\partial \psi _{+}}{\partial x}-i\frac{\partial \psi _{+}}{%
\partial y}\right) \right] \right\} dxdy,  \label{ER}
\end{gather}%
\begin{equation}
\mathbf{P}=i\int \int \left( \psi _{+}^{\ast }\nabla \psi _{+}+\psi
_{-}^{\ast }\nabla \psi _{-}\right) dxdy.  \label{P}
\end{equation}

The consistent derivation of the effective 2D SOC\ model from the full 3D
system of GPEs may give rise to the 2D equations with nonpolynomial
nonlinearity, as a generalization of the cubic terms in Eq. (\ref{R2D}) \cite%
{Wesley}. Such a generalized system also creates stable solitons in the 2D
free space. Another relevant generalization addresses a model of a dual-core
nonlinear coupler in optics, where SOC is emulated by temporal dispersion of
the linear inter-core coupling \cite{Kart}. It was demonstrated that this
model gives rise to stable 2D spatiotemporal solitons.

The 3D model is taken here with the SOC of the Weyl type \cite{HP}:%
\begin{gather}
\left[ i\frac{\partial }{\partial t}+\frac{1}{2}\nabla ^{2}+i\lambda \nabla
\cdot {\boldsymbol{\sigma }}\right.  \notag \\
\left. +\left(
\begin{array}{cc}
|\psi _{+}|^{2}+\eta |\psi _{-}|^{2} & 0 \\
0 & |\psi _{-}|^{2}+\eta |\psi _{+}|^{2}%
\end{array}%
\right) \right] \left(
\begin{array}{c}
\psi _{+} \\
\psi _{-}%
\end{array}%
\right) =0,  \label{3D}
\end{gather}%
where $\lambda $ is again the SOC coefficient, and the 3D matrix vector is ${%
\boldsymbol{\sigma =}}\left\{ \sigma _{x},\sigma _{y},\sigma _{z}\right\} $.
The 3D system conserves the norm and linear momentum, along with the
Hamiltonian,%
\begin{gather}
E_{\mathrm{tot}}=E_{\mathrm{kin}}+E_{\mathrm{int}}+E_{\mathrm{SOC}}\,,
\label{eq1} \\
E_{\mathrm{kin}}=\frac{1}{2}\int \int \int \,\left( |\nabla \psi
_{+}|^{2}+|\nabla \psi _{-}|^{2}\right) dxdydz,~  \notag \\
E_{\mathrm{int}}=-\frac{1}{2}\int \int \int \,\left( |\psi _{+}|^{4}+|\psi
_{-}|^{4}+2\eta |\psi _{+}\psi _{-}|^{2}\right) dxdydz\,  \notag \\
E_{\mathrm{SOC}}=-i\lambda \int \int \int \,\Psi ^{\dag }\left( \nabla \cdot
{\boldsymbol{\sigma }}\right) \Psi dxdydz.  \notag
\end{gather}

\subsection{Stable 2D solitons: quiescent and mobile semi-vortices and mixed
modes}

\subsubsection{2D semi-vortices}

Unlike the models considered above, which give rise to fundamental states
with zero vorticity, a vortical component is inherently present in any
self-trapped state generated by the nonlinear SOC systems. First, Eq. (\ref%
{R2D}) admits stationary solutions with real chemical potential $\mu $,
written in terms of the polar coordinates, $\left( r,\theta \right) $:%
\begin{equation}
\psi _{+}\left( x,y,t\right) =e^{-i\mu t}f_{1}(r),~\psi _{-}\left(
x,y,t\right) =e^{-i\mu t+i\theta }rf_{2}(r),  \label{frf}
\end{equation}%
where real functions $f_{1,2}\left( r\right) $ take finite values and have
zero derivatives at $r=0$, and feature the following asymptotic form at $%
r\rightarrow \infty $:%
\begin{equation}
f_{1}\approx Fr^{-1/2}e^{-\sqrt{-2\mu -\lambda ^{2}}r}\cos \left( \lambda
r+\delta \right) ,~f_{2}\approx -Fr^{-3/2}e^{-\sqrt{-2\mu -\lambda ^{2}}%
r}\sin \left( \lambda r+\delta \right) ,  \label{asympt}
\end{equation}%
with constants $F$ and $\delta $. As it follows from Eq. (\ref{asympt}), the
solutions may be exponentially localized at
\begin{equation}
\mu <-\lambda ^{2}/2.  \label{mulambda}
\end{equation}

Solutions (\ref{frf})\ are built as bound states of a fundamental
(zero-vorticity, $S_{+}=0$) soliton in component $\psi _{+}$ and a solitary
vortex, with vorticity $S_{-}=1$, in $\psi _{-}$, therefore composite modes
of this type are called \textit{semi-vortices} (SVs) \cite{we}. The
invariance of Eq. (\ref{GPE}) with respect to transformation
\begin{equation}
\psi _{{\pm }}\left( r,\theta \right) \rightarrow \psi _{{\mp }}\left( r,\pi
-\theta \right)  \label{transform}
\end{equation}%
gives rise to a conjugate semi-vortex, which is a mirror image of (\ref{frf}%
), with $\left( S_{+}=0,S_{-}=1\right) $ replaced by $\left(
S_{+}=-1,S_{-}=0\right) $:%
\begin{equation}
\psi _{+}\left( x,y,t\right) =-e^{-i\mu t-i\theta }rf_{2}(r),~\psi
_{-}=e^{-i\mu t}f_{1}(r).  \label{mirror}
\end{equation}

\begin{figure}[b]
\begin{center}
\includegraphics[height=3.5cm]{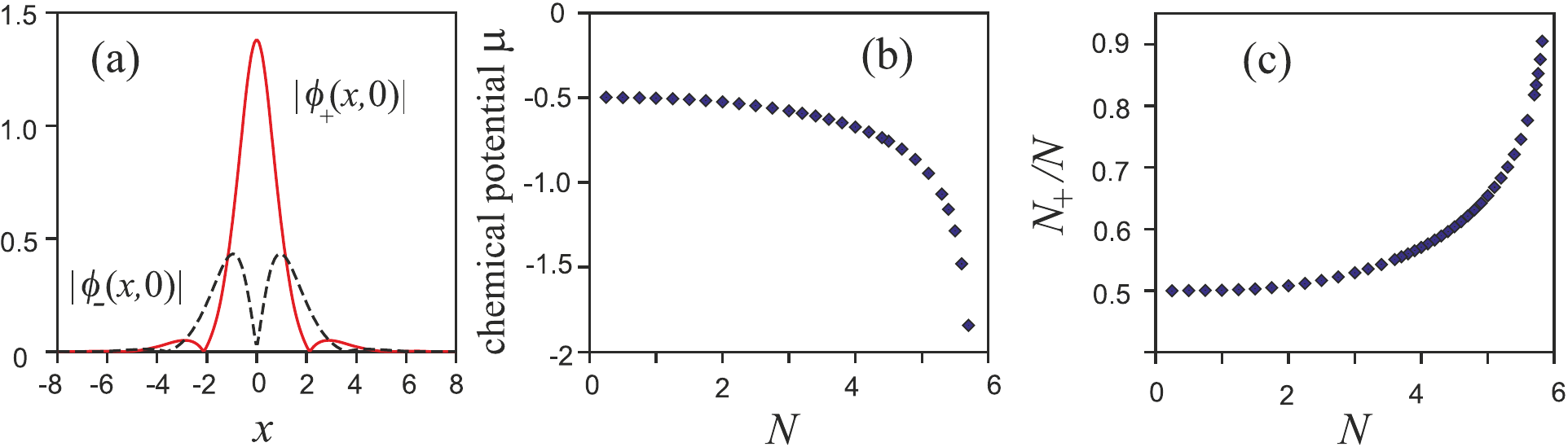}
\end{center}
\caption{(a) Cross sections of the fundamental, $\left\vert \protect\psi %
_{+}\left( x,y\right) \right\vert $, and vortical, $\left\vert \protect\psi %
_{-}\left( x,y\right) \right\vert $, components of a 2D semi-vortex (SV)
with $N=5$, along axis $y=0$, are shown by continuous and dashed lines,
respectively, as per Refs. \protect\cite{we} and \protect\cite{Sherman2}.
(b) Chemical potential $\protect\mu $ vs. norm $N$ for the SV\ family. (c)
The share of the norm of the zero-vorticity component, $N_{+}$, in the total
SV's norm $N$, as a function of $N$. In this figure, $\protect\lambda =1$
and $\protect\eta =0$ are fixed in Eq. (\protect\ref{R2D}).}
\label{SV}
\end{figure}

Numerically, stable SVs can be readily generated, as solutions to Eq. (\ref%
{GPE}), by means of imaginary-time simulations \cite{im-time}-\cite{im-time4}%
, starting from the Gaussian input,
\begin{equation}
\psi _{+}^{(0)}=A_{1}\exp \left( -\alpha _{1}r^{2}\right) ,\;\psi
_{-}^{(0)}=A_{2}r\exp \left( i\theta -\alpha _{2}r^{2}\right) ,  \label{00}
\end{equation}%
where $A_{1,2}$ and $\alpha _{1,2}>0$ are real constants. A typical example
of the SV is displayed, by means of cross sections of its components, in
Fig. \ref{SV}(a).

Further, Fig. \ref{SV}(b) represents the family of the SVs, showing their
chemical potential as a function of the norm. Note that the $\mu (N)$
dependence satisfies the VK criterion, $d\mu /dN<0$, which is the
above-mentioned necessary condition for the stability of solitary modes
supported by the self-attractive nonlinearity \cite{VK,Berge,Gadi}. The
family of the SV solitons exists precisely in the interval of norms (\ref%
{Nmax}), which, as said above, should secure their stability against the
critical collapse. It is also worthy to note that there is no finite minimum
(threshold) value of $N$ necessary for the existence of the SVs in the free
space. In the limit of $\mu \rightarrow -\infty $, the vortex component of
the SV vanishes, while the fundamental one degenerates into the usual TS,
with $N=N_{\mathrm{TS}}$, as shown by means of the dependence of ratio $%
N_{+}/N$ on $N$ in Fig. \ref{SV}(c).

Systematic real-time simulations confirm the stability of the whole SV
family at $\eta \leq 1$, while they are unstable at $\eta >1$ \cite{we},
where, however, there is another family of stable solitons in the form of
\textit{mixed modes} (MMs), see below. In fact, the SVs at $\eta \leq 1$ and
MMs at $\eta \geq 1$ are the first ever found examples of stable solitons
supported by the cubic self-attractive nonlinearity in the free 2D space.

The study of 2D\ SV solitons was extended to a system with long-range
anisotropic cubic interactions, mediated by dipole-dipole forces in a
bosonic gas composed of atoms carrying magnetic moments \cite{Raymond}. An
essentially novel feature found in the nonlocal model is spontaneous shift
of the pivot of the vortical component ($\psi _{-}$) with respect to its
zero-vorticity counterpart ($\psi _{+}$). Those solitons also demonstrate a
mobility scenario which is essentially different from the one outlined below
for the present model: they respond to an applied kick by drift in the
opposite direction (i.e., with an effective negative mass) along a spiral
trajectory.

\subsubsection{2D mixed modes}

Aside from the SVs, the same SOC system (\ref{R2D}) gives rise to another
type of vorticity-carrying solitons, in the form of MMs, which combine terms
with zero and unitary vorticities, $\left( S=0,S=-1\right) $ and $\left(
S=0,S=+1\right) $, in the spin-up and spin-down components, $\psi _{+}$ and $%
\psi _{-}$. Numerically, the MM can be produced by imaginary-time
simulations initiated by the following input:
\begin{eqnarray}
\psi _{+}^{(0)} &=&A_{1}\exp \left( -\alpha _{1}r^{2}\right) -A_{2}r\exp
\left( -i\theta -\alpha _{2}r^{2}\right) ,  \notag \\
\psi _{-}^{(0)} &=&A_{1}\exp \left( -\alpha _{1}r^{2}\right) +A_{2}r\exp
\left( i\theta -\alpha _{2}r^{2}\right) .  \label{mixed}
\end{eqnarray}%
In fact, the MM may be considered as a superposition of the SV\ (\ref{frf})
and its mirror image (\ref{mirror}). Accordingly, symmetry reflection (\ref%
{transform}) transforms the MM into itself.

\begin{figure}[tbp]
\begin{center}
\includegraphics[height=3.5cm]{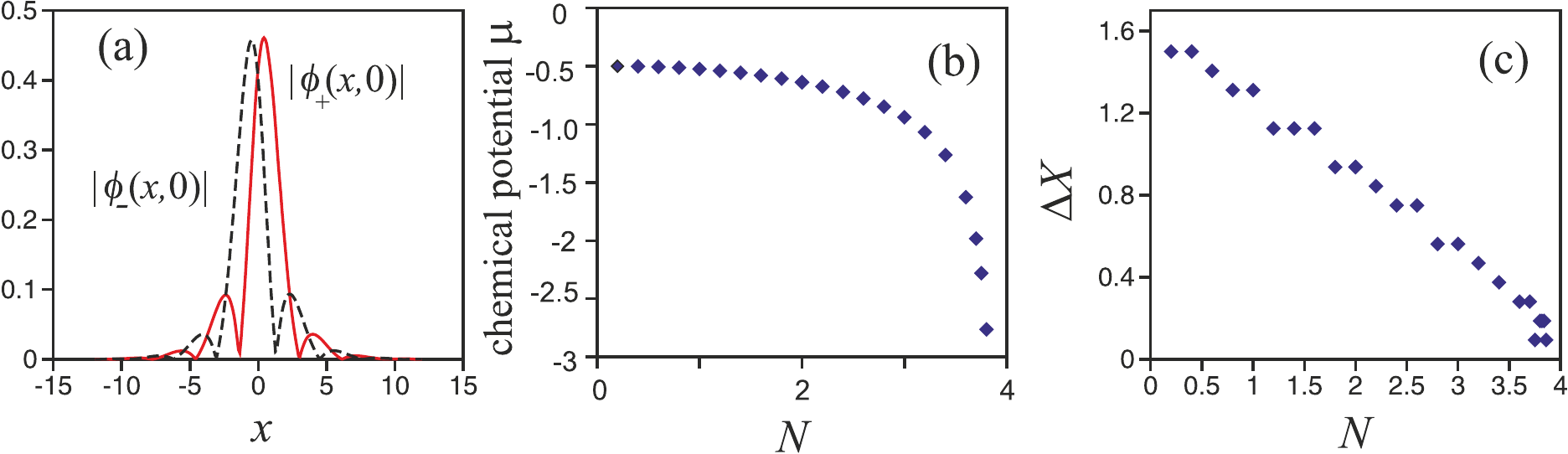}
\end{center}
\caption{Panels (a) and (b) have the same meaning as in Fig. \protect\ref{SV}%
, but for a 2D mixed-mode (MM) soliton with $\protect\eta =2$ and $\protect%
\lambda =1$ in Eq. (\protect\ref{R2D}). The norm of the soliton shown in
panel (a) is $N=5$. (c) Separation $\mathrm{DX}$ between peak positions of
components $|\protect\psi _{+}|$ and $|\protect\psi _{-}|$ vs. $N$. The
results are borrowed from Refs. \protect\cite{we} and \protect\cite{Sherman2}%
.}
\label{MM}
\end{figure}

A typical example of the MM and the respective $\mu (N)$ dependence are
displayed in Figs. \ref{MM}(a) and (b). Note that peak positions of the two
components, $|\psi _{+}\left( x,y\right) |$ and $|\psi _{-}\left( x,y\right)
|$, in this state are separated along $x$, Fig. \ref{MM}(c) showing the
separation ($\mathrm{DX}$) as a function of the norm. For a small amplitude
of the vortex component, $A_{2}$, Eq. (\ref{mixed}) yields $\mathrm{DX}%
\approx A_{2}/\left( \alpha _{1}A_{1}\right) $.

The $\mu (N)$ dependence for the MM family shows in Fig. \ref{MM}(b) that
the VK criterion holds in this case too, and, as well as SVs, the MMs do not
have any threshold value of $N$ necessary for their existence. The family
exists in the interval of $N<\tilde{N}_{\mathrm{TS}}(\eta )=2N_{\mathrm{TS}%
}/(1+\eta )$, where $N_{\mathrm{TS}}$ is the same critical (TS)\ norm as in
Eq. (\ref{Nmax}). In the limit of $N\rightarrow \tilde{N}_{\mathrm{TS}}(\eta
)$ the vortex components vanish in the MM, and it degenerates into a
two-component TS, cf. the above-mentioned degeneration of the SV in the
limit of $N\rightarrow N_{\mathrm{TS}}$. Separation $\mathrm{DX}$ between
peaks of the two components vanishes in this limit too, see Fig. \ref{MM}(c).

Direct simulations demonstrate that the MMs are unstable at $\eta <1$, and
stable at $\eta \geq 1$, i.e., precisely in the regions where the SVs are,
severally, stable and unstable \cite{we} (exactly at $\eta =1$, both the SV
and MM solutions are stable \cite{Fukuoka2}). The stability switch between
the SV and MM is explained by the comparison of energy (\ref{ER}) for them
at equal values of the norm: the energy is smaller for the SV at $\eta <1$,
and for the MM at $\eta >1$ \cite{we}. Accordingly, the SV and MM realize
the system's stable GS, respectively, at $\eta <1$ and $\eta >1$.

Lastly, it is relevant to mention that 2D solitons of the MM type,
additionally stabilized by the LHY correction to the model of the binary
condensate with SOC, were considered too \cite{SOC-LHY}.

\subsubsection{Mobility of stable 2D solitons}

Although the underlying system (\ref{R2D}) conserves the momentum (\ref{P}),
the SOC terms break the Galilean invariance of the original NLSEs. For this
reason, generating moving solitons from quiescent ones, which were
considered above, is a nontrivial problem. As shown in Ref. \cite{we}, the
system gives rise to the mobility of 2D solitons of the MM type along the $y$
axis, but not along $x$ (note that the essential anisotropy of the MM modes
with respect to the $x$ and $y$ directions manifests itself by the splitting
of peaks of the two components along the $x$ direction, see Figs. \ref{MM}%
(a) and (c)). The respective solutions moving at velocity $v_{y}$ can be
looked for as
\begin{equation}
\psi _{\pm }=\exp \left( iv_{y}y-\frac{i}{2}v_{y}^{2}t\right) \phi _{\pm
}(x;y^{\prime }\equiv y-v_{y}t;t)  \label{phi}
\end{equation}%
(Eq. (\ref{phi}) produces the Galilean transform of the wave functions $\psi
_{\pm }$ in Galilean-invariant systems). The substitution of ansatz (\ref%
{phi}) into Eq. (\ref{R2D}) leads to the coupled GPEs in the moving
reference frame, which differ from Eqs. (\ref{R2D}) by the presence of
linear mixing between the two components \cite{we}:
\begin{eqnarray}
i\frac{\partial \phi _{+}}{\partial t} &=&-\frac{1}{2}\nabla ^{2}\phi
_{+}-(|\phi _{+}|^{2}+\eta |\phi _{-}|^{2})\phi _{+}^{\prime }+\lambda
\left( \frac{\partial \phi _{-}}{\partial x}-i\frac{\partial \phi _{-}}{%
\partial y^{\prime }}\right) +\lambda v_{y}\phi _{-},  \notag \\
i\frac{\partial \phi _{-}}{\partial t} &=&-\frac{1}{2}\nabla ^{2}\phi
_{-}-(|\phi _{-}|^{2}+\eta |\phi _{+}|^{2})\phi _{-}-\lambda \left( \frac{%
\partial \phi _{+}}{\partial x}+i\frac{\partial \phi _{+}}{\partial
y^{\prime }}\right) +\lambda v_{y}\phi _{+}  \label{mix}
\end{eqnarray}%
(here, $\nabla ^{2}\equiv \partial ^{2}/\partial x^{2}+\partial
^{2}/\partial \left( y^{\prime }\right) ^{2}$).

\begin{figure}[tbp]
\begin{center}
\includegraphics[height=3.5cm]{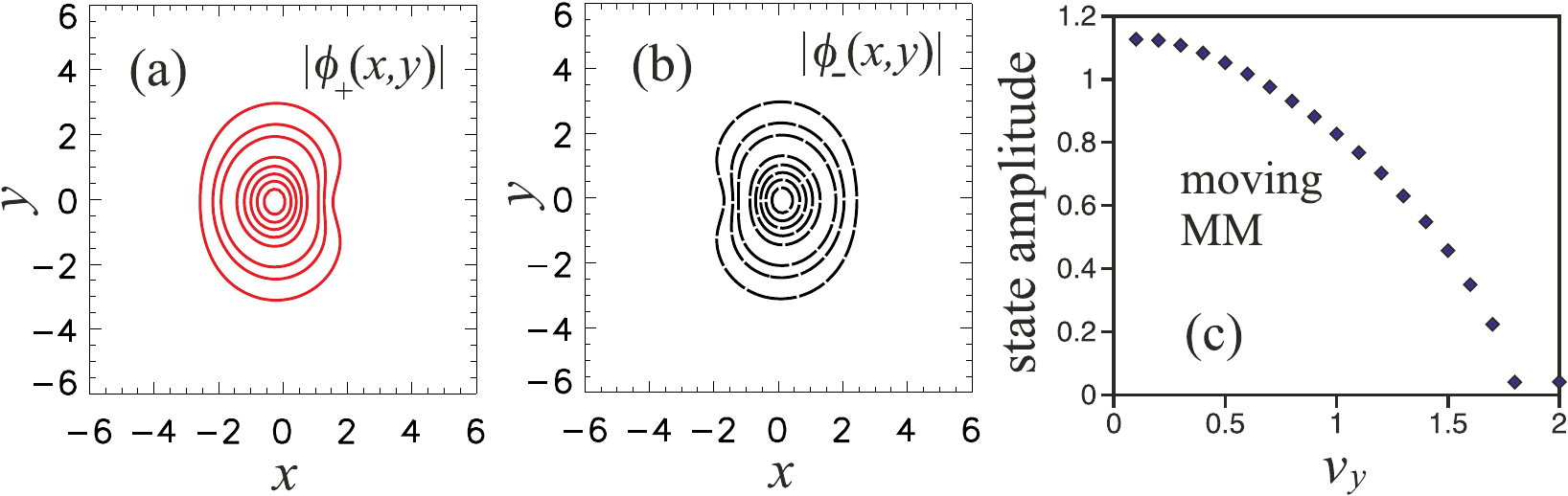}
\end{center}
\caption{Contour plots of $|\protect\psi _{+}\left( x,y\right) |$ (a) and $|%
\protect\psi _{-}\left( x,y\right) |$ (b) of the 2D stable MM (mixed-mode)
soliton with norm $N=3.1$, moving at velocity $v_{y}=0.5$ , for $\protect%
\eta =2$ and $\protect\lambda =1$, as per Refs. \protect\cite{we} and
\protect\cite{Sherman2}. (c) The amplitude of the moving solitons as a
function of $v_{y}$.}
\label{moving}
\end{figure}

Stationary solutions to equations (\ref{mix}) can be obtained, as well as in
the case of Eqs. (\ref{R2D}), by means of the imaginary-time-evolution
method. In particular, at $\eta =2$, when the GS is represented by the
quiescent MM soliton, its moving version, which is displayed in Figs. \ref%
{moving}(a,b) for $N=3.1$ and $v_{y}=0.5$, exists and is stable too. As well
as its quiescent counterpart, this mode features the mirror symmetry between
the profiles of $|\phi _{+}\left( x,y\right) |$ and $|\phi _{-}\left(
x,y\right) |$. Figure \ref{moving}(c) shows the amplitude of the moving
soliton, $A=\sqrt{|\phi _{+}(x=0,y^{\prime }=0)|^{2}+|\phi
_{-}(x=0,y^{\prime }=0)|^{2}}$, as a function of $v_{y}$. The amplitude
monotonously decreases with the growth of the velocity, vanishing at
\begin{equation}
v_{y}=\left( v_{y}\right) _{\max }^{(\mathrm{MM})}\approx 1.8,  \label{Vmax}
\end{equation}%
i.e., the mobile solitons exist in the limited interval of the velocities
\cite{we}.

The SV solitons may also be made mobile, but in a very narrow interval of
velocities -- e.g., at $v_{y}<\left( v_{y}\right) _{\max }^{(\mathrm{SV}%
)}\approx 0.03$ for $\eta =0,\lambda =1$, and $N=3.7$, cf. the limit
velocity given by Eq. (\ref{Vmax}) for the MMs. At $v_{y}>0.03$, the
imaginary-time solution of Eq. (\ref{mix}) with the SV input converges to
stable MM solitons, instead of the SV \cite{we}.

\subsection{3D metastable semi-vortices and mixed modes}

\subsubsection{Analytical considerations}

\begin{figure}[tbp]
\centering\includegraphics[width=0.5\columnwidth]{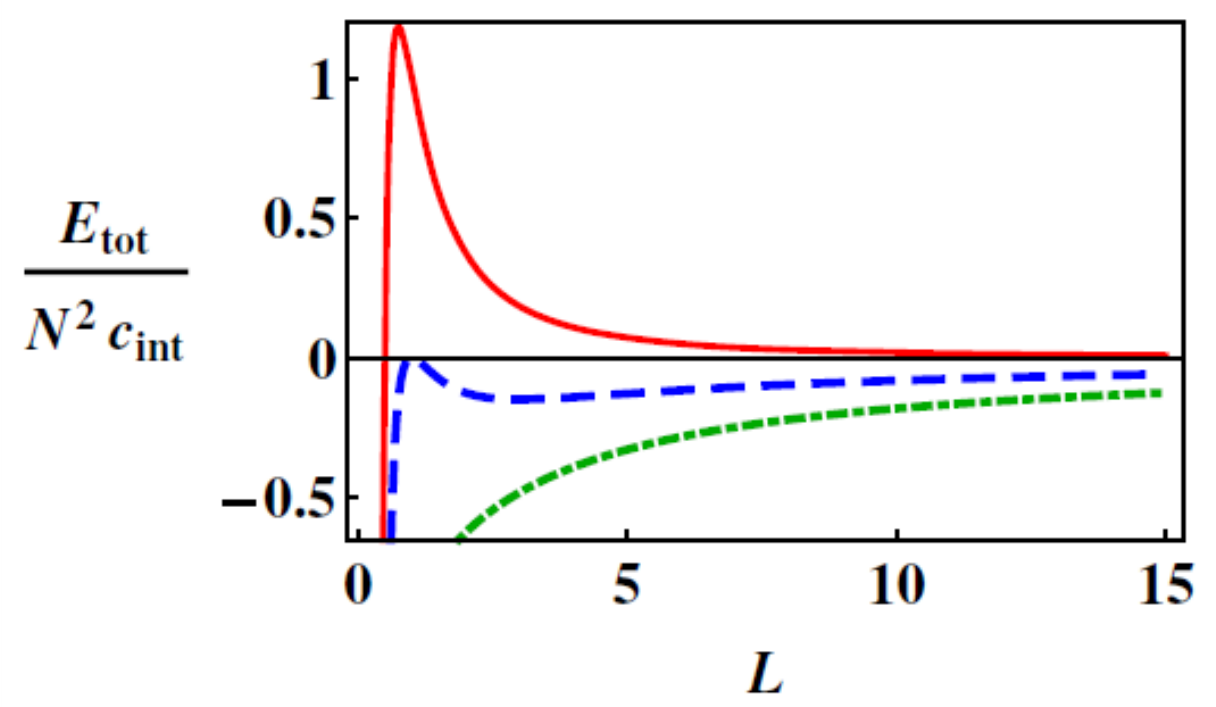}
\caption{(Color online) Energy $E_{\mathrm{tot}}$ of the 3D solitons in the
SOC system, as a function of condensate's size $L$, predicted by Eq. (%
\protect\ref{eq2}) (as per Ref. \protect\cite{HP}), with $c_{\mathrm{int}%
}\equiv c_{\mathrm{int}}^{\mathrm{(self)}}+c_{\mathrm{int}}^{\mathrm{(cross)}%
}\protect\eta $. The red solid, blue dashed, and green dashed-dotted lines
represent the variation of the energy for $\protect\lambda =0$, and for $%
\protect\lambda >0$ which does or does not satisfy condition (\protect\ref%
{eq3}), respectively.}
\label{scaling}
\end{figure}

The creation of metastable 3D solitons in the model based on Eq. (\ref{3D})
can be predicted, starting from evaluation of scaling of different terms in
the respective energy functional (\ref{eq1}). Assuming that a localized
state has characteristic size $L$ and norm $N$, an estimate for the
amplitudes of the wave function is $A\sim \sqrt{N}L^{-3/2}$. Accordingly,
the three terms in Eq. (\ref{eq1}) scale with $L$ as
\begin{equation}
E_{\mathrm{tot}}/N\sim c_{\mathrm{kin}}L^{-2}-c_{\mathrm{soc}}\lambda
L^{-1}-\left( c_{\mathrm{int}}^{\mathrm{(self)}}+c_{\mathrm{int}}^{\mathrm{%
(cross)}}\eta \right) NL^{-3}\,,  \label{eq2}
\end{equation}%
with positive coefficients $c_{\mathrm{kin}}$, $c_{\mathrm{soc}}$, and $c_{%
\mathrm{int}}^{\mathrm{(self/cross)}}$. As shown in Fig.~\ref{scaling}, Eq.~(%
\ref{eq2}) gives rise to a local minimum of $E_{\mathrm{tot}}(L)$ at finite $%
L$, provided that
\begin{equation}
0<\lambda N<{c_{\mathrm{kin}}^{2}}/\left[ {3\left( c_{\mathrm{int}}^{\mathrm{%
(self)}}+c_{\mathrm{int}}^{\mathrm{(cross)}}\eta \right) c_{\mathrm{soc}}}%
\right] \,.  \label{eq3}
\end{equation}%
Although this minimum cannot represent the GS, which formally corresponds to
$E_{\mathrm{tot}}\rightarrow -\infty $ at $L\rightarrow 0$ in the collapsed
state, as is suggested by Fig. \ref{scaling} too, the local minimum
corresponds to a self-trapped state which should be stable against small
perturbations. Condition (\ref{eq3}) suggests that metastable 3D solitons
may exist when the SOC term is present, while its strength $\lambda $ is not
too large, $N$ and $\eta $ being not too large either \cite{HP}.

Like the 2D system, its 3D counterpart admits the existence of solitons of
the SV and MM types. Also similar to the 2D case, for $\eta <1$ the energy
of the SV is lower than that for the MM, and vice versa for $\eta >1$.
Another similarity to the 2D system is that there is no threshold (minimum
norm) necessary for the existence of the metastable 3D solitons, while their
existence is bounded by a certain maximum norm, $N<N_{\max }$ \cite{HP}.

\begin{figure}[tbh]
\centering\includegraphics[width=0.40\columnwidth]{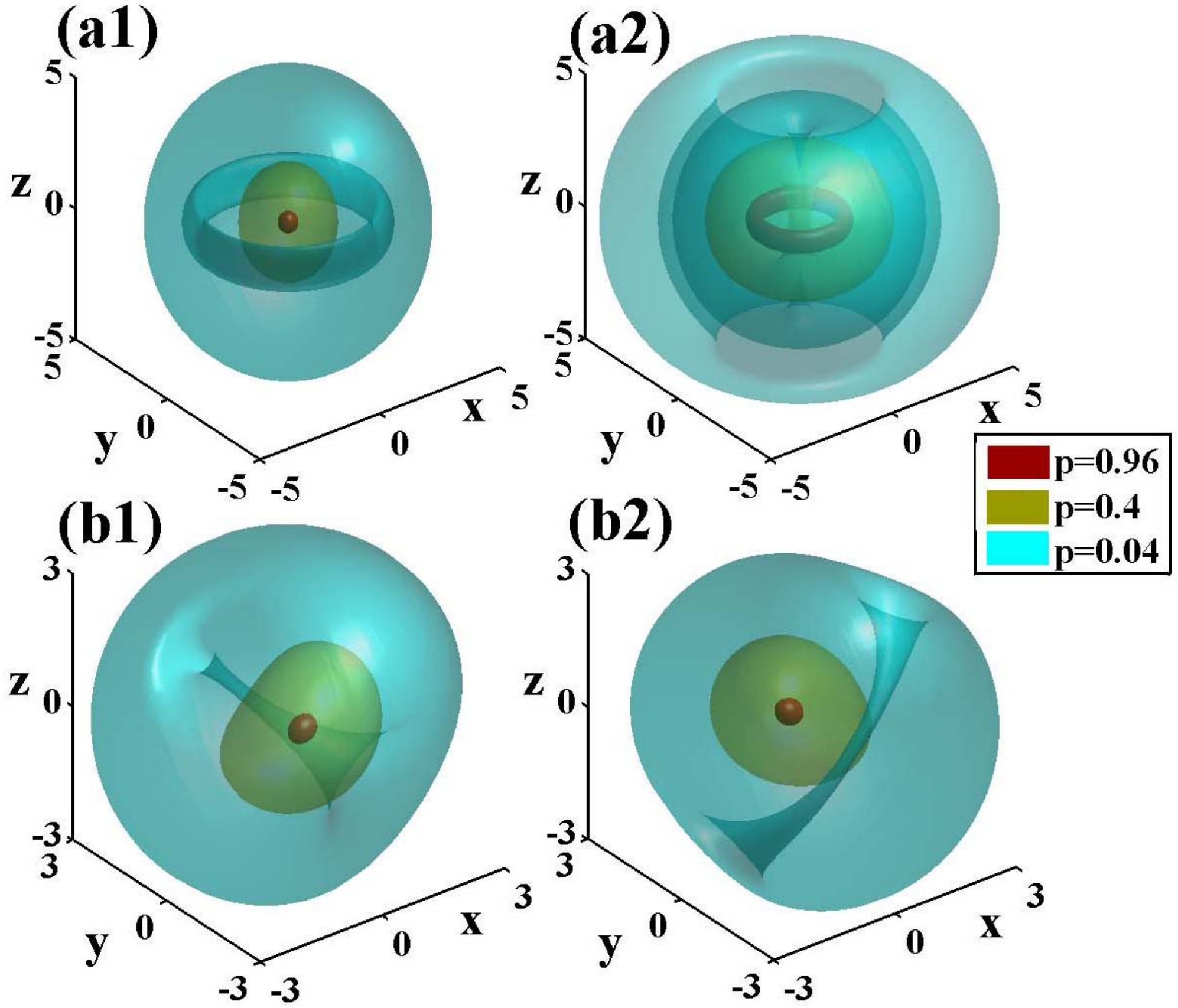}
\caption{(Color online) Density profiles of stable 3D solitons numerically
generated in the SOC system for $N=8$ and $\protect\lambda =1$, as per Ref.
\protect\cite{HP}. (a) A semi-vortex for $\protect\eta =0.3$, whose
fundamental and vortical components, $|\protect\psi _{+}|$ and $|\protect%
\psi _{-}|$, are plotted in (a1) and (a2), respectively. (b) A mixed mode
for $\protect\eta =1.5$, with (b1), (b2) displaying $|\protect\psi _{+}|$
and $|\protect\psi _{-}|$, respectively. In each subplot, different colors
represent constant-magnitude surfaces, $|\protect\psi _{\pm }\left(
x,y,z\right) |=\left( 0.96,0.4,0.04\right) \times |\protect\psi _{\pm }|_{%
\mathrm{max}}$.}
\label{3D-SOC}
\end{figure}

In the numerical form, stationary 3D solitons were produced by means of the
imaginary-time method applied to Eq. (\ref{3D}), and their stability against
small perturbations was verified in real-time simulations \cite{HP}. Typical
examples of density profiles of stable SV and MM solitons are displayed in
Fig.~\ref{3D-SOC}. Naturally, the MM states exhibit a more sophisticated
profile. Lastly, similar to what is shown above for the 2D system in Fig. %
\ref{moving}, it was found that stable 3D solitons can be set in motion in a
limited interval of velocities, cf. Eq. (\ref{Vmax}).

The use of two-component BEC\ is the simplest possibility to emulated BEC\
in bosonic gases. More complex realizations were elaborated in models of
three-component (spin-1) condensates. In particular, vortex solitons
supported by SOC in spin-1 condensates have been numerically constructed too
\cite{SKA-SOCvort1,SKA-SOCvort2}.

\section{Giant microwave-coupled vortex rings in binary BEC}

\subsection{The model}

Another possibility to create stable 2D vortex solitons with arbitrarily
large values of the vorticity was recently predicted in the hybrid model
based on a system of two GPEs coupled by a microwave field through a
magnetic-dipole transition. Two components of the matter waves, which are
governed by the coupled GPEs, represent a BEC mixture of two different
hyperfine states of the same atomic species, the microwave field being
generated by the transition between the levels, which gives rise to the
feedback of the matter waves on the microwave radiation. This physical model
was introduced, in the 1D form, in Ref. \cite{Jieli1}.

The 2D system of the coupled GPEs for two components, $\phi _{\uparrow
\downarrow }$, which form a pseudo-spinor wave function of the binary BEC,
was derived in Ref. \cite{Jieli2}. In the scaled form, the system takes the
form
\begin{gather}
i\frac{\partial \phi _{\downarrow }}{\partial t}=\left( -\frac{1}{2}\nabla
^{2}+H_{0}-\beta \left\vert \phi _{\uparrow }\right\vert ^{2}\right) \phi
_{\downarrow }  \notag \\
+\frac{\gamma \phi _{\uparrow }}{2\pi }\int \!\!\ln \left( \left\vert
\mathbf{r}-\mathbf{r^{\prime }}\right\vert \right) \phi _{\downarrow }\left(
\mathbf{r^{\prime }}\right) \phi _{\uparrow }^{\ast }\left( \mathbf{%
r^{\prime }}\right) \!d\mathbf{r^{\prime }},  \label{eq:final-down}
\end{gather}%
\begin{gather}
i\frac{\partial \phi _{\uparrow }}{\partial t}=\left( -\frac{1}{2}\nabla
^{2}+H_{0}-\beta \left\vert \phi _{\downarrow }\right\vert ^{2}\right) \phi
_{\uparrow }  \notag \\
+\frac{\gamma \phi _{\downarrow }}{2\pi }\int \!\!\ln \left( \left\vert
\mathbf{r}-\mathbf{r^{\prime }}\right\vert \right) \phi _{\downarrow }^{\ast
}\left( \mathbf{r^{\prime }}\right) \phi _{\uparrow }\left( \mathbf{%
r^{\prime }}\right) \!d\mathbf{r^{\prime }}.  \label{eq:final-up}
\end{gather}%
Here, the integral terms represent the action of the microwave field on the
matter waves, the field itself being replaced by the solution of the
respective Poisson equation, produced by means of the 2D Green's function, $%
\gamma $ is the strength of the field-matter interaction (which appears as
the effective long-range interaction in Eqs. (\ref{eq:final-down}) and (\ref%
{eq:final-up})), and $H_{0}$ is a background magnetic field. Further, $\beta
$ is the coefficient of the additional contact (collisional) interaction
between the components of the wave function, which may be present too, $%
\beta >0$ implying the attractive interactions.

For identical components, $\phi _{\downarrow }=\phi _{\uparrow }\equiv \phi
\exp \left( -iH_{0}t\right) $, Eqs. (\ref{eq:final-down}) and (\ref%
{eq:final-up}) reduce to a single equation,%
\begin{equation}
i\frac{\partial \phi }{\partial t}=\left[ -\frac{1}{2}\nabla ^{2}-\beta
\left\vert \phi \right\vert ^{2}+\frac{\gamma }{2\pi }\int \int \!\!\ln
\left( \left\vert \mathbf{r}-\mathbf{r^{\prime }}\right\vert \right)
\left\vert \phi \left( \mathbf{r^{\prime }}\right) \right\vert
^{2}\!dx^{\prime }dxy^{\prime }\right] \phi ,  \label{single}
\end{equation}%
where $\mathbf{r}^{\prime }=\left( x^{\prime },y^{\prime }\right) $, and its
soliton solutions with vorticity $S$ and chemical potential $\mu $ can be
looked for in the usual form,
\begin{equation}
\phi =\exp \left( -i\mu \tau +iS\theta \right) \Phi _{S}\left( r\right)
\mathbf{,}  \label{phiPhi}
\end{equation}%
\textbf{\ }cf. Eq. (\ref{vortex}). The Hamiltonian of Eq. (\ref{single}),
written in terms of the radial wave function $\Phi _{S}(r)$, defined as per
Eq. (\ref{phiPhi}), is%
\begin{gather}
H=2\pi \int_{0}^{\infty }rdr\left[ (\Phi _{S}^{\prime })^{2}+r^{-2}S^{2}\Phi
_{S}^{2}-\beta \Phi _{S}^{4}\right]  \notag \\
+\frac{\gamma }{2\pi }\int \int d\mathbf{r}_{1}d\mathbf{r}_{2}\ln \left(
\left\vert \mathbf{r}_{1}-\mathbf{r}_{2}\right\vert \right) \Phi _{S}^{2}(%
\mathbf{r}_{1})\Phi _{S}^{2}(\mathbf{r}_{1}).  \label{HPhi}
\end{gather}%
The following analysis is presented for the fixed normalization of the wave
function,
\begin{equation}
\int \int \left\vert \phi \left( x,y\right) \right\vert ^{2}dxdy\equiv 2\pi
\int_{0}^{\infty }\Phi ^{2}(r)rdr=\frac{1}{2}  \label{1/2}
\end{equation}%
(i.e., the total norm of the binary state is $1$).

\subsection{Basic results for vortex solitons}

\begin{figure}[tbp]
\centering\includegraphics[width=0.60\columnwidth]{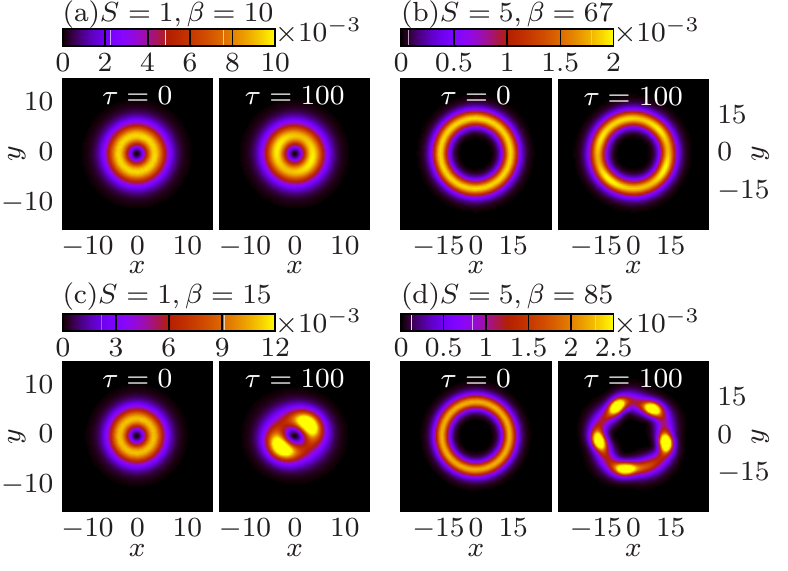}
\caption{Top and bottom panels display, severally, examples of the stable
and unstable perturbed evolution of the vortex solitons with indicated
values of $S$ and $\protect\beta $, in the framework of Eq. (\protect\ref%
{single}), as per Ref. \protect\cite{Jieli2}. The necklace-shaped set,
observed in the latter case, remains confined (keeping the same overall
radius) in the course of subsequent evolution. In this figure, time $t$ is
denoted $\protect\tau $.}
\label{fig-Jieli}
\end{figure}

In the presence of the self-focusing contact interaction, $\beta >0$,
solitons cannot be created by Eq. (\ref{single}) (in other words, by
Hamiltonian (\ref{HPhi}) beyond a point of the onset of the collapse, $\beta
=\beta _{\mathrm{collapse}}$). A simple analytical approximation makes it
possible to predict this point quite accurately for all vortex solitons,
with $S\geq 1$ \cite{Jieli2}. To this end, one should make use of the fact
that, according to numerical results, for $S\geq 2$ and $\beta $ large
enough, vortex solitons generated by the present model assume the shape of
narrow annuli, see an example for $S=5$ in Fig. \ref{fig-Jieli}(b). In the
radial direction, it may be approximated by the usual 1D soliton ansatz.
With regard to normalization (\ref{1/2}), it is written as
\begin{equation}
\Phi _{S}(r)=\sqrt{\beta }/\left( 8\pi R\right) \mathrm{sech}[\beta \left(
r-R\right) /\left( 8\pi R\right) ],  \label{narrow}
\end{equation}%
where $R$ is the radius of the narrow annulus. The substitution of this
approximation in Eq. (\ref{HPhi}) yields%
\begin{equation}
H(R)=\left[ S^{2}-\frac{\beta ^{2}}{3\left( 8\pi \right) ^{2}}\right] \frac{1%
}{2R^{2}}+\frac{\gamma }{8\pi }\ln R.  \label{Eeff}
\end{equation}%
Then, the equilibrium value of $R$ is selected as a point of the energy
minimum: $dE/dR=0$, i.e.,
\begin{equation}
R_{\mathrm{eq}}^{2}=(8\pi /\gamma )\left[ S^{2}-(1/3)\left( \beta /8\pi
\right) ^{2}\right] ,  \label{R^2}
\end{equation}%
and the analytically predicted collapse point, $\beta _{\mathrm{collapse}}^{%
\mathrm{(analyt)}}$, is one at which $R_{\mathrm{eq}}^{2}$ vanishes, i.e.,
the annulus collapses onto the center:
\begin{equation}
\beta _{\mathrm{collapse}}^{\mathrm{(analyt)}}=8\sqrt{3}\pi S\approx 43.5S.
\label{beta_max}
\end{equation}%
As seen in Table II, this analytical prediction is very close to its
numerically found counterparts at $S\geq 2$, and is rather close even at $%
S=1 $, when the form of the vortex soliton is not actually close to a narrow
ring, see Fig. \ref{fig-Jieli}(a).

It is worthy to note that prediction (\ref{beta_max}) does not depend on
strength $\gamma $ of the effective long-range interaction, because this
coefficient appears only as an overall factor in expression (\ref{R^2}), and
it remains valid in the limit of $\gamma \rightarrow 0$, i.e., for the usual
two-dimensional NLSE with the local cubic self-attractive term,
\begin{equation}
i\partial \phi /\partial t=-\left[ (1/2)\nabla ^{2}+\beta \left\vert \phi
\right\vert ^{2}\right] \phi .  \label{simple}
\end{equation}%
To explain this fact, we note that, at the eventual stage of the collapse,
when the shrinking 2D vortex ring becomes extremely narrow, Eq. (\ref{single}%
) becomes asymptotically tantamount to the local NLSE, therefore the
condition for the onset of the collapse is identical in both equations (\ref%
{single}) and (\ref{simple}). However, the latter equation has fundamental
and vortex-soliton solutions solely at $\beta =\beta _{\max }$, which are
completely unstable, while the effective long-range interaction in Eq. (\ref%
{single}) creates stable fundamental solitons and vortices for all $S$ (see
below). Actually, the approximate analytical result given by Eq. (\ref%
{beta_max}) provides an explanation for the numerical findings that were
reported in Ref. \cite{Minsk2} and many other works. In this connection, it
also relevant to mention that, taking into regard normalization condition (%
\ref{1/2}), value $\beta _{\mathrm{collapse}}(S=1)=48.3$ in Table II is
tantamount to the collapse threshold for $S=1$ written above in the form of
Eq. (\ref{S=1}).

\begin{table}[tbp]
\centering%
\begin{tabular}{|l|l|l|l|l|l|l|l|}
\hline
$S$ & $\beta _{\mathrm{collapse}}$ & $\beta _{\mathrm{collapse}}^{\mathrm{%
(analyt)}}$ & $\beta _{\mathrm{stability}}$ & $S$ & $\beta_{\mathrm{collapse}%
}$ & $\beta _{\mathrm{collapse}}^{\mathrm{(analyt)}}$ & $\beta _{\mathrm{%
stability}} $ \\ \hline
$0$ & $11.8$ & $\mathrm{n/a}$ & $\equiv \beta _{\mathrm{collapse}}$ & $3$ & $%
132.5$ & $130.6$ & $41$ \\ \hline
$1$ & $48.3$ & $43.5$ & $11$ & $4$ & $175.5$ & $174.1$ & $57$ \\ \hline
$2$ & $89.7$ & $87.0$ & $28$ & $5$ & $218.5$ & $217.7$ & $70$ \\ \hline
\end{tabular}%
\caption{Table II. $\protect\beta _{\mathrm{collapse}}$ and $\protect\beta _{%
\mathrm{collapse}}^{\mathrm{(analyt)}}$ are numerically obtained and
analytically predicted (see Eq. (\protect\ref{beta_max})) values of the
contact-interaction strength, $\protect\beta $, up to which the fundamental
and vortex solitons exist, in the framework of Eq. (\protect\ref{single}). $%
\protect\beta _{\mathrm{stability}}$ is the numerically identified stability
boundary of the vortex solitons.}
\end{table}

The stability of the solitons was identified by real-time simulations of
Eqs. (\ref{eq:final-down}) and (\ref{eq:final-up}) with random perturbations
added to the initial conditions (independent perturbations were taken for $%
\phi _{\uparrow }$\ and $\phi _{\uparrow }$, to verify the stability against
breaking the equality of the two components) \cite{Jieli2}. The fundamental
solitons ($S=0$) are stable in their entire existence region, $\beta <\beta
_{\mathrm{collapse}}(S=0)\approx 11.8$ (in the present notation, it is
tantamount to the TS norm, see Eq. (\ref{Nmax})).

For the vortex solitons with $S\geq 1$, the numerical analysis reveals an
internal stability boundary, $\beta _{\mathrm{stability}}(S)<\beta _{\mathrm{%
collapse}}(S)$, the vortices being stable at $\beta <\beta _{\mathrm{%
stability}}(S)$, see Table 1. In the interval of $\beta _{\mathrm{stability}%
}(S)<\beta <\beta _{\mathrm{collapse}}(S)$, they are broken by azimuthal
perturbations into rotating necklace-shaped clusters of fragments, which
resembles the initial stage of the instability development of localized
vortices in usual models; however, unlike those models \cite%
{necklace1,necklace2,necklace3}, the necklace does not expand, remaining
confined under the action of the effective nonlocal interaction. Typical
examples of the stable and unstable evolution of the vortex solitons in the
present model are displayed in Figs. \ref{fig-Jieli}.

The stability boundary, $\beta _{\mathrm{stability}}(S)$, can be found in an
approximate analytical form too \cite{Jieli2}. To this end, the wave
function of an azimuthally perturbed vortex ring is approximated by%
\begin{equation}
\phi \approx \exp \left( -i\mu t+iS\theta \right) A(\theta )\Phi _{S}(r),
\label{ansatz}
\end{equation}%
cf. Eq. (\ref{phiPhi}). An evolution equation for the modulation amplitude, $%
A$, is derived by substituting ansatz (\ref{ansatz}) in Eq. (\ref{single})
and averaging it in the radial direction:
\begin{equation}
i\frac{\partial A}{\partial \tau }=-\frac{1}{2R^{2}}\frac{\partial ^{2}A}{%
\partial \theta ^{2}}+\left[ \frac{\gamma \ln R}{4\pi R}-\frac{2\beta ^{2}}{%
3\left( 8\pi R\right) ^{2}}\right] |A|^{2}A  \label{NLS}
\end{equation}%
(recall $R$ is considered as the constant large radius of the narrow vortex
annulus). In this approximation, the stability analysis amounts to studying
the modulational stability of the solution with $A=1$ against azimuthal
perturbations $\sim \exp \left( ip\theta \right) $, with its own integer
winding numbers $p$, in the framework of the effectively one-dimensional
NLSE (\ref{NLS}). A simple result is that the stability is maintained under
condition $p^{2}\geq \left( 8/3\right) \left( \beta /8\pi \right) ^{2}$, if
the term $\sim \beta ^{2}$ dominates in Eq. (\ref{NLS}). Further, numerical
results demonstrate that, as usual, the critical instability corresponds to $%
p^{2}=S^{2}$ (for instance, the appearance of five fragments in Fig. \ref%
{fig-Jieli}(d) demonstrates that, for $S=5$, the dominant splitting mode has
$p=5 $). Thus, it is expected that the vortex soliton remains stable at
\begin{equation}
\beta <\beta _{\mathrm{stability}}^{(\mathrm{analyt})}(S)=2\sqrt{6}\pi
S\approx \allowbreak 15.4S.  \label{analyt}
\end{equation}%
On the other hand, the numerically found stability boundary, presented in
Table II, follows an empirical formula, $\beta _{\mathrm{stability}%
}(S)\approx \allowbreak 15S-4$. Thus, the analytical approximation given by
Eq. (\ref{analyt}) is accurate enough for $S\geq 2$.

It follows from these results that the giant vortex rings, with higher
values of $S$, are much more robust than their counterparts with smaller $S$%
. This feature is opposite to what was previously found in those (few)
models which are able to produce stable vortex solitons with $S>1$, cf.
Refs. \cite{Pego,Brtka,3DLHY,GZ-ln}. It is also relevant to mention that,
while the fundamental soliton is the system's GS at $\beta <\beta _{\mathrm{%
stability}}(S=0)$, the ground state does not exist at $\beta >\beta _{%
\mathrm{stability}}(S=0)$, due to the possibility of the collapse. Thus, the
vortex solitons with winding number $S$ cannot represent the GS in the case
of $\beta _{\mathrm{stability}}(S=0)<\beta <\beta _{\mathrm{stability}}(S)$
(e.g., $11.8<\beta <70$ for $S=5$, as per Table II). Nevertheless, the
vortex solitons exist in this region as metastable states, cf. the 3D
solitons in the SOC system based on Eq. (\ref{3D}), as illustrated above by
Fig. \ref{scaling}. In that system, the vortical solitons of the SV and MM
types also exist as metastable states, although the system does not have a
GS, due to the possibility of the supercritical collapse \cite{HP}.

Lastly, in the case of the self-repulsive contact interactions, $\beta <0$,
Eqs. (\ref{eq:final-down}) and (\ref{eq:final-up}) give rise to stable
solitons for all values of $S$ (at least, up to $5$) and indefinitely large
values of $|\beta |$, as shown in work \cite{Jieli2}.

\subsection{A brief outline: vortex solitons in other systems with nonlocal
nonlinearity}

The nonlocality is an essential feature of Eqs. (\ref{eq:final-down}) and (%
\ref{eq:final-up}). Other forms of long-range interactions naturally occur
in many other physically significant models. In optics, the nonlocality
naturally emerges in the propagation equation for beams in liquid crystals
\cite{LC1,LC2}, as well as in the case of the thermal nonlinearity \cite%
{thermal}. A typical model is based on the following two-dimensional NLSE
\cite{Briedis}:%
\begin{equation}
i\frac{\partial u}{\partial z}=\left[ -\frac{1}{2}\nabla ^{2}-\frac{1}{\pi
\sigma ^{2}}\int \!\!\int \exp \left( -\frac{\left\vert \mathbf{r}-\mathbf{%
r^{\prime }}\right\vert ^{2}}{\sigma ^{2}}\right) \left\vert u\left( \mathbf{%
r^{\prime }}\right) \right\vert ^{2}\!dx^{\prime }dy^{\prime }\right] u,
\label{uu}
\end{equation}%
cf. Eq. (\ref{single}), where $\sigma $ is a spatial scale of the long-range
interaction.

It was demonstrated that, unlike the self-focusing local cubic (Kerr)
nonlinearity, its nonlocal version may support stable vortex solitons, with
unitary ($S=1$) and higher-order ($S\geq 2$) vorticities, as well as
higher-order states characterized by zeros in the radial direction \cite%
{Briedis}. This fact is not surprising, because, in the limit of strong
nonlocality, i.e., with $\sigma $ much larger than a characteristic size of
solitons, Eq. (\ref{uu}) reduces to an effectively linear equation for $%
u\left( x,y,z\right) \equiv \tilde{u}\left( x,y,z\right) \exp \left(
iN/\left( \pi \sigma ^{2}\right) \right) $, with the two-dimensional
trapping HO\ potential:%
\begin{equation}
i\frac{\partial \tilde{u}}{\partial z}=\left( -\frac{1}{2}\nabla ^{2}+\frac{N%
}{\pi \sigma ^{2}}r^{2}\right) \tilde{u},  \label{uuu}
\end{equation}%
where $N$ is again the norm of the field (this linear equation, with the HO
potential proportional to $N$, is known as the model of \textquotedblleft
accessible solitons" \cite{Snyder}). Obviously, the linear Schr\"{o}dinger
equation (\ref{uuu}) admits a full set of stable states with all values of $%
S $, as well as other higher-order modes.

Experimentally, the creation of a stable two-component bound state, built of
a vortex soliton in one optical beam, and a fundamental soliton in another
one, carried by a different wavelength, in a nematic liquid crystal was
demonstrated in Ref. \cite{Izdeb}.

BEC\ made of dipolar atoms are described by GPEs with nonlocal anisotropic
terms which represent dipole-dipole interactions \cite{Pfau}. Stable 2D and
3D vortex solitons were predicted in various models including such
interactions \cite{Lashkin,Tikhon,SKA-dipolar,SKA-dipolar2,Goran}, although
no experimental results on this topic have been reported thus far.

\section{An outline of other topics and experimental studies}

\subsection{Conservative systems}

There are other aspects of the studies of vortex solitons which are not
included in the present review. As concerns the theory, a specific
possibility is provided by the use of spatially nonuniform self-repulsive
nonlinearity, with the local strength growing from the center to periphery
at any rate faster than $r^{D}$, where $r$ is the distance from the center,
and $D$ the spatial dimension. \cite{HS}. This model makes it possible to
create very robust solitons of diverse types, including fundamental ones and
solitary vortices \cite{ICFO}, \cite{further}-\cite{further7}, as well as
sophisticated 3D states -- notably, \textit{hopfions} (vortex tori with
inner twist, which feature two independent topological numbers) \cite%
{hopfion}. A challenging problem is finding still more general physically
relevant conditions for the creation of complex 3D modes, such as skyrmions
(which, similar to hopfions, carry two different topological charges) \cite%
{skyrmion1,skyrmion2,skyrmion3}, monopoles \cite{field-theory}, linked
vortex rings, globally coupled vortex clusters \cite{clusters}, and others.

Another relevant ramification is the investigation of discrete vortex
solitons in lattice media \cite{discrvort,Pelin,big,Thaw}, which essentially
generalize the above-mentioned simplest discrete NLSE (\ref{DNLSE}). Various
forms of stable topological modes were created in virtual photonic lattices
induced in strongly birefringent photorefractive materials (using the
experimental technique proposed in Ref. \cite{MSegev}), see original works
\cite{discrvort1,discrvort2,discrvort3,Jena-vortex} and a review in \cite%
{big}.

\subsection{Dissipative systems}

\subsubsection{Free-space models}

A separate vast research area is the study of multidimensional solitons with
embedded vorticity in dissipative media, see, e.g., a review in Ref. \cite%
{Dum-diss} and recent theoretical works \cite%
{NNR1,NNR2,NNR3,DimaHS,DimaThaw2,DimaThaw1} and references therein. A basic
class of models for the generation of dissipative vortex solitons is based
on CGLEs \cite{Kramer} with the CQ or, more generally, saturable
nonlinearity. In particular, a model for the evolution of amplitude $u\left(
x,y,z\right) $ of electromagnetic fields in laser cavities \cite%
{Lugiato,NNR0,Mandel} may be based on the following form of CGLE, which was
first introduced, on phenomenological grounds, by Petviashvili and Sergeev
\cite{Petvia}:%
\begin{equation}
i\frac{\partial u}{\partial z}=-i\gamma u-\left( \frac{1}{2}-i\beta \right)
\nabla ^{2}u-\left( 1-i\varepsilon \right) |u|^{2}u+\left( \nu -i\mu \right)
|u|^{4}u,  \label{GL}
\end{equation}%
where positive coefficients $\gamma $, $\mu $, $\varepsilon $, and $\nu $
account, severally, for the linear and quintic losses, cubic gain, and
coefficient of the quintic self-defocusing. The presence of the linear loss
is necessary to make the zero background stable, thus admitting the
existence of stable dissipative solitons. The gain is provided by the cubic
term, while the quintic loss is necessary for the overall stability of the
model. The presence of the effective viscosity term with $\beta >0$ is
necessary for stability of vortex solitons \cite{Lucian}. This term may
represent, e.g., diffusion of free carriers (electrons and holes) generated
by the electromagnetic fields in the laser cavity.

Solutions to Eq. (\ref{GL}) for vortex dissipative solitons with winding
number $S$ are looked for, in terms of the polar coordinates $\left(
r,\theta \right) $, as%
\begin{equation}
u=\exp \left( ikz+iS\theta +i\Phi (r)\right) A(r),  \label{Phi}
\end{equation}%
where real $k$ is a propagation constant, $\Phi (r)$ is a real phase, and $%
A(r)$ is a real amplitude function. Unlike conservative models, such as the
one based on the NLSE with the CQ nonlinearity, see Eq. (\ref{CQ}), in which
solitons exist in the form of continuous families parametrized by $k$, CGLEs
give rise only to pairs of dissipative solitons with two discrete values of
\textit{eigenvalues} $k$ for each integer $S$, starting from $S=0$. One of
these solutions, which features a larger amplitude, may be stable, as an
\textit{attractor} in the dissipative system, while the additional solution
plays the role of a separatrix between \textit{attraction basins} of the
attractors represented by the zero solution and stable dissipative soliton
\cite{PhysicaDD}.

\begin{figure}[tbp]
\centering\includegraphics[width=0.60\columnwidth]{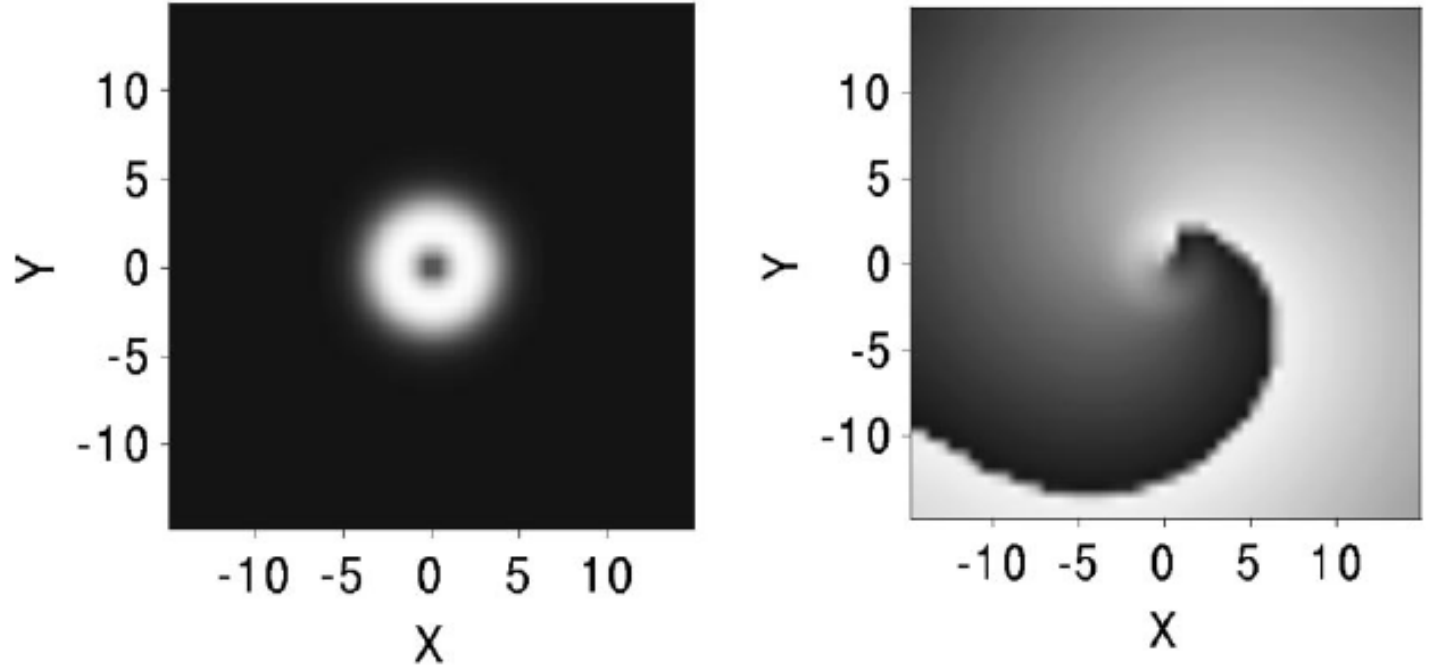}
\caption{The left and right panels display the amplitude and spiral-phase
shapes of a stable dissipative soliton with vorticity $S=1$, produced by Eq.
(\protect\ref{GL}) with $\protect\gamma =\protect\beta =0.5$, $\protect%
\varepsilon =2.5$, $\protect\mu =1$, and $\protect\nu =0.1$, as per Ref.
\protect\cite{Lucian}.}
\label{spiral}
\end{figure}

Unlike solitary vortices in conservative media, the phase structure of their
dissipative counterparts assumes the spiral shape \cite{Lucian}. Indeed, the
asymptotic form of solution (\ref{Phi}) at $r\rightarrow \infty $ is%
\begin{equation}
u\sim r^{-1/2}\exp \left( ikz+iS\theta +iqr-\lambda r\right) ,
\label{asymptotic}
\end{equation}%
with a real radial wavenumber, $q$, and $\lambda >0$ accounting for the
exponential localization of the dissipative soliton. In particular, $%
q=\gamma /\lambda $ in the case of $\beta =0$. The spiral structure of the
phase is produced by the combination of terms $S\theta $ and $qr$ in Eq. (%
\ref{asymptotic}), see an example for $S=1$ in Fig. \ref{spiral}. The model
based on CGLE (\ref{GL}) with $\beta >0$ gives rise to stable spiral
solitons with vorticities $S=1$ and $S=2$ \cite{Lucian}.

\begin{figure}[tbp]
\centering\includegraphics[width=0.67\columnwidth]{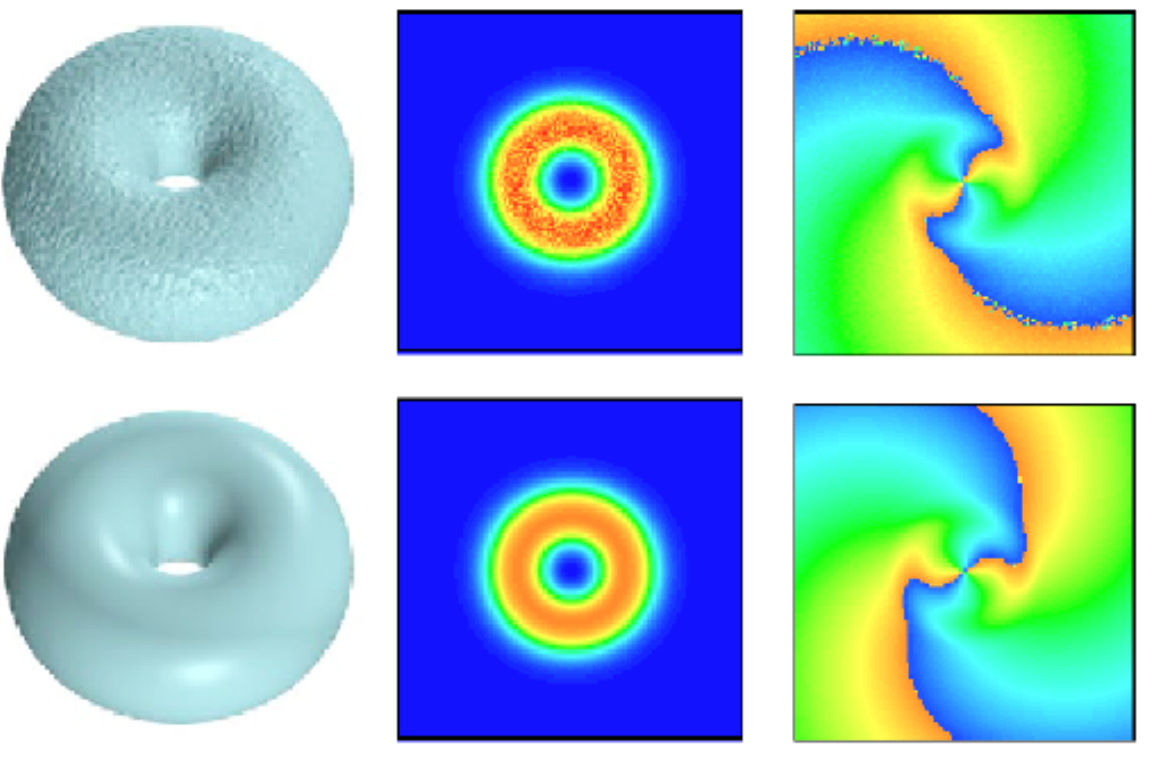}
\caption{The bottom row displays the result of the self-cleaning of an
initially perturbed 3D dissipative vortex (shown in the top row) with
embedded vorticity $S=2$, as per simulations of the three-dimensional CGLE
reported in Ref. \protect\cite{CGL-3Dvort}. The left, middle, and right
panels display, respectively, the 3D amplitude profile, its 2D cross
section, and the phase profile in the same 2D plane.}
\label{3D-CGL}
\end{figure}

The 3D version of the CGLE with the CQ nonlinearity also makes it possible
to generate stable 3D vortex solitons, in the form of \textquotedblleft
donuts", with embedded vorticities $S=1$ and $2$ \cite{CGL-3Dvort}, provided
that the quasi-2D viscosity term, with $\beta >0$, is kept in the 3D version
of Eq. (\ref{GL}). An example of stable evolution of an initially perturbed
3D vortex solitons with $S=2$ is displayed in Fig. \ref{3D-CGL}.

\subsubsection{Dissipative systems with the trapping potential}

\begin{figure}[tbp]
\centering\includegraphics[width=0.5\columnwidth]{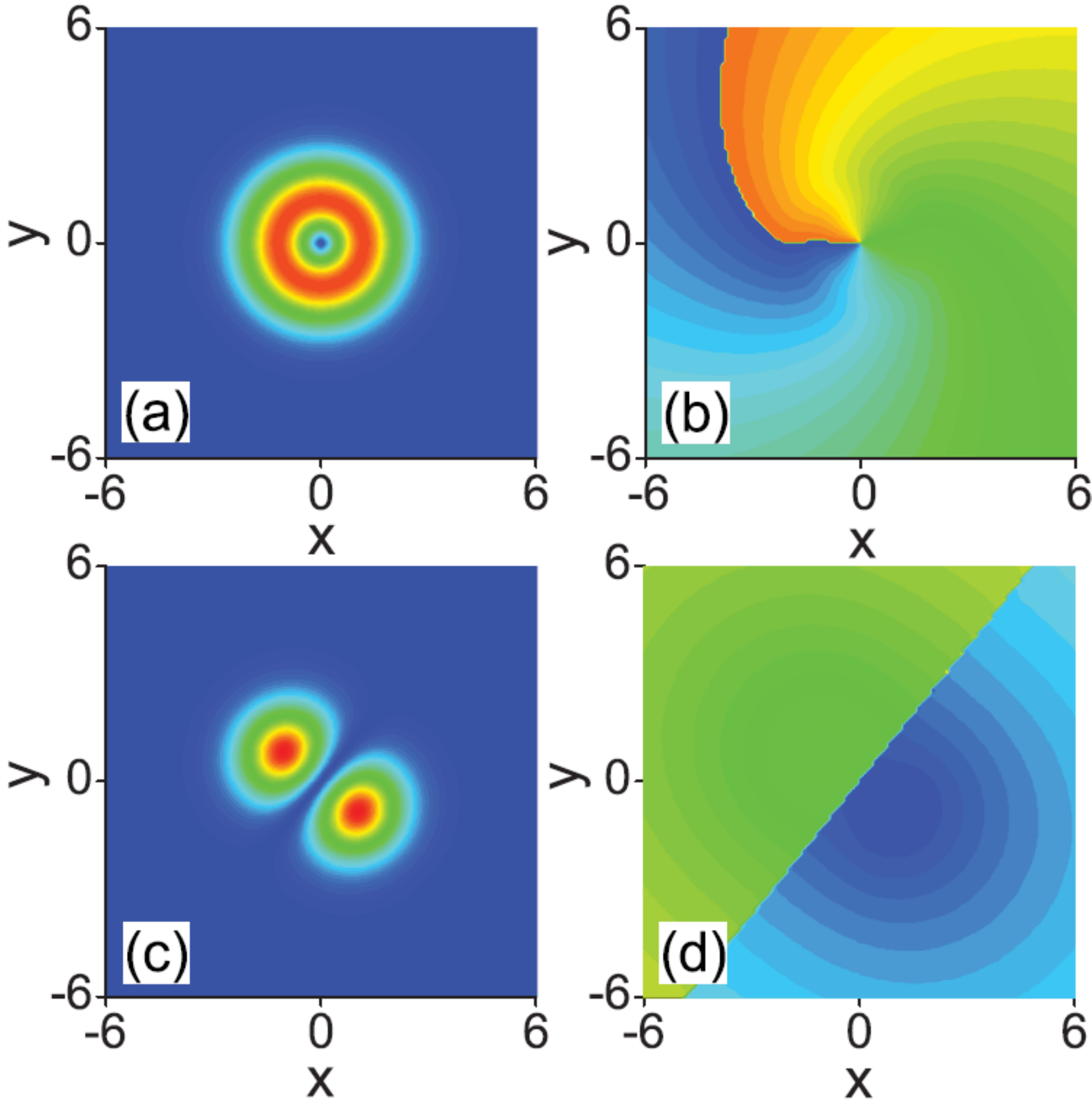}
\caption{Spontaneous transformation of an unstable dissipative vortex
soliton with winding number $S=1$ into a stably rotating dipole state,
produced by simulations of Eq. (\protect\ref{GL2}), as per Ref. \protect\cite%
{Skarka0}. The parameters are $\protect\gamma =0.5$, $\protect\varepsilon %
=1.8$, $\protect\mu =1$, $\protect\nu =0.1$, and $\Omega ^{2}=0.25$.}
\label{dipole}
\end{figure}

In most cases, models of laser cavities do not contain the diffusion term $%
\sim \beta $ in Eq. (\ref{GL}), because photons are not subject to
diffusion. As mentioned above, dissipative vortex solitons (unlike ones with
$S=0$) cannot be stable, in the free space, in the absence of this term.
However, they can be readily stabilized by trapping potentials \cite{Skarka0}%
. In particular, the respectively modified CGLE,
\begin{equation}
i\frac{\partial u}{\partial z}=-i\gamma u-\frac{1}{2}\nabla ^{2}u-\left(
1-i\varepsilon \right) |u|^{2}u+\left( \nu -i\mu \right) |u|^{4}u+\frac{1}{2}%
\Omega ^{2}r^{2}u,  \label{GL2}
\end{equation}%
where $\Omega ^{2}$ is the strength of the HO trapping potential, supports
stable vortex modes with $S=1$, while all the states with $S\geq 2$ are
unstable against spontaneous splitting, similar to what is demonstrated
above for the 2D version of the conservative NLSE/GPE (\ref{GPE}) with the
same HO trapping potential. Also similar to the conservative model, those
vortex modes with $S=1$ which are unstable suffer splitting in the framework
of Eq. (\ref{GL2}). However, in contrast with the case of Eq. (\ref{GPE}),
in the latter case fragments produced by the splitting do not perform cycles
of periodic recombination-splitting evolution, see Fig. \ref{periodic}, nor
are they destroyed by the intrinsic collapse. Instead, they form a stably
rotating dipole, as shown in Fig. \ref{dipole}. Similarly, an unstable
dissipative vortex soliton with $S=2$ splits in a set of three fragments,
which form a steadily rotating \textquotedblleft tripole" \cite{Skarka0}.

\begin{figure}[tbp]
\centering\includegraphics[width=0.54\columnwidth]{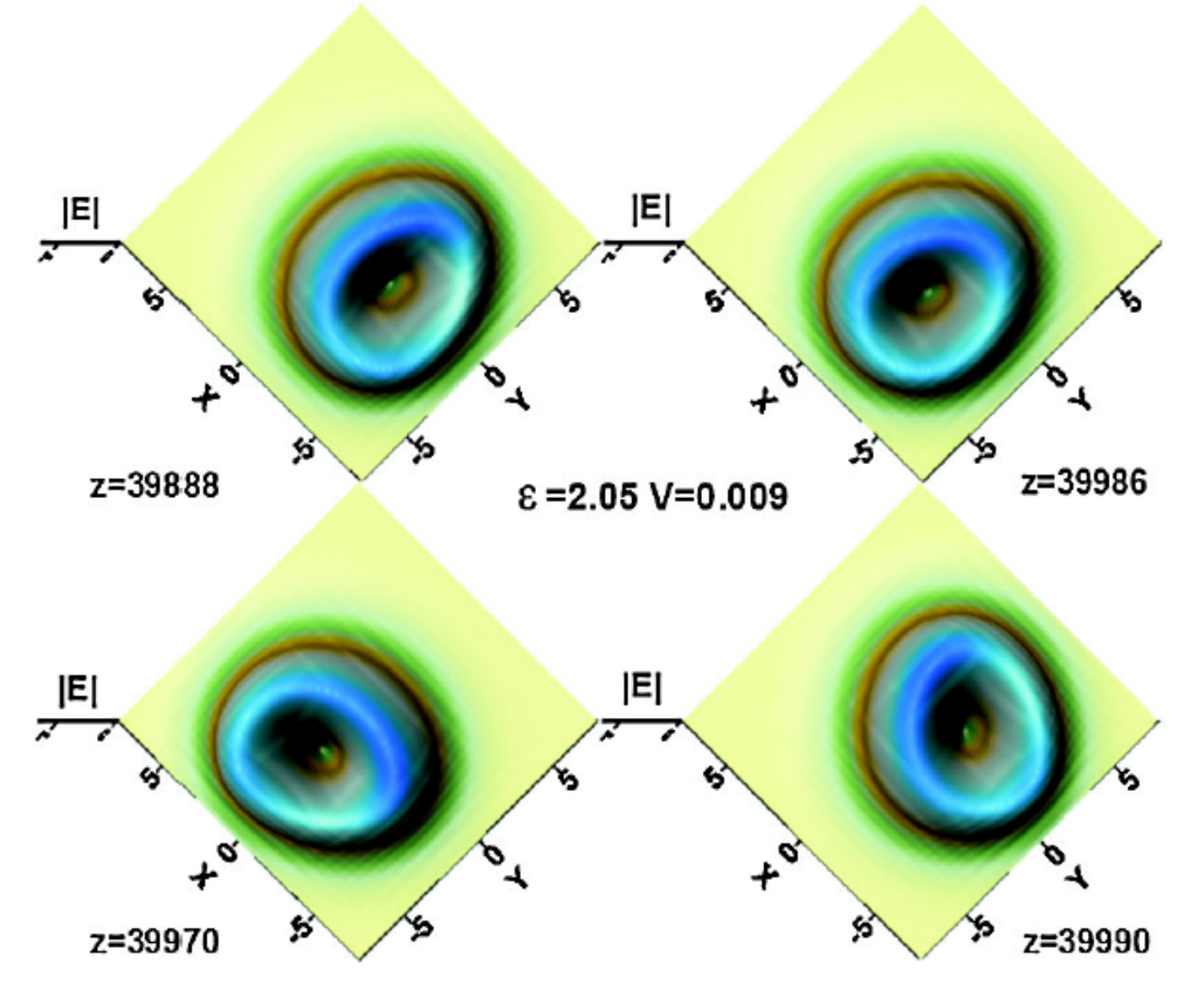}
\caption{An example of a stable dynamical mode with winding number $S=1$, in
the form of the ring-shaped vortex which features elliptic deformation and
eccentricity. The solution is generated by the model based on Eqs. (\protect
\ref{GL}) (without the diffusion term, $\protect\beta =0$) and (\protect\ref%
{gamma}), as per Ref. \protect\cite{Skarka1}. Four configurations of the
intensity profile, $\left\vert u\left( x,y,z\right) \right\vert\equiv |E|$,
displayed at different values of $z$, demonstrate that it performs orbital
motion around the origin and synchronous rotation around itself, with period
$Z=4$.}
\label{eccentric}
\end{figure}

Two-dimensional vortex patterns in dissipative media may be also supported
by the linear gain applied in an appropriately defined spatially confined
area, such as a ring-shaped one \cite{ring} (experimentally, a similar
structure was created for pumping a spin-orbit-coupled exciton-polariton
condensate in a semiconductor microcavity \cite{Amo}). In a similar context,
many vortex modes have been found in the framework of the two-dimensional
CGLE with the CQ nonlinearity, taken in the form similar to Eq. (\ref{GL}),
and with a radially-dependent linear-loss coefficient, which has a minimum
at $r=0$ \cite{Skarka1}:%
\begin{equation}
\gamma (r)=\gamma _{0}+\gamma _{2}r^{2}.  \label{gamma}
\end{equation}%
Even without the inclusion of the diffusion term, $\beta =0$ (whose physical
origin is not straightforward), this model gives rise to a great variety of
stable vortex states, such as ordinary stationary vortex solitons,
elliptically deformed and crescent-shaped steadily rotating ones, and
eccentrically deformed vortices, which combine steady rotation and orbiting
around the origin, see an example in Fig. \ref{eccentric}.

A great variety of dynamical self-trapped patters, initiated by inputs with
embedded vorticity, is produced by CGLE (\ref{GL}) with linear gain, instead
of the linear loss, placed around $r=0$, which corresponds to%
\begin{equation}
\gamma (r)=-\gamma _{0}+\gamma _{2}r^{2}  \label{gamma2}
\end{equation}%
with $\gamma _{0}>0$ (cf. Eq. (\ref{gamma})), i.e., the gain area is
introduced at $0\leq r^{2}<\gamma _{0}/\gamma _{2}$. As reported in Ref.
\cite{Skarka2}, this model readily gives rise to spontaneous breaking of the
underlying axial symmetry and thus generates rotating and periodically
transmuting patterns shaped, roughly, as four-, five-, six-, seven-, eight-,
nine-, and ten-pointed stars. While these rotating modes carry the orbital
angular momentum, they lose the original vorticity (the phase singularity is
expelled from the original pattern).

\begin{figure}[tbp]
\centering\includegraphics[width=0.77\columnwidth]{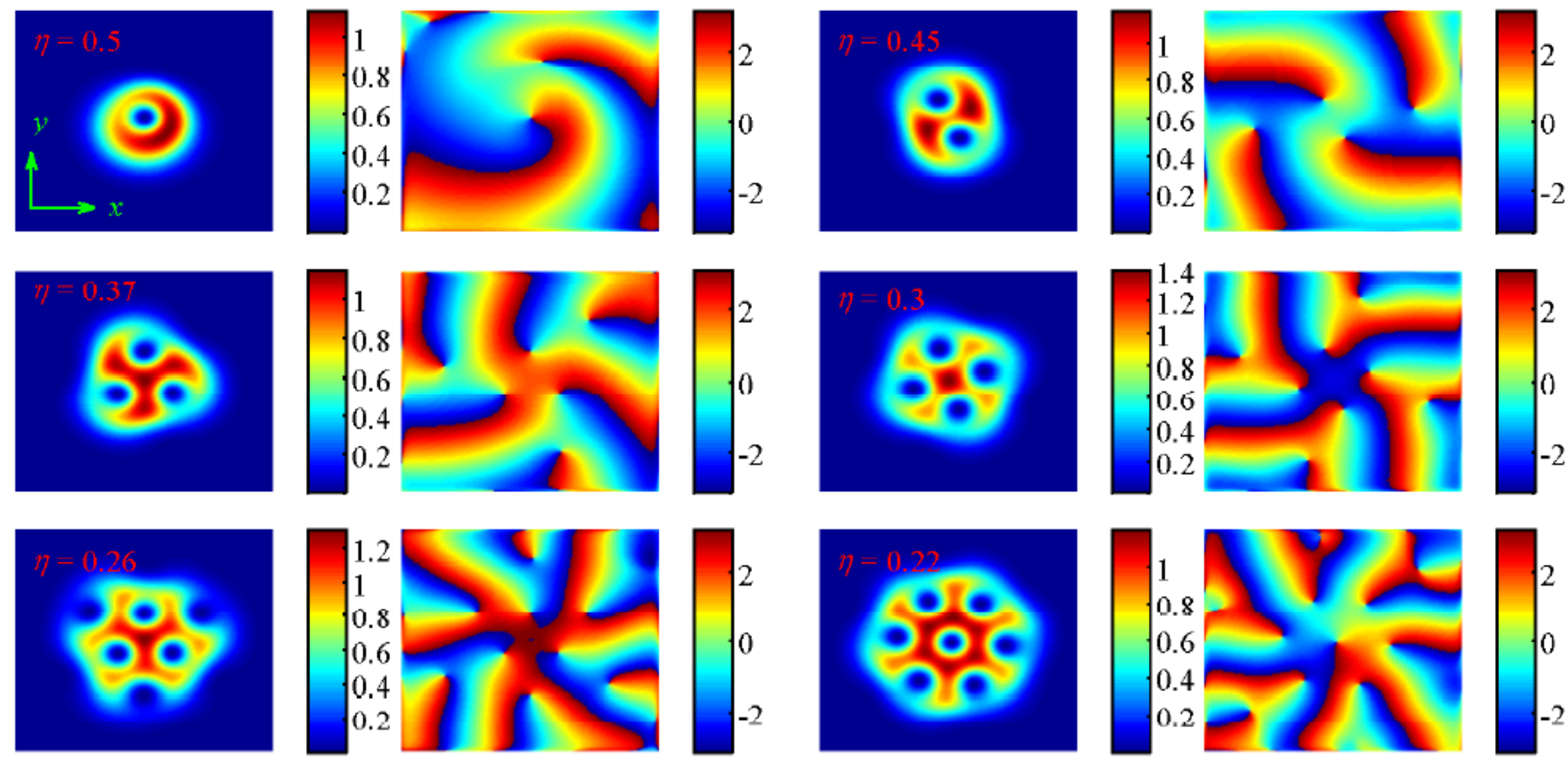}
\caption{A sequence of stably rotating 2D states produced by simulations of
Eq. (\protect\ref{Thaw}), as per Ref. \protect\cite{DimaThaw2}, for $\protect%
\gamma =\protect\varepsilon =2.5$, $\Omega ^{2}=2$, for decreasing values of
diffusivity $\protect\eta $, which are indicated in the panels: a crescent
vortex ($S=1$), and multi-vortex complexes with $S=2,3,4,6,$ and $7$,
respectively.}
\label{multiS}
\end{figure}

A recent ramification of studies of vortex solitons under the action of the
trapping potential is represented by the model with the uniform linear gain,
$\gamma >0$ (and still including the diffusion term), cubic loss ($%
\varepsilon >0$), and HO trap \cite{DimaThaw2}. The model may apply to laser
cavities and exciton-polariton condensates:%
\begin{equation}
i\frac{\partial \psi }{\partial t}=-\frac{1}{2}(1-i\eta )\nabla ^{2}\psi
-\sigma |\psi |^{2}\psi +i(\gamma -\varepsilon |\psi |^{2})\psi +\frac{1}{2}%
\Omega ^{2}r^{2}\psi  \label{Thaw}
\end{equation}%
(here the diffusion coefficient is denoted $\eta $ instead of $\beta $, cf.
Eq. (\ref{GL2})). The decrease of $\eta $ in this model gives rise to a
variety of stably rotating multi-vortex states, with $S=1,2,3,4,6,7$, as
shown in Fig. \ref{multiS}. For the same values of other parameters and $%
\eta <$ $0.22$, the model generates a turbulent state, while at $\eta >0.5$
it gives rise to stable axisymmetric vortex with $S=1$.

A similar model was elaborated for the model of the spinor (two-component)
exciton-polariton condensate under the action of the SOC \cite{DimaHS},%
\begin{gather}
i\frac{\partial \psi _{+}}{\partial t}=-{\frac{1}{2}}(1-i\eta )\nabla
^{2}\psi _{+}+(|\psi _{+}|^{2}+\alpha |\psi _{-}|^{2})\psi _{+}  \notag \\
+\beta \left( \frac{\partial }{\partial x}-i\frac{\partial }{\partial y}%
\right) ^{2}\psi _{-}+i(\gamma -\varepsilon |\psi _{+}|^{2})\psi _{+}+\left(
\frac{1}{2}r^{2}+\Omega \right) \psi _{+},  \notag \\
\label{HSDima} \\
i\frac{\partial \psi _{-}}{\partial t}=-{\frac{1}{2}}(1-i\eta )\nabla
^{2}\psi _{-}+(|\psi _{-}|^{2}+\alpha |\psi _{+}|^{2})\psi _{-}  \notag \\
+\beta \left( \frac{\partial }{\partial x}+i\frac{\partial }{\partial y}%
\right) ^{2}\psi _{+}+i(\gamma -\varepsilon |\psi _{-}|^{2})\psi _{-}+\left(
\frac{1}{2}r^{2}-\Omega \right) \psi _{-},  \notag
\end{gather}%
where $\beta $ is the SOC strength (note that in exciton-polariton
condensates SOC is represented by the second-order differential operators,
unlike the first-order ones in the model of the atomic BEC, cf. Eq. (\ref%
{R2D})), $\alpha $ is the coefficient of the nonlinear interaction between
the components, the strength of the HO trapping potential is fixed to be $1$
by rescaling, and $\Omega $ represents the Zeeman splitting between the
components (if any).

\begin{figure}[tbp]
\centering\includegraphics[width=0.60\columnwidth]{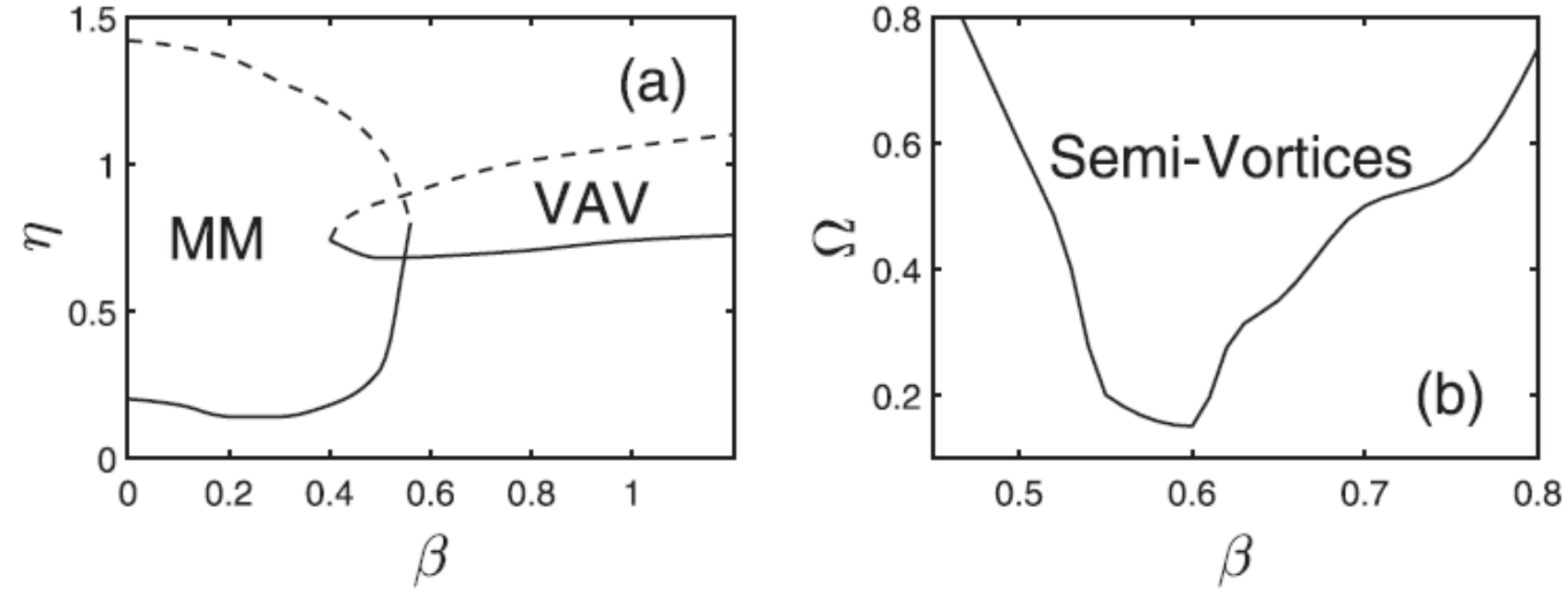}
\caption{(a) Stability areas of the VAV and MM states in the parameter plane
of the SOC strength ($\protect\beta $) and diffusion coefficient ($\protect%
\eta $) of the model based on Eqs. (\protect\ref{HSDima}), with $\protect%
\gamma =0.6$, $\protect\varepsilon =0.3$, and $\Omega =0$. The SV state is
taken as per Eq. (\protect\ref{VAV}). (b) The stability area for the SV
mode, defined as per Eq, (\protect\ref{ZeemanSV}), in the plane of $\protect%
\beta $ and the strength of the Zeeman splitting ($\Omega $) at the same
parameters as mentioned above, and $\protect\eta =0.4$. The plots are
borrowed from Ref. \cite{DimaHS}.}
\label{map}
\end{figure}

The analysis of Eqs. (\ref{HSDima}) has demonstrated that, in the absence of
the Zeeman terms ($\Omega =0$), the system may give rise to stable
vortex-antivortex (VAV) complexes, in the form of
\begin{equation}
\left( \psi _{\pm }\right) _{\mathrm{VAV}}=\exp \left( -i\mu t\pm i\theta
\right) \phi (r),  \label{VAV}
\end{equation}
written in polar coordinates $\left( r,\theta \right) $, with a common
amplitude function $\phi (r)$, and stable MM (mixed-mode) states, see Fig. %
\ref{map}(a). The figure demonstrates a small bistability region in which
the VAV and MM species coexist as stable ones. In addition to that, Fig. \ref%
{map}(b) shows that the system with $\Omega >0$ may support stable SV
(semi-vortex) states, in the form of%
\begin{equation}
\psi _{+}=\exp (-i\mu t)\phi _{+}(r),\psi _{-}=\exp \left( -i\mu t+2i\theta
\right) \phi _{-}(r).  \label{ZeemanSV}
\end{equation}%
In comparison with the SV defined as per Eqs. (\ref{frf}) and (\ref{mirror}%
), the second-order SOC operator in Eqs. (\ref{HSDima}) imparts winding
number $S=2$ to the vortex component of the SV, instead of $S=1$.

\subsection{$\mathcal{PT}$-symmetric systems}

A specific variety of the dissipative systems is represented by ones subject
to the condition of the parity-time ($\mathcal{PT}$) symmetry, i.e.,
settings with spatially separated and mutually balanced gain and loss
elements. They maintain a real spectrum of energy eigenvalues (i.e., avoid
instability, which is often driven in dissipative systems by linear gain),
provided that the strength of the gain-loss terms does not exceed a certain
critical level \cite{PT1,PT2}. In the combination with nonlinearity, $%
\mathcal{PT}$-symmetric systems readily create solitons \cite{PTsol1,PTsol2}%
. While these solitons are usually considered in the 1D form, they may be
found in multidimensional models as well. In particular, in some specific
systems, such as one with a $\mathcal{PT}$-symmetric lattice potential \cite%
{Optica}, and another one with a specially designed profiles of spatially
inhomogeneous nonlinearity and gain-loss terms \cite{Luz}, stable vortex
solitons were predicted too. Stable \textquotedblleft dark" vortices in a $%
\mathcal{PT}$-symmetric medium, supported by a finite-amplitude background
field, were studied too \cite{Panos-PTvort}.

\subsection{A brief outline of experimental results}

In terms of experiments, the field of multidimensional solitons remains a
challenging one. As concerns vortex solitons, which is the central topic of
the present review, none of them was experimentally created in a stationary
form. Transient modes in the form of semi-discrete \textquotedblleft
bullets" with embedded vorticity were observed in arrays of optical
waveguides with the Kerr nonlinearity, which form a hexagonal lattice in the
transverse cross section \cite{Jena-vortex}, see an illustration in Fig. \ref%
{Jena}. Spatial 2D solitons with embedded vorticity, which were effectively
stabilized, as transient states, by nonlinear (cubic) losses in the medium
with the CQ nonlinearity (liquid carbon disulfide), were reported too \cite%
{Cid3}, as shown in Fig. \ref{Cid}. Thus, the current state of the topic of
vortex solitons calls for performing new experiments.
\begin{figure}[tbp]
\includegraphics[height=4.0cm]{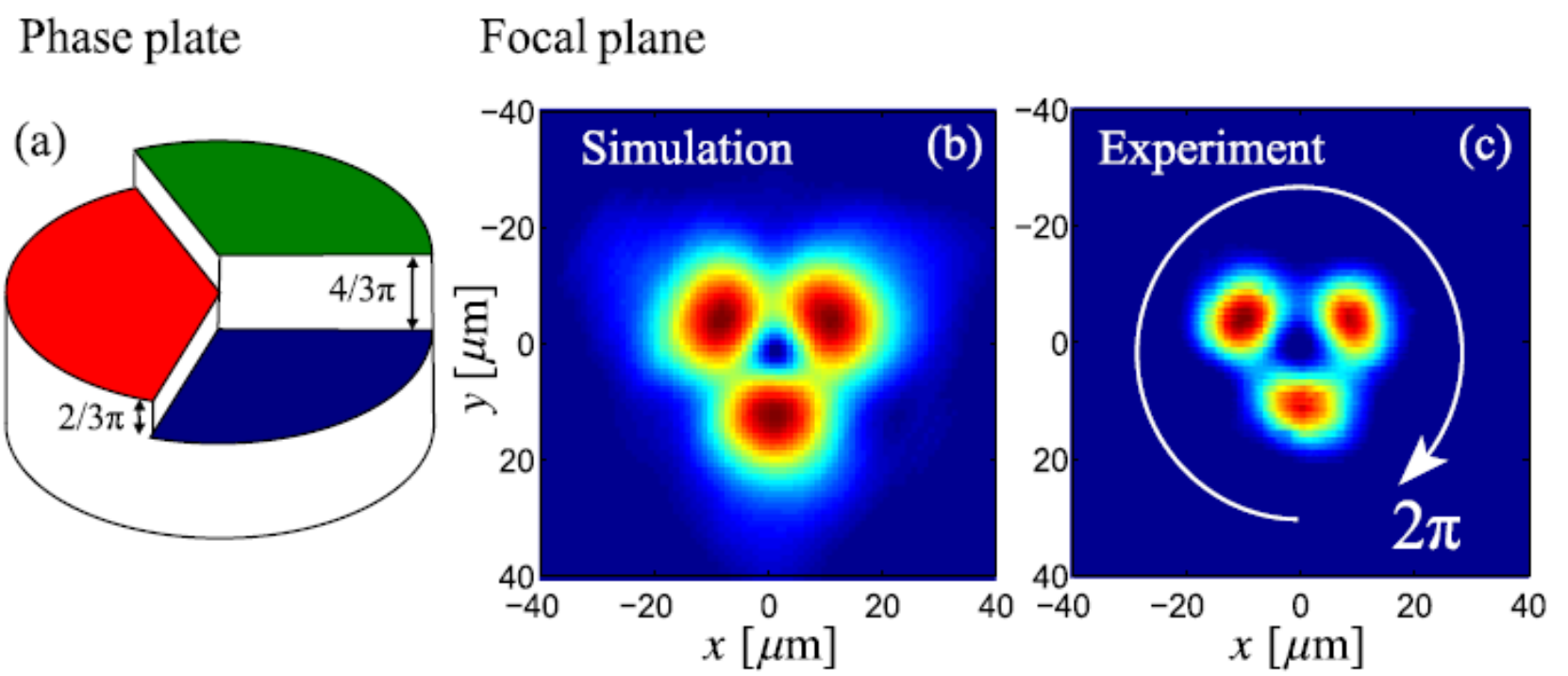}
\caption{A semi-discrete soliton with embedded vorticity $S=1$, created in
Ref. \protect\cite{Jena-vortex} as a transient state, in a hexagonal array
of waveguides made in bulk silica. (a) A phase plate used for imprinting the
vortex structure into the input beam. Panels (b) and (c) display,
respectively, the numerically simulated and experimentally observed
intensity distributions in the transverse plane, with phase shifts $2\protect%
\pi /3$ between adjacent peaks.}
\label{Jena}
\end{figure}
\begin{figure}[tbp]
\includegraphics[height=9.5cm]{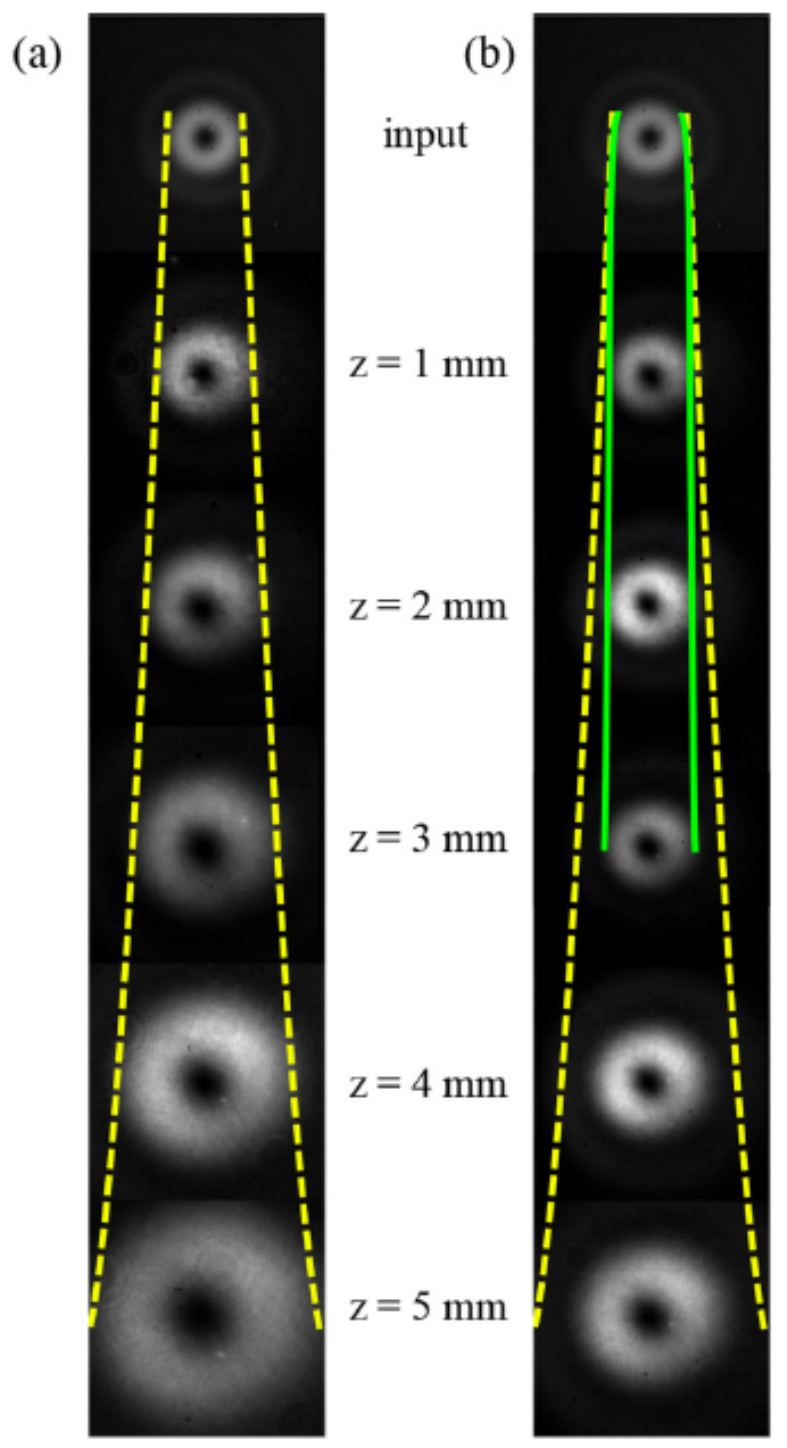}
\caption{Transverse intensity distributions in vortex light beams (with $S=1$%
), launched into the CQ medium, with essential cubic losses, as per Ref.
\protect\cite{Cid3}. (a) expansion of the quasi-linear beam, created with a
low total power ($1$ GW/cm{$^{2}$}); (b) transient stability of the
nonlinear vortex beam, with initial total power $9$ GW/cm{$^{2}$}, up to
propagation distance $z=3$ mm. At larger distances, the beams starts to
expand, as its power decreases due the cubic losses.}
\label{Cid}
\end{figure}

\section{Conclusion: A summary of the review}

This article aims to present a relatively brief review of the broad area of
multidimensional solitons, focusing on ones with embedded vorticity. The
multidimensional solitons and solitary vortices find most important physical
realizations in BEC and nonlinear optics, the crucially important problem
being search for physically relevant settings which provide for the
stabilization of these states against the collapse and splitting, that tend
to destroy the fundamental (zero-vorticity) and vortical solitons,
respectively, in media with the most common cubic self-attractive
nonlinearity.

The article reviews, first, some well-established topics, which remain
relevant to the current studies. One topic is the stabilization of the 3D
and 2D states with vorticities $S=0$ and $S\geq 1$ by competing attractive
and repulsive nonlinearities. This part includes the newest addition, in the
form of the 3D and 2D GPEs with the LHY\ (Lee-Huang-Yang) corrections, which
is related to the recent breakthrough in theoretical and experimental
studies of self-trapped states in atomic BEC, \textit{viz}., QDs (quantum
droplets). The other well-established topic deals with the use of trapping
HO (harmonic-oscillator) and spatially periodic (lattice) potentials,
combined with the cubic self-attraction, for the creation of stable vortex
solitons featuring the multi-peak structure.

Two other sections of the article address two recently elaborated schemes
for the stabilization of multidimensional vortex solitons in BEC with cubic
nonlinearities. One scheme predicts the creation of stable 2D and 3D
solitons, which mix terms with $S=0$ and $1$ (SVs, i.e., semi-vortices) or $%
S=0$ and $\pm 1$ (MMs, mixed modes), in a two-component system which
realizes the SOC in the atomic condensates. The second novel scheme predicts
giant vortex rings (which may be stable with indefinitely large values of $S$%
) in 2D two-component BEC\ coupled by microwave radiation. A remarkable
peculiarity of the latter model is that virtually all the essential results,
concerning the existence and stability of the vortex solitons in it, can be
produced in an approximate analytical form.

The analysis of the two novel schemes reveals a drastic difference between
the 2D and 3D settings. In the former case, the stabilization mechanism
creates the GS (ground state), that did not exist otherwise. The total norm
of the GSs takes values below the threshold necessary for the onset of the
critical collapse, driven by the cubic attractive nonlinearity in the 2D
space. In the 3D settings, the supercritical collapse does not allow the
creation of a GS, but, nevertheless, an appropriate mechanism may create
metastable solitons.

The review also summarizes, in a brief form, some other results relevant to
the general topic of vortex solitons. In particular, these are relatively
old and new findings for self-trapped vortex states in dissipative media.
Relevant experimental findings are briefly outlined two.

To conclude, it is relevant to stress that there remain vast room for
further theoretical and, especially, experimental studies of vortex solitons
and objects related to them in photonics, BEC, and other quickly developing
areas of physics.

\section*{Acknowledgments}

I would like to thank my collaborators in original works that were dealing
with various topics related to multidimensional solitons (most of those
works are cited in this review): C. B. de Ara\'{u}jo, B. B. Baizakov, V.
Besse, O. V. Borovkova, G. Boudebs, C. M. Brtka, W. B. Cardoso, R.
Carretero-Gonz\'{a}lez, Zhaopin Chen, Zhigang Chen, J. Cuevas, G. Dong, N.
Dror, Z. Fan, D. J. Frantzeskakis, A. Gammal, G. Gligori\'{c}, L. Had\v{z}%
ievski, X. Jiang, Y. V. Kartashov, P. G. Kevrekidis, V. V. Konotop, H.
Leblond, B. Li, Y. Li, V. E. Lobanov, A. Maluckov, D. Mazilu, D. Mihalache,
W. Pang, H. Pu, J. Qin, A. S. Reyna, H. Sakaguchi, L. Salasnich, M. Salerno,
E. Ya. Sherman, Ya. Shnir, L. Tarruell, F. Toigo, L. Torner, F. Wise, Y.-C.
Zhang, and Z.-W. Zhou. I also thank Prof. Victor P\'{e}rez-Garc\'{\i}a, the
editor of Physica D, for the invitation to write a review article for the
journal.

Parts of this work were supported by grants No. 2010239 from the Binational
(US-Israel) Science Foundation (BSF), and No. 2015616 from the joint program
in physics between the National Science Foundation (US) and BSF.


\begin{thebibliography}{999}
\bibitem{ZK} N. J. Zabuski and M. D. Kruskal, Interaction of ``solitons" in
a collisionless plasma and recurrence of initial states, Phys. Rev. Lett.
\textbf{15}, 240 (1965).

\bibitem{KA} Y. S. Kivshar and G. P. Agrawal, \textit{Optical Solitons: From
Fibers to Photonic Crystals} (Academic Press, San Diego, 2003).

\bibitem{Peyrard} T. Dauxois and M. Peyrard, \textit{Physics of Solitons}
(Cambridge University Press, Cambridge, 2006).

\bibitem{Silb} Y. Silberberg, Collapse of optical pulses, Opt. Lett. \textbf{%
15}, 1282-1284 (1990).

\bibitem{Brazh} V. A. Brazhnyi and V. V. Konotop, Theory of nonlinear matter
waves in optical lattices, Mod. Phys. Lett. B \textbf{18}, 627-651 (2004).

\bibitem{BEC-sol1} F. Kh. Abdullaev, A. Gammal, A. M. Kamchatnov, and L.
Tomio, Dynamics of bright matter-wave solitons in a Bose-Einstein
condensate, Int. J. Mod. Phys. B \textbf{19}, 3415-3473 (2005).

\bibitem{Morsch} O. Morsch and M. Oberthaler, Dynamics of Bose--Einstein
condensates in optical lattices, Rev. Mod. Phys. \textbf{78}, 179-215 (2006).

\bibitem{PhysicaD} V. M. P\'{e}rez-Garc\'{\i}a, N. G. Berloff, P. G.
Kevrekidis, V. V. Konotop, and B. A. Malomed, Nonlinear phenomena in
degenerate quantum gases, Physica D \textbf{238}, 1289-1298 (2009).

\bibitem{BEC-sol2} D. J. Frantzeskakis, Dark solitons in atomic
Bose-Einstein condensates: from theory to experiments, J. Phys. A: Math.
Theor. \textbf{43}, 213001 (2010).

\bibitem{BEC-sol3} L. Salasnich, Bright solitons in ultracold atoms, Opt.
Quant. Electron. \textbf{49}, 409 (2017).

\bibitem{old-review} B. A. Malomed, D. Mihalache, F. Wise, and L. Torner,
Spatiotemporal optical solitons, J. Optics B: Quant. Semicl. Opt. \textbf{7}%
, R53-R72 (2005); Viewpoint: On multidimensional solitons and their legacy
in contemporary Atomic, Molecular and Optical physics, J. Phys. B: At. Mol.
Opt. Phys. \textbf{49}, 170502 (2016).

\bibitem{Kruskal} C. S. Gardner, J. M. Greene, M. D. Kruskal, and R. M.
Miura, Method for solving the Korteweg - de Vries equation, Phys. Rev. Lett.
\textbf{19}, 1095-1097 (1967).

\bibitem{ZS} V. E. Zakahrov and A. B. Shabat, Exact theory of
two-dimensional self-focusing and one-dimensional self-modulation of waves
in nonlinear media, Zhurn. Eksp. Teor. Fiz. \textbf{61}, 118-134 (1971)
[English translation: Sov. Phys. JETP \textbf{34}, 62 (1972)].

\bibitem{AKNS} M. J. Ablowitz, D. J. Kaup, A. C. Newell, and H. Segur,
Nonlinear-evolution equations of physical significance, Phys. Rev. Lett.
\textbf{31}, 125-127 (1973).

\bibitem{Borovik} A. E. Borovik, N-soliton solutions of the nonlinear
Landau-Lifshitz equation, Pis'ma Zh. Eksp. Teor. Fiz. \textbf{28}, 629
(1978) [English translation: JETP Lett. \textbf{28}, 581-584 (1978)].

\bibitem{Mikhailov} A. V. Mikhailov, The Landau-Lifshitz equation and
Riemann boundary-value problem on a torus, Phys. Lett. \textbf{92}, 51-54
(1982).

\bibitem{Rodin} Yu. L. Rodin, The Riemann boundary-value problem on Riemann
surfaces and the inverse scattering problem for the Landau-Lifshitz
equation, Physica D 11, 90-108 (1984).

\bibitem{Dryuma} V. S. Dryuma, On the analytical solution of the
two-dimensional Korteweg-de Vries equation, Sov. Phys. JETP Lett. \textbf{19}%
, 753-757 (1974).

\bibitem{1989} Yu. S. Kivshar and B. A. Malomed, Dynamics of solitons in
nearly integrable systems, Rev. Mod. Phys. \textbf{61}, 763-915 (1989).

\bibitem{Zakharov} V. E. Zakharov, S. V. Manakov, S. P. Novikov, and L. P.
Pitaevskii, \textit{Solitons: The Inverse Scattering Method} (Nauka
Publishers: Moscow, 1980; English translation: Consultants Bureau, New York,
1984).

\bibitem{Ablowitz} M. Ablowitz and H. Segur, \textit{Solitons and Inverse
Scattering Method} (SIAM, Philadelphia, 1981).

\bibitem{Newell} A. C. Newell, \textit{Solitons in Mathematics and Physics}
(SIAM, Philadelphia, 1985).

\bibitem{Askaryan} G. A. Askar'yan, Cherenkov radiation and transition
radiation from electromagnetic waves, Sov. Phys. JETP\ \textbf{15}, 943-946
(1962).

\bibitem{Townes} R. Y. Chiao, E. Garmire, and C. H. Townes, Self-trapping of
optical beams, Phys. Rev. Lett. \textbf{13}, 479-482 (1964).

\bibitem{Berge} L. Berg\'{e}, Wave collapse in physics: principles and
applications to light and plasma waves, Phys. Rep. \textbf{303}, 259-370
(1998).

\bibitem{SulemSulem} C. Sulem and P. L. Sulem, \textit{The nonlinear Schr%
\"{o}dinger equation: self-focusing and wave collapse} (Springer: Berlin,
1999).

\bibitem{Gadi} G. Fibich, \textit{The Nonlinear Schr\"{o}dinger Equation:
Singular Solutions and Optical Collapse} (Springer: Heidelberg, 2015).

\bibitem{Gaeta} L. T. Vuong, T. D. Grow, A. Ishaaya, A. L. Gaeta, G. W. 't
Hooft, E. R. Eliel, and G. Fibich, Collapse of optical vortices, Phys. Rev.
Lett. \textbf{96}, 133901 (2006).

\bibitem{pull} H. Sakaguchi and B. A. Malomed, Suppression of the
quantum-mechanical collapse by repulsive interactions in a quantum gas,
Phys. Rev. A 83, 013607 (2011).

\bibitem{skyrmion1} A. Hosaka and H. Toki, Chiral bag model for the nucleon,
Phys. Rep. \textbf{227}, 65-188 (1996).

\bibitem{skyrmion2} T. Sakai and S. Sugimoto, Low energy hadron physics in
holographic QCD, Prog. Theor. Phys. \textbf{113}, 843-882 (2005).

\bibitem{skyrmion3} R. A. Battye, N. S. Manton, and P. M. Sutcliffe,
Skyrmions and the alpha-particle model of nuclei, Proc. Roy. Soc. A 463,
261-279 (2007).

\bibitem{Minsk1} V. I. Kruglov and R. A. Vlasov, Spiral self-trapping
propagation of optical beams, Phys. Lett. A \textbf{111}, 401-404 (1985).

\bibitem{Minsk2} V. I. Kruglov, Yu. A. Logvin, and V. M. Volkov, The theory
of spiral laser beams in nonlinear media, J. Mod. Opt. \textbf{39},
2277-2291(1992).

\bibitem{Yankauskas} Z. K. Yankauskas, Radial field distributions in a
self-focusing light beam, Sov. Radiophys. \textbf{9}, 261-263 (1966).

\bibitem{VPG-management} A. Alexandrescu, G. D. Montesinos, and V. M. P\'{e}%
rez-Garc\'{\i}a, Stabilization of high-order solutions of the cubic
nonlinear Schr\"{o}dinger equation, Phys. Rev. E \textbf{75}, 046609 (2007).

\bibitem{semi1} N. Kumada, A. Sawada, Z. F. Ezawa, S. Nagahama, H. Azuhata,
K. Muraki, T. Saku, and Y. Hirayama, Doubly enhanced skyrmions in $\nu=2$
bilayer quantum Hall states, J. Phys. Soc. Jpn. \textbf{69}, 3178 (2000).

\bibitem{semi2} W. Munzer, A. Neubauer, T. Adams, S. Muhlbauer, C. Franz, F.
Jonietz, R. Georgii, P. B\"{o}ni, B. Pedersen, M. Schmidt, A. Rosch, and C.
Pfleiderer, Skyrmion lattice in the doped semiconductor Fe$_{1-x}$Co$_x$Si,
Phys. Rev. B \textbf{81}, 041203 (2010).

\bibitem{VPG-skyrmion} F. Pinsker, N. G. Berloff, and V. M. P\'{e}rez-Garc%
\'{\i}a, Nonlinear quantum piston for the controlled generation of vortex
rings and soliton trains, Phys. Rev. A \textbf{87}, 053624 (2013).

\bibitem{hopfion} H. Aratyn, L. A. Ferreira, and A. H. Zimerman, Exact
static soliton solutions of (3+1)-dimensional integrable theory with nonzero
Hopf numbers, Phys. Rev. Lett. \textbf{83}, 1723-1726 (1999).

\bibitem{Yasha} Y. V. Kartashov, B. A. Malomed, Y. Shnir, and L. Torner,
Twisted toroidal vortex-solitons in inhomogeneous media with repulsive
nonlinearity, Phys. Rev. Lett. \textbf{113}, 264101 (2014).

\bibitem{super} E. Babaev, Dual neutral variables and knot solitons in
triplet superconductors, Phys. Rev. Lett. \textbf{88}, 177002 (2002).

\bibitem{Sutcliffe} P. Sutcliffe, Knots in the Skyrme-Faddeev model, Proc.
Roy. Soc. A 463, 3001-3020 (2007).

\bibitem{Radu} E. Radu and M. S. Volkov, Stationary ring solitons in field
theory - knots and vortons, Phys. Rep. \textbf{468}, 101-151 (2008).

\bibitem{field-theory} B. Kleihaus, J. Kunz, and Y. Shnir, Monopoles,
antimonopoles, and vortex rings, Phys. Rev. D \textbf{68}, 101701 (2003).

\bibitem{ferro} N. R. Cooper, Propagating magnetic vortex rings in
ferromagnets, Phys. Rev. Lett. \textbf{82}, 1554 (1999).

\bibitem{ferro-Sutcliffe} P. Sutcliffe, Vortex rings in ferromagnets:
Numerical simulations of the time-dependent three-dimensional
Landau-Lifshitz equation, Phys. Rev. B \textbf{76}, 184439 (2007).

\bibitem{BECtheory-Sk1} J. Ruostekoski and J. R. Anglin, Creating vortex
rings and three-dimensional skyrmions in Bose-Einstein condensates, Phys.
Rev. Lett. \textbf{86}, 3934 (2001).

\bibitem{BECtheory-Sk2} R. A. Battye, N. R. Cooper and P. M. Sutcliffe,
Stable skyrmions in two-component Bose-Einstein condensates, Phys. Rev.
Lett. \textbf{88}, 080401 (2002).

\bibitem{BECtheory-Sk3} C. M. Savage and J. Ruostekoski, Energetically
stable particlelike skyrmions in a trapped Bose-Einstein condensate, Phys.
Rev. Lett. \textbf{91}, 010403 (2003).

\bibitem{BECtheory-Sk4} J. Ruostekoski and J. R. Anglin, Monopole core
instability and Alice rings in spinor Bose-Einstein condensates, Phys. Rev.
Lett. \textbf{91}, 190402 (2003).

\bibitem{BECexper-Sk1} L. S. Leslie, A. Hansen, K. C. Wright, B. M. Deutsch,
and N. P. Bigelow, Phys. Rev. Lett. \textbf{103}, Creation and detection of
skyrmions in a Bose-Einstein condensate, 250401 (2009).

\bibitem{BECexper-Sk2} J. Y. Choi, W. J. Kwon, and Y. I. Shin, Observation
of topologically stable 2D skyrmions in an antiferromagnetic spinor
Bose-Einstein condensate, Phys. Rev. Lett. \textbf{108}, 035301 (2012).

\bibitem{BEChopfion} Y. M. Bidasyuk, A. V. Chumachenko, O. O. Prikhodko, S.
I. Vilchinskii, M. Weyrauch, and A. I. Yakimenko, Stable Hopf solitons in
rotating Bose-Einstein condensates, Phys. Rev. A \textbf{92}, 053603 (2015).

\bibitem{Dum} D. Mihalache, Linear and nonlinear light bullets: recent
theoretical and experimental studies, Rom. J. Phys. \textbf{57}, 352-371
(2012).

\bibitem{Dum-select} D. Mihalache, Multidimensional localized structures in
optics and Bose-Einstein condensates: A selection of recent studies, Rom.
Journ. Phys. \textbf{59}, 295-315 (2014)

\bibitem{me} B. A. Malomed, Multidimensional solitons: Well-established
results and novel findings, Eur. Phys. J. Special Topics \textbf{225},
2507-2532 (2016).

\bibitem{NatureRev} Y. Kartashov, G. Astrakharchik, B. Malomed, and L.
Torner, Frontiers in multidimensional self-trapping of nonlinear fields and
matter, Nature Reviews Physics, https://doi.org/10.1038/ s42254-019-0025-7.

\bibitem{Manolo} M. Quiroga-Teixeiro and H. Michinel, Stable azimuthal
stationary state in quintic nonlinear optical media, J. Opt. Soc. Am. B
\textbf{14}, 2004-2009 (1997).

\bibitem{Pego} R. L. Pego and H. A. Warchall, Spectrally stable encapsulated
vortices for nonlinear Schr\"{o}dinger equations, J. Nonlinear Sci. \textbf{%
12}, 347-394 (2002).

\bibitem{Fetter} A. L. Fetter, Rotating trapped Bose-Einstein condensates,
Rev. Mod. Phys. \textbf{81}, 657-691 (2009).

\bibitem{nine} D. Mihalache, D. Mazilu, L.-C. Crasovan, I. Towers, A. V.
Buryak, B. A. Malomed, L. Torner, J. P. Torres, and F. Lederer, Stable
spinning optical solitons in three dimensions, Phys. Rev. Lett. \textbf{88},
073902 (2002).

\bibitem{CQ2comp2D} D. Mihalache, D. Mazilu, B. A. Malomed, and F. Lederer,
Stable vortex solitons in a vectorial cubic-quintic model, J. Opt. B \textbf{%
6}, S341-S350 (2004).

\bibitem{FWM} D. Mihalache, D. Mazilu, I. Towers, B. A. Malomed, and F.
Lederer, Stable two-dimensional spinning solitons in a bimodal cubic-quintic
model with four-wave mixing, J. Optics A: Pure and Applied Optics \textbf{4}%
, 615-623 (2002).

\bibitem{S=1and2chi2chi32D} I. Towers, A. V. Buryak, R. A. Sammut, and B. A.
Malomed, Stable localized vortex solitons, Phys. Rev. E \textbf{63},
055601(R) (2001).

\bibitem{S=3and4chi2chi32D} D. Mihalache, D. Mazilu, B. A. Malomed, and F.
Lederer, Stable vortex solitons supported by competing quadratic and cubic
nonlinearities. Phys. Rev. E \textbf{69}, 066614 (2004).

\bibitem{chi2chi33D} D. Mihalache, D. Mazilu, L.-C. Crasovan, I. Towers, B.
A. Malomed, A. V. Buryak, L. Torner, and F. Lederer, Stable
three-dimensional spinning optical solitons supported by competing quadratic
and cubic nonlinearities, Phys. Rev. E \textbf{66}, 016613 (2002).

\bibitem{interaction} B. A. Malomed, Potential of interaction between two-
and three-dimensional solitons, Phys. Rev. E \textbf{58}, 7928-7933 (1998).

\bibitem{SKA-int1} S. K. Adhikari, Mean-field model of interaction between
bright vortex solitons in Bose-Einstein condensates, New J. Phys. \textbf{5}%
, 137 (2003).

\bibitem{SKA-int2} S. K. Adhikari, Elastic collision and breather formation
of spatiotemporal vortex light bullets in a cubic-quintic nonlinear medium,
Laser Phys. Lett. \textbf{14}, 065402 (2017).

\bibitem{2D} F. Dalfovo and S. Stringari, Bosons in anisotropic traps:
Ground state and vortices, Phys. Rev. A \textbf{53}, 2477-2485 (1996).

\bibitem{2D2} R. J. Dodd, J. Res. Natl. Inst. Stand. Technol. \textbf{101},
545 (1996).

\bibitem{2D3} T. J. Alexander and L. Berg\'{e}, Ground states and vortices
of matter-wave condensates and optical guided waves, Phys. Rev. E \textbf{65}%
, 026611 (2002).

\bibitem{2D4} L. D. Carr and C. W. Clark, Vortices in attractive
Bose-Einstein condensates in two dimensions, Phys. Rev. Lett. \textbf{97},
010403 (2006).

\bibitem{Sadhan} S. K. Adhikari, Collapse of attractive Bose-Einstein
condensed vortex states in a cylindrical trap, Phys. Rev. E \textbf{65},
016703 (2001).

\bibitem{Ueda} H. Saito and M. Ueda, Split instability of a vortex in an
attractive Bose-Einstein condensate, Phys. Rev. Lett. \textbf{89}, 190402
(2002).

\bibitem{Ueda2} H. Saito and M. Ueda, Split-merge cycle, fragmented
collapse, and vortex disintegration in rotating Bose-Einstein condensates
with attractive interactions, Phys. Rev. A \textbf{69}, 013604 (2004).

\bibitem{Dum2D} D. Mihalache, D. Mazilu, B. A. Malomed, and F. Lederer,
Vortex stability in nearly-two-dimensional Bose-Einstein condensates with
attraction, Phys. Rev. A \textbf{73}, 043615 (2006).

\bibitem{Dum3D} B. A. Malomed, F. Lederer, D. Mazilu, and D. Mihalache, On
stability of vortices in three-dimensional self-attractive Bose-Einstein
condensates, Phys. Lett. B \textbf{36}1, 336-340 (2007).

\bibitem{we} H. Sakaguchi, B. Li, and B. A. Malomed, Creation of
two-dimensional composite solitons in spin-orbit-coupled self-attractive
Bose-Einstein condensates in free space, Phys. Rev. E \textbf{89}, 032920
(2014).

\bibitem{Sherman2} H. Sakaguchi, B. Li, E. Ya. Sherman, and B. A. Malomed,
Composite solitons in two-dimensional spin-orbit coupled self-attractive
Bose-Einstein condensates in free space, Romanian Reports in Physics \textbf{%
70}, 502 (2018).

\bibitem{HP} Y.-C. Zhang, Z.-W. Zhou, B. A. Malomed, and H. Pu, Stable
solitons in three dimensional free space without the ground state:
Self-trapped Bose-Einstein condensates with spin-orbit coupling, Phys. Rev.
Lett. \textbf{115}, 253902 (2015).

\bibitem{SKA-SOCvort1} S. Gautam and S. K. Adhikari, Vortex-bright solitons
in a spin-orbit coupled spin-1 condensate, Phys. Rev. A \textbf{95}, 013608
(2017).

\bibitem{SKA-SOCvort2} S. Gautam and S. K. Adhikari, Three-dimensional
vortex-bright solitons in a spin-orbit coupled spin-1 condensate, Phys. Rev.
A \textbf{97}, 013629 (2017).

\bibitem{Jieli2} J. Qin, G. Dong, and B. A. Malomed, Stable giant vortex
annuli in microwave-coupled atomic condensates, Phys. Rev. A \textbf{94},
053611 (2016).

\bibitem{Coullet} P. Coullet, L. Gil, and F. Rocca, Optical vortices, Opt.
Commun. \textbf{73}, 403-408 (1989).

\bibitem{Neu} J. C. Neu, Vortices in complex scalar fields, Physica D
\textbf{43}, 385-406 (1990).

\bibitem{Swartz1} G. A. Swartzlander and C. T. Law, Optical vortex solitons
observed in Kerr nonlinear media \textbf{69}, 2503-2506 (1992).

\bibitem{Swartz2} D. Rozas, C. T. Law, and G. A. Swartzlander, Propagation
dynamics of optical vortices, J. Opt. Soc. Am. B \textbf{14}, 3054-3065
(1997).

\bibitem{Soskin} I. V. Basistiy, V. Yu. Bazhenov, M. S. Soskin, and M. V.
Vasnetsov, Optics of light beams with screw dislocations, Opt. Commun. 103,
422-428 (1993).

\bibitem{vort-review-early} A. S. Desyatnikov, L. Torner, and Y. S. Kivshar,
Optical vortices and vortex solitons, Progr. Opt. \textbf{47}, 1-60 (2005).

\bibitem{Vyslo} Y. V. Kartashov, V. A. Vysloukh, and L. Torner, Stable
ring-profile vortex solitons in Bessel optical lattices, Phys. Rev. Lett.
\textbf{94}, 043902 (2005).

\bibitem{Bessel-beam-DiTrapani} C. L. Arnold, S. Akturk, A. Mysyrowicz, V.
Jukna, A. Couairon, T. Itina, R. Stoian, C. Xie, J. M. Dudley, F.
Courvoisier, S. Bonanomi, O. Jedrkiewicz, and P. Di Trapani, Nonlinear
Bessel vortex beams for applications, J. Phys. B: At. Mol. Opt. Phys.
\textbf{48}, 094006 (2015).

\bibitem{vort-in-liq-cryst} R. Barboza, U. Bortolozzo, M. G. Clerc, S.
Residori, and E. Vidal-Henriquez, Optical vortex induction via light-matter
interaction in liquid-crystal media, Adv. Opt. Phot. 7, 635-683 (2015).

\bibitem{vort-in-communications} J. Wang, Advances in communications using
optical vortices, Photon. Res. \textbf{4}, B14-B28 (2016).

\bibitem{polaritons} I. Carusotto and C. Ciuti, Quantum fluids of light,
Rev. Mod. Phys. \textbf{85} (2013).

\bibitem{polaritons1} B. Deveaud-Pl\'{e}dran, On the condensation of
polaritons, J. Opt. Soc. Am. B \textbf{29}, A138-A145 (2012).

\bibitem{polaritons2} T. Byrnes, N. Y. Kim, and Y. Yamamoto,
Exciton-polariton condensates, Nature Phys. \textbf{10}, 803-813 (2014).

\bibitem{Aftalion} A. Aftalion, \textit{Vortices in Bose-Einstein Condensates%
} (Birkh\"{a}user,Boston,2006)

\bibitem{Cornell-vortex} M. R. Matthews, B. P. Anderson, P. C. Haljan, D. S.
Hall, C. E. Weiman, and E. A. Cornell, Vortices in a Bose-Einstein
condensate, Phys. Rev. Lett. \textbf{83}, 2498-2501 (1999).

\bibitem{Cornell2} B. P. Anderson, P. C. Haljan, C. E. Wieman, and E. A.
Cornell, Vortex precession in Bose-Einstein condensates: Observations with
filled and empty cores, Phys. Rev. Lett. 85, 2857-2860 (2000).

\bibitem{Cornell3} S. Tung, V. Schweikhard, and E. A. Cornell, Observation
of vortex pinning in Bose-Einstein condensates, Phys. Rev. Lett. \textbf{97}%
, 240402 (2006).

\bibitem{Hall} D. V. Freilich, D. M. Bianchi, A. M. Kaufman, T. K. Langin,
and D. S. Hall, Real-Time dynamics of single vortex lines and vortex dipoles
in a Bose-Einstein condensate, Science \textbf{329}, 1182-1185 (2010).

\bibitem{TsuKasa} M. Tsubota and K. Kasamatsu, Dynamics of quantized
vortices in superfluid helium and rotating Bose-Einstein condensates, J. Low
Temp. Phys. \textbf{138}, 471-480 (2005).

\bibitem{BEC-exper} R. Srinivasan, Vortices in Bose-Einstein condensates: A
review of the experimental results, Pramana \textbf{66}, 3-30 (2006).

\bibitem{Fermi} M. W. Zwierlein, J. R. Abo-Shaeer, A. Schirotzek, C. H.
Schunck, and W. Ketterle, Vortices and superfluidity in a strongly
interacting Fermi gas, Nature \textbf{435}, 1047-1051 (2005).

\bibitem{Fermi-review} S. Giorgini, L. P. Pitaevskii, and S. Stringari,
Theory of ultracold atomic Fermi gases, Rev. Mod. Phys. \textbf{80},
1215-1274 (2008).

\bibitem{black} P. G. Kevrekidis and D. J. Frantzeskakis, and R.
Carretero-Gonz\'{a}lez, \textit{The Defocusing Nonlinear Schr\"{o}dinger
Equation: From Dark Solitons to Vortices and Vortex Rings} (SIAM,
Philadelphia, 2015).

\bibitem{Editorial} M. Soskin, S. V. Boriskina, Y. Chong, M. R Dennis, and
A. Desyatnikov, Singular optics and topological photonics, J. Opt. B \textbf{%
19}, 010401 (2017).

\bibitem{Gbur} G. J. Gbur, \textit{Singular Optics} (CRC Press, Boca Raton
(FL), 2017).

\bibitem{Allen} L. Allen, M. W. Beijersbergen, R. J. C. Spreeuw, and J. P.
Woerdman, Orbital angular momentum of light and the transformation of
Laguerre-Gaussian laser modes, Phys. Rev. A \textbf{45}, 8185-8189 (1992).

\bibitem{opt-angular-mom} S. Franke-Arnold, L. Allen, and M. Padgett,
Advances in optical angular momentum, Laser \& Photon. Rev. \textbf{2},
299-313 (2008).

\bibitem{Bliokh-Nori} K. Y. Bliokh and F. Nori, Transverse and longitudinal
angular momenta of light, Phys. Rep. \textbf{592}, 1-38 (2015).

\bibitem{angular-momentum-storage} B.-S. Shi, D.-S. Ding, and W. Zhang,
Quantum storage of orbital angular momentum entanglement in cold atomic
ensembles, J. Phys. B: At. Mol. Opt. Phys. \textbf{51}, 032004 (2018).

\bibitem{Svidzinsky} A. A. Svidzinsky and A. L. Fetter, Dynamics of a vortex
in a trapped Bose-Einstein condensate, Phys. Rev. A \textbf{62}, 063617
(2000).

\bibitem{bending} A. Aftalion and T. Riviere, Vortex energy and vortex
bending for a rotating Bose-Einstein condensate, Phys. Rev. A \textbf{64},
043611 (2001).

\bibitem{bending2} A. Aftalion and I. Danaila, Three-dimensional vortex
configurations in a rotating Bose-Einstein condensate, Phys. Rev. A \textbf{%
68}, 023603 (2003).

\bibitem{Dalibard-bending} P. Rosenbusch, V. Bretin, and J. Dalibard,
Dynamics of a single vortex line in a Bose-Einstein condensate, Phys. Rev.
Lett. \textbf{89}, 200403 (2002).

\bibitem{Bagnato-bending} E. A. L. Henn, J. A. Seman, E. R. F. Ramos, M.
Caracanhas, P. Castilho, E. P. Ol\'{\i}mpio, G. Roati, D. V. Magal\~{a}es,
K. M. F. Magal\~{a}es, and V. S. Bagnato, Observation of vortex formation in
an oscillating trapped Bose-Einstein condensate, Phys. Rev. A \textbf{79},
043618 (2009).

\bibitem{Berloff} C. Yin, N. G. Berloff, V. M. P\'{e}rez-Garc\'{\i}a, D.
Novoa, A. V. Carpentier, H. Michinel, Coherent atomic soliton molecules for
matter-wave switching, Phys. Rev. A \textbf{83}, 051605 (2011).

\bibitem{chi1} G. I. Stegeman, D. J. Hagan, and L. Torner, Cascading
phenomena and their applications to all-optical signal processing,
mode-locking, pulse compression and solitons, Opt. Quant. Elect. \textbf{28}%
, 1691-1740 (1996).

\bibitem{chi2} C. Etrich, F. Lederer, B. A. Malomed, T. Peschel, and U.
Peschel, Optical solitons in media with a quadratic nonlinearity, Prog. Opt.
\textbf{41}, 483-568 (2000).

\bibitem{chi3} A. V. Buryak, P. Di Trapani, D. V. Skryabin, and S. Trillo,
Optical solitons due to quadratic nonlinearities: from basic physics to
futuristic applications, Phys. Rep. 370, 63-235 (2002).

\bibitem{chi4} M. Colin, L. Di Menza, and J. C. Saut, Solitons in quadratic
media, Nonlinearity \textbf{29}, 1000-1035 (2016).

\bibitem{Rubi} A. A. Kanashov and A. M. Rubenchik, On diffraction and
dispersion effect on three-wave interaction, Physica D \textbf{4}, 122-134
(1981).

\bibitem{HaoHe} B. A. Malomed, P. Drummond, H. He, A. Berntson, D. Anderson,
and M. Lisak, Spatio-temporal solitons in optical media with a quadratic
nonlinearity. Phys. Rev. E \textbf{56}, 4725-4735 (1997).

\bibitem{Frank1} X. Liu, L. J. Qian, and F. W. Wise, Generation of Optical
Spatiotemporal Solitons, Phys. Rev. Lett. \textbf{82}, 4631-4634 (1999).

\bibitem{Frank2} X. Liu, K. Beckwitt, and F. Wise, Two-dimensional optical
spatiotemporal solitons in quadratic media, Phys. Rev. E \textbf{62},
1328-1340 (2000).

\bibitem{Dima1} W. J. Firth and D. V. Skryabin, Optical solitons carrying
orbital angular momentum, Phys. Rev. Lett. \textbf{79}, 2450-2453 (1997).

\bibitem{Petrov1} L. Torner and D. V. Petrov, Azimuthal instabilities and
self-breaking of beams into sets of solitons in bulk second-harmonic
generation, Electr. Lett. \textbf{33}, 608-610 (1997).

\bibitem{Petrov2} L. Torner and D. V. Petrov, Splitting of light beams with
spiral phase dislocations into solitons in bulk quadratic nonlinear media,
J. Opt. Soc. Am. B \textbf{14}, 2017-2023 (1997).

\bibitem{Dima2} D. V. Skryabin and W. J. Firth, Dynamics of self-trapped
beams with phase dislocation in saturable Kerr and quadratic nonlinear
media, Phys. Rev. E \textbf{58}, 3916-3930 (1998).

\bibitem{3wave} J. P. Torres, J. M. Soto-Crespo, L. Torner, and D. V.
Petrov, Solitary-wave vortices in type II second-harmonic generation, Opt.
Commun. \textbf{149}, 77-83 (1998).

\bibitem{Herve-3waves} H. Leblond, B. A. Malomed, and D. Mihalache,
Quasistable two-dimensional solitons with hidden and explicit vorticity in a
medium with competing nonlinearities, Phys. Rev. E \textbf{71}, 036608
(2005).

\bibitem{Petrov} D. V. Petrov, L. Torner, J. Martorell, R, Vilaseca, J. P.
Torres, and C. Cojocaru, Observation of azimuthal modulational instability
and formation of patterns of optical solitons in a quadratic nonlinear
crystal, Opt. Lett. 23, 1444-1446 (1998).

\bibitem{stimulated} S. Minardi, G. Molina-Terriza, P. Di Trapani, J. P.
Torres, and L. Torner, Soliton algebra by vortex-beam splitting, Opt. Lett.
\textbf{26}, 1004-1006 (2001).

\bibitem{SKA} S. Gautam and S. K. Adhikari, Self-trapped quantum balls in
binary Bose-Einstein condensates, J. Phys. B: At. Mol. Opt. Phys. \textbf{52}%
, 055302 (2019).

\bibitem{dip1} I. Ferrier-Barbut, H. Kadau, M. Schmitt, M. Wenzel, and T.
Pfau, Observation of quantum droplets in a strongly dipolar Bose gas, Phys.
Rev. Lett. \textbf{116}, 215301 (2016).

\bibitem{dip2} L. Chomaz, S. Baier, D. Petter, M.\thinspace J. Mark, F. W%
\"{a}chtler, L. Santos, and F. Ferlaino, Quantum-fluctuation-driven
crossover from a dilute Bose-Einstein condensate to a macrodroplet in a
dipolar quantum fluid, Phys. Rev. X \textbf{6}, 041039 (2016).

\bibitem{Macri} A. Cidrim, F. E. A. dos Santos, E. A. L. Henn, and T. Macr%
\'{\i}, Vortices in self-bound dipolar droplets, Phys. Rev. A \textbf{98},
023618 (2018).

\bibitem{Rypdal} J. J. Rasmussen and K. Rypdal, Blow-up in nonlinear Schr%
\"{o}dinger equations. 1. A general review, Phys. Scripta \textbf{33},
481-497 (1986).

\bibitem{Tikho} V. Tikhonenko, J. Christou, and B. Luther-Daves, Spiraling
bright spatial solitons formed by the breakup of an optical vortex in a
saturable self-focusing medium, J. Opt. Soc. Am. B \textbf{12}, 2046-2052
(1995).

\bibitem{Cid-robust} L. Edilson, L. Falc\~{a}o-Filho, C. B. de Ara\'{u}jo,
G. Boudebs, H. Leblond, and V. Skarka, Robust two-dimensional spatial
solitons in liquid carbon disulfide, Phys. Rev. Lett. \textbf{110}, 013901
(2013).

\bibitem{Cid-review} A. S. Reyna and C. B. de Ara\'{u}jo, High-order optical
nonlinearities in plasmonic nanocomposites -- a review, Adv. Opt. Phot.
\textbf{9}, 720-774 (2017).

\bibitem{Cid2} A. S. Reyna, K. C. Jorge, and C. B. de Ara\'{u}jo,
Two-dimensional solitons in a quintic-septimal medium, Phys. Rev. A \textbf{%
90}, 063835 (2014).

\bibitem{Pushkarovs} Kh. I. Pushkarov, D. I. Pushkarov, and I. V. Tomov,
Self-action of light beams in nonlinear media: soliton solutions, Opt.
Quantum Electron. \textbf{11}, 471-478 (1979).

\bibitem{Enns} S. Cowan, R. H. Enns, S. S. Rangnekar, and S. S. Sanghera,
Quasi-soliton and other behavior of the nonlinear cubic-quintic Schr\"{o}%
dinger equation, Can. J. Phys. \textbf{64}, 311-315 (1986).

\bibitem{Drits} V. I. Kruglov, V. M. Volkov, R. A. Vlasov, and V. V. Drits,
Auto-waveguide propagation and the collapse of spiral light beams in
non-linear media, J. Phys. A: Math. Gen. \textbf{21}, 4381-4395 (1988).

\bibitem{spatiotemp-vort1} N. Dror and B. A. Malomed, Symmetric and
asymmetric solitons and vortices in linearly coupled two-dimensional
waveguides with the cubic-quintic nonlinearity, Physica D \textbf{240},
526-541 (2011).

\bibitem{spatiotemp-vort2} N. Jhajj, I. Larkin, E. W. Rosenthal, S.
Zahedpour, J. K. Wahlstrand, and H. M. Milchberg, Spatiotemporal Optical
Vortices, Phys. Rev. X \textbf{6}, 031037 (2016).

\bibitem{NonlinDyn} Y. Y. Wang, L. Chen, C. Q. Dai, J. Zheng, and Y. Fan,
Exact vector multipole and vortex solitons in the media with spatially
modulated cubic-quintic nonlinearity. Nonlin. Dynamics \textbf{90},
1269-1275 (2017).

\bibitem{NonlinDyn2} C. Q. Dai, G. Q. Zhou, R. P. Chen, X. J. Lai, and J.
Zheng, Vector multipole and vortex solitons in two-dimensional Kerr media,
Nonlin. Dynamics \textbf{88}, 2629-2635 (2017).

\bibitem{LeeHY} T. D. Lee, K. Huang, and C. N. Yang, Eigenvalues and
eigenfunctions of a Bose system of hard spheres and its low-temperature
properties, Phys. Rev. \textbf{106}, 1135--1145 (1957).

\bibitem{Petrov-QD} D. S. Petrov, Quantum mechanical stabilization of a
collapsing Bose-Bose mixture. Phys. Rev. Lett. \textbf{115}, 155302 (2015).

\bibitem{Astra} D. S. Petrov and G. E. Astrakharchik, Ultradilute
low-dimensional liquids, Phys. Rev. Lett. \textbf{117}, 100401 (2016).

\bibitem{Tarr1} C. R. Cabrera, L. Tanzi, J. Sanz, B. Naylor, P. Thomas, P.
Cheiney, and L. Tarruell, Quantum liquid droplets in a mixture of
Bose-Einstein condensates, Science \textbf{359}, 301-304 (2018).

\bibitem{Tarr2} P. Cheiney, C. R. Cabrera, J. Sanz, B. Naylor, L. Tanzi, and
L. Tarruell, Bright soliton to quantum droplet transition in a mixture of
Bose-Einstein condensates, Phys. Rev. Lett. \textbf{120}, 135301 (2018).

\bibitem{Ing} G. Semeghini, G. Ferioli, L. Masi, C. Mazzinghi, L. Wolswijk,
F. Minardi, M. Modugno, G. Modugno, M. Inguscio, and M. Fattori, Self-bound
quantum droplets in atomic mixtures, Phys. Rev. Lett. \textbf{120}, 235301
(2018).

\bibitem{Ing2} G. Ferioli, G. Semeghini, L. Masi, G. Giusti, G. Modugno, M.
Inguscio, A. Gallem\'{\i}, A. Recati, and M. Fattori, Collisions of
self-bound quantum droplets, Phys. Rev. Lett. \textbf{122}, 090401 (2019).

\bibitem{3DLHY} Y. V. Kartashov, B. A. Malomed, L. Tarruell, and L. Torner,
Three-dimensional droplets of swirling superfluids, Phys. Rev. A \textbf{98}%
, 013612 (2018).

\bibitem{GZ-ln} Y. Li, Z. Chen, Z. Luo, C. Huang, H. Tan, W. Pang, and B. A.
Malomed, Two-dimensional vortex quantum droplets, Phys. Rev. A \textbf{98},
063602 (2018).

\bibitem{necklace} Y. V. Kartashov, B. A. Malomed, and L. Torner,
Metastability of quantum droplet clusters, Phys. Rev. Lett., in press.

\bibitem{Viskol} Z. Chen, Y. Li, B. A. Malomed, and L. Salasnich,
Spontaneous symmetry breaking of fundamental states, vortices, and dipoles
in two and one-dimensional linearly coupled traps with cubic
self-attraction, Phys. Rev. A \textbf{96}, 033621 (2016).

\bibitem{Pit} L. P. Pitaevskii and S. Stringari, \textit{Bose-Einstein
Condensation} (Oxford University Press, Oxford, 2003).

\bibitem{low-dim} B. B. Baizakov, B. A. Malomed and M. Salerno,
Multidimensional solitons in a low-dimensional periodic potential, Phys.
Rev. A \textbf{70}, 053613 (2004).

\bibitem{low-dim2} D. Mihalache, D. Mazilu, F. Lederer, Y. V. Kartashov,
L.-C. Crasovan, and L. Torner, Stable three-dimensional spatiotemporal
solitons in a two-dimensional photonic lattice, Phys. Rev. E \textbf{70},
055603(R) (2004).

\bibitem{Herve-lowdim} H. Leblond, B. A. Malomed, and D. Mihalache,
Three-dimensional vortex solitons in quasi-two-dimensional lattices, Phys.
Rev. E \textbf{76}, 026604 (2007).

\bibitem{RMP} Y. V. Kartashov, B. A. Malomed, and L. Torner, Solitons in
nonlinear lattices, Rev. Mod. Phys. \textbf{83}, 247-306 (2011).

\bibitem{pancake} L. Chomaz, L. Corman, T. Bienaime, R. Desbuquois, C.
Weitenberg, S. Nascimb\`{e}ne, J. Beugnon, and J. Dalibard, Emergence of
coherence via transverse condensation in a uniform quasi-two-dimensional
Bose gas, Nature Comm. \textbf{6}, 6162 (2015).

\bibitem{Randy-NJP} K. E. Strecker, G. B. Partridge, A. G. Truscott, and R.
G. Hulet, Bright matter wave solitons in Bose--Einstein condensates, New J.
Phys. \textbf{5}, 73.1 (2003).

\bibitem{PhCr} J. D. Joannopoulos, S. G. Johnson, J. N. Winn, and R. D.
Meade, Photonic Crystals: Molding the Flow of Light (Princeton University
Press, Princeton, 2008).

\bibitem{JY} M. Skorobogatiy and J. Yang, \textit{Fundamentals of Photonic
Crystal Guiding} (Cambridge University Press, Cambridge, 2008).

\bibitem{Kriz} E. A. Cerda-Mendez, D. Sarkar, D. N. Krizhanovskii, S. S.
Gavrilov, K. Biermann, M. S. Skolnick, and P. V. Santos, Exciton-polariton
gap solitons in two-dimensional lattices, Phys. Rev. Lett. \textbf{111},
146401 (2013).

\bibitem{Jena} A. Szameit and S. Nolte, Discrete optics in
femtosecond-laser-written photonic structures, J. Phys. B: At. Mol. Opt.
Phys. \textbf{43}, 163001 (2010).

\bibitem{MSegev} N. K. Efremidis, S. Sears, D. N. Christodoulides, J. W.
Fleischer, and M. Segev, Discrete solitons in photorefractive optically
induced photonic lattices, Phys. Rev. E \textbf{66}, 046602 (2002).

\bibitem{Yang-book} J. Yang, \textit{Nonlinear Waves in Integrable and
Nonintegrable Systems} (SIAM, Philadelphia, 2010).

\bibitem{lattice1} B. B. Baizakov, B. A. Malomed, and M. Salerno,
Multidimensional solitons in periodic potentials, Europhys. Lett. \textbf{63}%
, 642-648 (2003).

\bibitem{lattice2} J. Yang and Z. H. Musslimani, Fundamental and vortex
solitons in a two-dimensional optical lattice, Opt. Lett. \textbf{28},
2094-2096 (2003).

\bibitem{lattice3} Z. H. Musslimani and J. Yang, Self-trapping of light in a
two-dimensional photonic lattice, J. Opt. Soc. Am. \textbf{21}, 973-981
(2004).

\bibitem{replication} J. R. Salgueiro, M. Zacar\'{e}s, H. Michinel, and A.
Ferrando, Vortex replication in Bose-Einstein condensates trapped in
double-well potentials, Phys. Rev. A \textbf{79}, 033625 (2009).

\bibitem{Asymm} T. J. Alexander, A. A. Sukhorukov, and Y. S. Kivshar,
Asymmetric vortex solitons in nonlinear periodic lattices, Phys. Rev. Lett.
93, 063901 (2004).

\bibitem{Progress} B. A. Malomed, Variational methods in nonlinear fiber
optics and related fields, Progr. Optics \textbf{43}, 71-193 (2002).

\bibitem{VK} N. G. Vakhitov and A. A. Kolokolov, Stationary solutions of the
wave equation in a medium with nonlinearity saturation, Radiophys. Quant.
Electron. \textbf{16}, 783-789 (1973).

\bibitem{Luca} L. Salasnich, B. A. Malomed, and F. Toigo, Matter-wave
vortices in cigar-shaped and toroidal waveguides, Phys. Rev. A \textbf{76},
063614 (2007).

%\bibitem{Rb85} A. L. Marchant, T. P. Billam, T. P. Wiles, M. M. H. Yu, S. A.
%Gardiner, and S. L. Cornish,
%Controlled formation and reflection of a bright solitary matter-wave,
%Nature Comm. \textbf{4}, 1865 (2013).

%\bibitem{Rb85-2}
%J. Everitt, M. A. Sooriyabandara, G. D. McDonald, K. S. Hardman, C. Quinlivan,
%P. Manju, P. Wigley, J. E. Debs, J. D. Close, C. C. N. Kuhn, and N. P.
%Robins, arXiv:1509.06844.

\bibitem{Weiman} S. L. Cornish, S. T. Thompson, and C. E. Wieman, Formation
of bright matter-wave solitons during the collapse of attractive
Bose-Einstein condensates, Phys. Rev. Lett. \textbf{96}, 170401 (2006).

\bibitem{Radik} R. Driben, V. V. Konotop, B. A. Malomed, and T. Meier,
Dynamics of dipoles and vortices in nonlinearly coupled three-dimensional
field oscillators, Phys. Rev. E \textbf{94}, 012207 (2016).

\bibitem{Brtka} M. Brtka, A. Gammal, and B. A. Malomed, Hidden vorticity in
binary Bose-Einstein condensates, Phys. Rev. A \textbf{82}, 053610 (2010).

\bibitem{VPG-mixed} J. J. Garc\'{\i}a-Ripoll and V. M. P\'{e}rez-Garc\'{\i}%
a, Stable and unstable vortices in multicomponent Bose-Einstein condensates,
Phys. Rev. Lett. \textbf{84}, 4264-4267 (2000).

\bibitem{big} F. Lederer, G. I. Stegeman, D. N. Christodoulides, G. Assanto,
M. Segev, and Y. Silberberg, Discrete solitons in optics, Phys. Rep. \textbf{%
463}, 1-126 (2008).

\bibitem{Vyslo1} Y. V. Kartashov, V. A. Vysloukh, and L. Torner, Soliton
shape and mobility control in optical lattices, Prog. Opt. \textbf{52},
63-148 (2009).

\bibitem{Vyslo2} Y. V. Kartashov, V. A. Vysloukh, and L. Torner, Solitons in
complex optical lattices, Eur. Phys. J. Special Topics \textbf{173}, 87-105
(2009).

\bibitem{Yukalov} V. I. Yukalov, Cold bosons in optical lattices, Laser
Physics \textbf{19}, 1-110 (2009).

\bibitem{Szameit} M. Heinrich, R. Keil, F. Dreisow, A. T\"{u}nnermann, A.
Szameit, and S. Nolte, Nonlinear discrete optics in femtosecond
laser-written photonic lattices, Appl. Phys. B \textbf{104}, 469-480 (2011).

\bibitem{Entropy} G. Watanabe, B. P. Venkatesh, and R. Dasgupta, Nonlinear
phenomena of ultracold atomic gases in optical lattices: Emergence of novel
Features in Extended States, Entropy \textbf{18}, 118 (2016).

\bibitem{PGK-book} P. G. Kevrekidis, \textit{The Discrete Nonlinear Schr\"{o}%
dinger Equation: Mathematical Analysis, Numerical Computations, and Physical
Perspectives} (Springer: Berlin and Heidelberg, 2009).

\bibitem{HS-EPL} H. Sakaguchi and B. A. Malomed, Higher-order vortex
solitons, multipoles, and supervortices on a square optical lattice.
Europhys. Lett. \textbf{72}, 698-704 (2005).

\bibitem{Radik2} R. Driben and B. A. Malomed, Stabilization of
two-dimensional solitons and vortices against supercritical collapse by
lattice potentials, Eur. Phys. J. D \textbf{50}, 317-323 (2008).

\bibitem{simulator} P. Hauke, F. M. Cucchietti, and L. Tagliacozzo, I.
Deutsch, and M. Lewenstein, Can one trust quantum simulators?, Rep. Prog.
Phys. \textbf{75}, 082401 (2012).

\bibitem{simulator2} T. H. Johnson, S. R. Clark, and D. Jaksch, What is a
quantum simulator?, EPJ Quantum Technology \textbf{1},10 (2014).

\bibitem{simulator3} E. Zohar, J. I. Cirac, and B. Reznik, Quantum
simulations of lattice gauge theories using ultracold atoms in optical
lattices, Rep. Prog. Phys. \textbf{79}, 014401 (2016).

\bibitem{Dresselhaus} G. Dresselhaus, Spin-orbit coupling effects in zinc
blende structures, Phys. Rev. \textbf{100}, 580-586 (1955).

\bibitem{Rashba} Y. A. Bychkov and E. I. Rashba, Oscillatory effects and the
magnetic-susceptibility of carriers in inverse-layers, J. Phys. C \textbf{17}%
, 6039-6045 (1984).

\bibitem{Campbell} D. L. Campbell, G. Juzeli\={u}nas, and I. B. Spielman,
Realistic Rashba and Dresselhaus spin-orbit coupling for neutral atoms,
Phys. Rev. A \textbf{84}, 025602 (2011).

\bibitem{socbec} Y. J. Lin, K. Jimenez-Garcia, and I. B. Spielman,
Spin-orbit-coupled Bose-Einstein condensates, Nature \textbf{471}, 83-86
(2011).

\bibitem{socbec2} J.-Y. Zhang, S.-C. Ji, Z. Chen, L. Zhang, Z.-D. Du, B.
Yan, G.-S. Pan, B. Zhao, Y. J. Deng, H. Zhai, S. Chen, and J.-W. Pan,
Collective dipole oscillations of a spin-orbit coupled Bose-Einstein
condensate, Phys. Rev. Lett. \textbf{109}, 115301 (2012).

\bibitem{socbec3} C. Hamner, C. Qu, Y. Zhang, J. Chang, M. Gong, C. Zhang,
and P. Engels, Dicke-type phase transition in a spin-orbit-coupled
Bose-Einstein condensate, Nature Commun. \textbf{5}, 4023 (2014).

\bibitem{socbec4} A. J. Olson, S.-J. Wang, R. J. Niffenegger, C.-H. Li, C.
H. Greene, and Y. P. Chen, Tunable Landau-Zener transitions in a
spin-orbit-coupled Bose-Einstein condensate, Phys. Rev. A \textbf{90},
013616 (2014).

\bibitem{theory-SOC} Y. Zhang, L. Mao, and C. Zhang, Mean-Field dynamics of
spin-orbit coupled Bose-Einstein condensates, Phys. Rev. Lett. \textbf{108},
035302 (2012).

\bibitem{theory-SOC2} Y. Li, L. P. Pitaevskii, and S. Stringari, Quantum
Tricriticality and Phase Transitions in Spin-Orbit Coupled Bose-Einstein
Condensates, Phys. Rev. Lett. \textbf{108}, 225301 (2012).

\bibitem{theory-SOC3} Y. Zhang, G. Chen, and C. Zhang, Tunable Spin-orbit
Coupling and Quantum Phase Transition in a Trapped Bose-Einstein Condensate,
Scientific Reports \textbf{3}, 1937 (2013).

\bibitem{theory-SOC4} D. A. Zezyulin, R. Driben, V. V. Konotop, and B. A.
Malomed, Nonlinear modes in binary bosonic condensates with
pseudo-spin-orbital coupling, Phys. Rev. A \textbf{88}, 013607 (2013).

\bibitem{theory-SOC5} Y.-C. Zhang, Sh.-W. Song, and W.-M. Liu, The
confinement induced resonance in spin-orbit coupled cold atoms with Raman
coupling, Scientific Reports \textbf{4}, 4992 (2014).

\bibitem{Konotop} V. Achilleos, D. J. Frantzeskakis, P. G. Kevrekidis, and
D. E. Pelinovsky, Matter-wave bright solitons in spin-orbit coupled
Bose-Einstein condensates, Phys. Rev. Lett. \textbf{110}, 264101 (2013).

\bibitem{Konotop2} Y. V. Kartashov, V. V. Konotop, and F. Kh. Abdullaev, Gap
solitons in a spin-orbit-coupled Bose-Einstein condensate, Phys. Rev. Lett.
\textbf{111}, 060402 (2013).

\bibitem{Konotop3} Y. Xu, Y. Zhang, and B. Wu, Bright solitons in
spin-orbit-coupled Bose-Einstein condensates, Phys. Rev. A \textbf{87},
013614 (2013).

\bibitem{Konotop4} L. Salasnich and B. A. Malomed, Localized modes in dense
repulsive and attractive Bose-Einstein condensates with spin-orbit and Rabi
couplings, Phys. Rev. A \textbf{87}, 063625 (2013).

\bibitem{Konotop5} Y. V. Kartashov, V. V. Konotop, and D. A. Zezyulin,
Bose-Einstein condensates with localized spin-orbit coupling: Soliton
complexes and spinor dynamics, Phys. Rev. A \textbf{90}, 063621 (2014).

\bibitem{gap-sol} V. E. Lobanov, Y. V. Kartashov, and V. V. Konotop,
Fundamental, multipole, and half-vortex gap solitons in spin-orbit coupled
Bose-Einstein condensates, Phys. Rev. Lett. \textbf{112}, 180403 (2014).

\bibitem{Fukuoka} S. Sinha, R. Nath, and L. Santos, Trapped two-dimensional
condensates with synthetic spin-orbit coupling, Phys. Rev. Lett. \textbf{107}%
, 270401 (2011).

\bibitem{Fukuoka1} C. J. Wu, I. Mondragon-Shem, and X.-F. Zhou,
Unconventional Bose-Einstein Condensations from Spin-Orbit Coupling, Chin.
Phys. Lett. \textbf{28}, 097102 (2011).

\bibitem{Fukuoka3} Y. Deng, J. Cheng, H. Jing, C. P. Sun, and S. Yi,
Spin-orbit-coupled dipolar Bose-Einstein condensates, Phys. Rev. Lett.
\textbf{108}, 125301 (2012).

\bibitem{Fukuoka4} Textures of $F=2$ spinor Bose-Einstein condensates with
spin-orbit coupling, T. Kawakami, T. Mizushima, and K. Machida, Phys. Rev. A
\textbf{84}, 011607 (2011).

\bibitem{Fukuoka5} Half-quantum vortex state in a spin-orbit-coupled
Bose-Einstein condensate, B. Ramachandhran, B. Opanchuk, X.-J. Liu, H. Pu,
P. D. Drummond, and H. Hu, Phys. Rev. A \textbf{85}, 023606 (2012).

\bibitem{Fukuoka6} G. J. Conduit, Line of Dirac monopoles embedded in a
Bose-Einstein condensate, Phys. Rev. A \textbf{86}, 021605(R) (2012).

\bibitem{Fukuoka7} E. Ruokokoski, J. A. M. Huhtam\"{a}ki, and M. M\"{o}tt%
\"{o}nen, Stationary states of trapped spin-orbit-coupled Bose-Einstein
condensates, Phys. Rev. A \textbf{86}, 051607 (2012).

\bibitem{Fukuoka8} H. Sakaguchi and B. Li, Vortex lattice solutions to the
Gross-Pitaevskii equation with spin-orbit coupling in optical lattices,
Phys. Rev. A \textbf{87}, 015602 (2013).

\bibitem{Fukuoka9} A. Fetter, Vortex dynamics in spin-orbit-coupled
Bose-Einstein condensates, Phys. Rev. A \textbf{89}, 023629 (2014).

\bibitem{Fukuoka10} A. Fetter, Vortex Dynamics in a Spin-Orbit-Coupled
Bose-Einstein Condensate, J. Low Temp. Phys. \textbf{180}, 37-52 (2015).

\bibitem{Wesley} L. Salasnich, W. B. Cardoso, and B. A. Malomed, Localized
modes in quasi-two-dimensional Bose-Einstein condensates with spin-orbit and
Rabi couplings, Phys. Rev. A \textbf{90}, 033629 (2014).

\bibitem{Fukuoka2} H. Sakaguchi and B. A. Malomed, Discrete and continuum
composite solitons in Bose-Einstein condensates with the Rashba spin-orbit
coupling in one and two dimensions, Phys. Rev. E \textbf{90}, 062922 (2014).

\bibitem{rf:1} J. Dalibard, F. Gerbier, G. Juzeli\={u}nas, and P. \"{O}%
hberg, Artificial gauge potentials for neutral atoms, Rev. Mod. Phys.
\textbf{83}, 1523-1543 (2011).

\bibitem{rf:12} V. Galitski and I. B. Spielman, Spin-orbit coupling in
quantum gases, Nature \textbf{494}, 49-54 (2013).

\bibitem{rf:13} X. Zhou, Y. Li, Z. Cai, and C. Wu, Unconventional states of
bosons with the synthetic spin-orbit coupling, J. Phys. B: At. Mol. Opt.
Phys. \textbf{46}, 134001 (2013).

\bibitem{rf:14} N. Goldman, G. Juzeli\={u}nas, P. \"{O}hberg, and I. B.
Spielman, Light-induced gauge fields for ultracold atoms, Rep. Progr. Phys.
\textbf{77}, 126401 (2014).

\bibitem{rf:15} H. Zhai, Degenerate quantum gases with spin--orbit coupling:
a review, Rep. Prog. Phys. \textbf{78}, 026001 (2015).

\bibitem{im-time} B. D. Esry, C. H. Greene, J. P. Burke, Jr., and J. L.
Bohn, Hartree-Fock theory for double condensates, Phys. Rev. Lett. \textbf{78%
}, 3594-3597 (1997).

\bibitem{im-time2} D. L. Feder, M. S. Pindzola, L. A. Collins, B. I.
Schneider, and C. W. Clark, Dark-soliton states of Bose-Einstein condensates
in anisotropic traps, Phys. Rev. A \textbf{62}, 053606 (2000).

\bibitem{im-time3} M. L. Chiofalo, S. Succi, and M. P. Tosi, Ground state of
trapped interacting Bose-Einstein condensates by an explicit imaginary-time
algorithm, Phys. Rev. E \textbf{62}, 7438-7444 (2000).

\bibitem{im-time4} W. Bao and Q. Du, Computing the ground state solution of
Bose-Einstein condensates by a normalized gradient flow, SIAM J. Sci.
Comput. \textbf{25}, 1674-1697 (2004).

\bibitem{Kart} Y. V. Kartashov, B. A. Malomed, V. V. Konotop, V. E. Lobanov,
and L. Torner, Stabilization of solitons in bulk Kerr media by dispersive
coupling, Opt. Lett. \textbf{40}, 1045-1048 (2015).

\bibitem{Raymond} X. Jiang, Z. Fan, Z. Chen, W. Pang, Y. Li, and B. A.
Malomed, Two-dimensional solitons in dipolar Bose-Einstein condensates with
spin-orbit coupling, Phys. Rev. A \textbf{93}, 023633 (2016).

\bibitem{SOC-LHY} Y. Li, Z. Luo, Y. Liu, Z. Chen, C. Huang, S. Fu, H. Tan,
and B. A. Malomed, Two-dimensional solitons and quantum droplets supported
by competing self- and cross-interactions in spin-orbit-coupled condensates,
New J. Phys. \textbf{19}, 113043 (2017).

\bibitem{Jieli1} J. Qin, G. Dong, and B. A. Malomed, Hybrid
matter-wave-microwave solitons produced by the local-field effect, Phys.
Rev. Lett. \textbf{115}, 023901 (2015).

\bibitem{necklace1} M. Solija\v{c}i\'{c}, S. Sears, and M. Segev,
Self-trapping of ``necklace" beams in self-focusing Kerr media, Phys. Rev.
Lett. \textbf{81}, 4851 (1998).

\bibitem{necklace2} A. S. Desyatnikov and Y. S. Kivshar, Necklace-Ring
Vector Solitons, Phys. Rev. Lett. \textbf{87}, 033901 (2001).

\bibitem{necklace3} G. D. Montesinos, V. M. P\'{e}rez-Garc\'{\i}a, H.
Michinel, and J. R. Salgueiro, Stabilized vortices in layered Kerr media,
Phys. Rev. E \textbf{71}, 036624 (2005).

\bibitem{clusters} L. C. Crasovan, G. Molina-Terriza, J. P. Torres, L.
Torner, V. M. P\'{e}rez-Garc\'{\i}a, and D. Mihalache, Globally linked
vortex clusters in trapped wave fields, Phys. Rev. E \textbf{66}, 036612
(2002).

\bibitem{HS} H. Sakaguchi, New models for multi-dimensional stable vortex
solitons, Frontiers of Physics \textbf{14}, 1230 (2019).

\bibitem{LC1} D. W. McLaughlin, D. J. Muraki, M. J. Shelley, and W. Xiao, A
paraxial model for optical self-focussing in a nematic liquid crystal,
Physica D \textbf{88}, 55-81 (1995).

\bibitem{LC2} G. Assanto and M. Peccianti, Spatial solitons in nematic
liquid crystals, IEEE J. Quantum Electron. \textbf{39}, 13-21 (2003).

\bibitem{thermal} S. Akhmanov, D. Krindach, A. Migulin, A. Sukhorukov, and
R. Khokhlov, Thermal self-actions of laser beams, IEEE J. Quantum Electron.
\textbf{4}, 568-575 (1968).

\bibitem{Briedis} D. Briedis, D. E. Petersen, D. Edmundson, W. Kr\'{o}%
likowski, and O. Bang, Ring vortex solitons in nonlocal nonlinear media,
Opt. Exp. \textbf{13}, 435-443 (2005).

\bibitem{Snyder} A. W. Snyder and D. J. Mitchell, Accessible solitons,
Science \textbf{276}, 1538-1541 (1997).

\bibitem{Izdeb} Y. Izdebskaya, G. Assanto, and W. Krolikowski, Observation
of stable-vector vortex solitons. Opt. Lett. \textbf{40}, 4182-4185 (2015).

\bibitem{Pfau} T. Lahaye, C. Menotti, L. Santos, M. Lewenstein, and T. Pfau,
The physics of dipolar bosonic quantum gases, Rep. Prog. Phys. \textbf{72},
126401 (2009).

\bibitem{Lashkin} V. M. Lashkin, Two-dimensional nonlocal vortices,
multipole solitons, and rotating multisolitons in dipolar Bose-Einstein
condensates, Phys. Rev. A \textbf{75}, 043607 (2007).

\bibitem{Tikhon} I. Tikhonenkov, B. A. Malomed, and A. Vardi. Vortex
solitons in dipolar Bose-Einstein condensates, Phys. Rev. A \textbf{78},
043614 (2008).

\bibitem{SKA-dipolar} L. E. Young-S., P. Muruganandam, and S. K. Adhikari,
Dynamics of quasi-one-dimensional bright and vortex solitons of a dipolar
Bose-Einstein condensate with repulsive atomic interaction, J. Phys. B
\textbf{44}, 101001 (2011).

\bibitem{SKA-dipolar2} S. K. Adhikari and P. Muruganandam, Two-dimensional
dipolar Bose-Einstein condensate bright and vortex solitons on
one-dimensional optical lattice, J. Phys. B \textbf{45}, 045301 (2012).

\bibitem{Goran} G. Gligori\'{c}, A. Maluckov, M. Stepi\'{c}, L. Had\v{z}%
ievski, and B. A. Malomed, Discrete vortex solitons in dipolar Bose-Einstein
condensates. J. Phys. B: At. Mol. Opt. Phys. \textbf{43}, 055303 (2010).

\bibitem{ICFO} O. V. Borovkova, Y. V. Kartashov, B. A. Malomed, and L.
Torner, %Algebraic bright and vortex solitons in defocusing media,
Opt. Lett. \textbf{36}, 3088-3090 (2011); O. V. Borovkova, Y. V. Kartashov,
L. Torner, and B. A. Malomed,
%Bright solitons from defocusing nonlinearities,
Phys. Rev. E \textbf{84}, 035602 (R) (2011).

\bibitem{further} Q. Tian, L. Wu, Y. Zhang, and J.-F. Zhang, Vortex solitons
in defocusing media with spatially inhomogeneous nonlinearity, Phys. Rev. E
\textbf{85}, 056603 (2012).

\bibitem{further2} Y. Wu, Q. Xie, H. Zhong, L. Wen, and W. Hai, Algebraic
bright and vortex solitons in self-defocusing media with spatially
inhomogeneous nonlinearity, Phys. Rev. A \textbf{87}, 055801 (2013).

\bibitem{further3} R. Driben, Y. V. Kartashov, B. A. Malomed, T. Meier, and
L. Torner, Soliton gyroscopes in media with spatially growing repulsive
nonlinearity, Phys. Rev. Lett. \textbf{112}, 020404 (2014).

\bibitem{further4} R. Driben, Y. Kartashov, B. A. Malomed, T. Meier, and L.
Torner, Three-dimensional hybrid vortex solitons, New J. Phys. \textbf{16},
063035 (2014).

\bibitem{further5} R. Driben, N. Dror, B. Malomed, and T. Meier, Multipoles
and vortex multiplets in multidimensional media with inhomogeneous
defocusing nonlinearity, New J. Phys. \textbf{17}, 083043 (2015).

\bibitem{further6} R. Driben, T. Meier, and B. A. Malomed, Creation of
vortices by torque in multidimensional media with inhomogeneous defocusing
nonlinearity, Sci. Rep. \textbf{5}, 9420 (2015).

\bibitem{further7} N. Dror and B. A. Malomed, Solitons and vortices in
nonlinear potential wells, J. Optics \textbf{16}, 014003 (2016).

\bibitem{discrvort} B. A. Malomed and P. G. Kevrekidis, Discrete vortex
solitons, Phys. Rev. E \textbf{64}, 026601 (2001).

\bibitem{Pelin} D. E. Pelinovsky, P. G. Kevrekidis, and D. J. Frantzeskakis,
Physica D \textbf{212}, 20-53 (2005).

\bibitem{Thaw} T. Mayteevarunyoo, B. A. Malomed, B. B. Baizakov, and M.
Salerno, Matter-wave vortices and solitons in anisotropic optical lattices,
Physica D \textbf{238}, 1439-1448 (2009).

\bibitem{discrvort1} D. N. Neshev, T. J. Alexander, E. A. Ostrovskaya, Y. S.
Kivshar, H. Martin, I. Makasyuk, Z. G. Chen, Observation of discrete vortex
solitons in optically induced photonic lattices, Phys. Rev. Lett. \textbf{92}%
, 123903 (2004).

\bibitem{discrvort2} J. W. Fleischer, G. Bartal, O. Cohen, O. Manela, M.
Segev, J. Hudock, D. N. Christodoulides, Observation of vortex-ring
``discrete" solitons in 2D photonic lattices, Phys. Rev. Lett. \textbf{92},
123904 (2004).

\bibitem{discrvort3} B. Terhalle, T. Richter, K. J. H. Law, D. G\"{o}ries,
P. Rose, T. J. Alexander, P. G. Kevrekidis, A. S. Desyatnikov, W.
Krolikowski, F. Kaiser, C. Denz, and Y. S. Kivshar, Observation of
double-charge discrete vortex solitons in hexagonal photonic lattices, Phys.
Rev. \textbf{79}, 043821 (2009).

\bibitem{Dum-diss} D. Mihalache, Three-dimensional dissipative optical
solitons, Cent. Eur. J. Phys. \textbf{6}, 582-587 (2008).

\bibitem{NNR1} N. A. Veretenov, N. N. Rosanov, and S. V. Fedorov, Rotating
and precessing dissipative-optical-topological-3D solitons, Phys. Rev. Lett.
\textbf{117}, 183901 (2016).

\bibitem{NNR2} N. A. Veretenov, S. V. Fedorov, and N. N. Rosanov,
Topological vortex and knotted dissipative optical 3D solitons generated by
2D vortex solitons, Phys. Rev. Lett. \textbf{119}, 263901 (2017).

\bibitem{NNR3} S. V. Fedorov, N. A. Veretenov, and N. N. Rosanov,
Irreversible hysteresis of internal structure of tangle dissipative optical
solitons, Phys. Rev. Lett. \textbf{122}, 023903 (2019).

\bibitem{DimaHS} H. Sakaguchi, B. A. Malomed, and D. V. Skryabin, Spin-orbit
coupling and nonlinear modes of the polariton condensate in a harmonic trap,
New J. Phys. \textbf{19}, 08503 (2017).

\bibitem{DimaThaw2} T. Mayteevarunyoo, B. A. Malomed, and D. V. Skryabin,
One- and two-dimensional modes in the complex Ginzburg-Landau equation with
a trapping potential, Opt. Exp. \textbf{26}, 8849-8865 (2018).

\bibitem{DimaThaw1} T. Mayteevarunyoo, B. Malomed, and D. Skryabin, Vortex
modes supported by spin-orbit coupling in a laser with saturable absorption,
New J. Phys. \textbf{20}, 113019 (2018).

\bibitem{Kramer} I. S. Aranson and L. Kramer, The world of the complex
Ginzburg-Landau equation, Rev. Mod. Phys. \textbf{74}, 99-143 (2002).

\bibitem{Lugiato} L. A. Lugiato, Transverse nonlinear optics --
Introductions and review, Chaos Sol. Fract. 4, 1251-1258 (1994).

\bibitem{NNR0} A. G. Vladimirov, S. V. Fedorov, N. A. Kaliteevskii, G. V.
Khodova, and N. N. Rosanov, Numerical investigation of laser localized
structures, J. Opt. B: Quantum Semiclass. Opt. \textbf{1}, 101-106 (1999).

\bibitem{Mandel} P. Mandel and M. Tlidi, Transverse dynamics in cavity
nonlinear optics (2000-2003), J. Opt. B -- Quant. Semicl. Opt. \textbf{6},
R60-R75 (2004).

\bibitem{Petvia} V. I. Petviashvili and A. M. Sergeev, Spiral solitons in
active media with excitation thresholds, Dokl. Akad. Nauk SSSR \textbf{276},
1380-1384 (1984) [Sov. Phys. Dokl. \textbf{29}, 493 (1984)].

\bibitem{Lucian} B. A. Malomed, L.-C. Crasovan, and D. Mihalache. Stability
of vortex solitons in the cubic-quintic model. Physica D \textbf{161},
187-201 (2002).

\bibitem{Skarka0} D. Mihalache, D. Mazilu, V. Skarka, B. A. Malomed, H.
Leblond, N. B. Aleksi\'{c}, and F. Lederer. Stable topological modes in
two-dimensional Ginzburg-Landau models with trapping potentials, Phys.\ Rev.
A \textbf{82}, 023813 (2010).

\bibitem{PhysicaDD} B. A. Malomed, Evolution of nonsoliton and
\textquotedblleft quasiclassical\textquotedblright\ wavetrains in nonlinear
Schr\"{o}dinger and Korteweg - de Vries equations with dissipative
perturbations, Physica D \textbf{29}, 155-172 (1987).

\bibitem{CGL-3Dvort} D. Mihalache, D. Mazilu, F. Lederer, Y. V. Kartashov,
L.-C. Crasovan, L. Torner, and B. A. Malomed, Stable vortex tori in the
three-dimensional cubic-quintic Ginzburg-Landau equation, Phys. Rev. Lett.
\textbf{97}, 073904 (2006).

\bibitem{ring} V. E. Lobanov, Y. V. Kartashov, V. A. Vysloukh, and L.
Torner, Stable radially symmetric and azimuthally modulated vortex solitons
supported by localized gain, Opt. Lett. \textbf{36}, 85-87 (2011).

\bibitem{Amo} V. G. Sala, D. D. Solnyshkov, I. Carusotto, T. Jacqmin, A. Lema%
\^{\i}tre, H. Ter\c{c}as, A. Nalitov, M. Abbarchi, E. Galopin, I. Sagnes, J.
Bloch, G. Malpuech, and A. Amo, Spin-orbit coupling for photons and
polaritons in microstructures, Phys. Rev. X \textbf{5}, 011034 (2015).

\bibitem{Skarka1} V. Skarka, N. B. Aleksi\'{c}, H. Leblond, B. A. Malomed,
and D. Mihalache, Varieties of stable vortical solitons in Ginzburg-Landau
media with radially inhomogeneous losses, Phys. Rev. Lett. \textbf{105},
213901 (2010).

\bibitem{Skarka2} V. Skarka, N. B. Aleksi\'{c}, M. Leki\'{c}, B. N. Aleksi%
\'{c}, B. A. Malomed, D. Mihalache, and H. Leblond, Formation of complex
two-dimensional dissipative solitons via spontaneous symmetry breaking,
Phys. Rev. A \textbf{90}, 023845 (2014).

\bibitem{PT1} A. Ruschhaupt, F. Delgado, and J. G. Muga, Physical
realization of $\mathcal{PT}$-symmetric potential scattering in a planar
slab waveguide, J. Phys. A:\ Math. Gen. 38, L171-L176 (2005).

\bibitem{PT2} R. El-Ganainy, K. G. Makris, D. N. Christodoulides, and Z. H.
Musslimani, Theory of coupled optical $\mathcal{PT}$-symmetric structures,
Opt. Lett. \textbf{32}, 2632-2634 (2007).

\bibitem{PTsol1} V. V. Konotop, J. Yang, and D. A. Zezyulin, Nonlinear waves
in $\mathcal{PT}$-symmetric systems, Rev. Mod. Phys. \textbf{88}, 035002
(2016).

\bibitem{PTsol2} S. V. Suchkov, A. A. Sukhorukov, J. H. Huang, S. V.
Dmitriev, C. Lee, and Y. S. Kivshar, Nonlinear switching and solitons in $%
\mathcal{PT}$-symmetric photonic systems, Laser Photonics Rev. \textbf{10},
177-213 (2016).

\bibitem{Optica} Y. V. Kartashov, C. Hang, G. X. Huang, and L. Torner,
Three-dimensional topological solitons in $\mathcal{PT}$-symmetric optical
lattices, Optica \textbf{3}, 1048-1055 (2016).

\bibitem{Luz} E. Luz, V. Lutsky, E. Granot, and B. A. Malomed, Robust $%
\mathcal{PT}$ symmetry of two-dimensional fundamental and vortex solitons
supported by spatially modulated nonlinearity, Sci. Rep. \textbf{9}, 4483
(2019).

\bibitem{Panos-PTvort} V. Achilleos, P. G. Kevrekidis, D. J. Frantzeskakis,
and R. Carretero-Gonz\'{a}lez, Dark solitons and vortices in $\mathcal{PT}$%
-symmetric nonlinear media: From spontaneous symmetry breaking to nonlinear $%
\mathcal{PT}$ phase transitions, Phys. Rev. A \textbf{86}, 013808 (2012).

\bibitem{Jena-vortex} F. Eilenberger, K. Prater, S. Minardi, R. Geiss, U. R%
\"{o}pke, J. Kobelke, K. Schuster, H. Bartelt, S. Nolte, A. T\"{u}nnermann,
and T. Pertsch, Observation of discrete, vortex light bullets, Phys. Rev. X
\textbf{3}, 041031 (2013).

\bibitem{Cid3} A. S. Reyna, G. Boudebs, B. A. Malomed, and C. B. de Ara\'{u}%
jo, Robust self-trapping of vortex beams in a saturable optical medium,
Phys. Rev. A \textbf{93}, 013840 (2016).
\end{thebibliography}
\end{document}